\documentclass[aps,floatfix,prd,nofootinbib,superscriptaddress,reprint,showpacs,10pt,preprintnumbers,longbibliography]{revtex4-2}

\usepackage[T1]{fontenc}
\usepackage[utf8]{inputenc}
\usepackage[pdftex]{graphicx}
\usepackage[USenglish]{babel}
\usepackage{float}
\usepackage{amsmath}
\usepackage{amssymb}
\usepackage{mathtools}
\usepackage{physics}
\usepackage{nicefrac}
\usepackage{bbold}
\usepackage{braket}
\usepackage{feynmp-auto}
\usepackage[binary-units=true]{siunitx}
\usepackage{amsfonts}
\usepackage{dsfont}
\usepackage{array}
\usepackage{bm}
\usepackage{mathrsfs}
\usepackage{pifont}
\usepackage{multirow}
\usepackage{upgreek}
\usepackage[dvipsnames]{xcolor}
\usepackage{verbatim}
\usepackage{slashed}
\usepackage{ucs}
\usepackage{color}
\usepackage{subfigure}
\usepackage{tikz}
\usepackage{xspace}
\usepackage[pdftex,
  pdftitle={Simulating both parity sectors of the Hubbard Model with Tensor Networks},
  pdfauthor={Manuel Schneider, Johann Ostmeyer, Karl Jansen, Thomas Luu, Carsten Urbach},
  bookmarks,
  colorlinks,
  linkcolor=myblue,
  citecolor=mymagenta,
  menucolor=black,
  urlcolor=myblue,
  plainpages=false,
  pdfpagelabels,
  hypertexnames=false]{hyperref}
\usepackage{cleveref}
\usepackage{oplotsymbl}

\graphicspath{{figures/},{inkscape/},{plots/},{tikz/}}

\DeclareMathOperator{\ord}{\mathcal{O}}

\newcommand{\pdagger}{{\phantom{\dagger}}}
\newcommand{\matr}[2]{\left(\begin{array}{#1}#2\end{array}\right)}

\newcommand{\eto}[1]{\ensuremath{\mathrm{e}^{#1}}}

\newcommand{\ordnung}[1]{\ensuremath{\ord\left(#1\right)}}
\newcommand{\erwartung}[1]{\ensuremath{\left\langle#1\right\rangle}}

\newcommand{\tn}{tensor network\xspace}

\newcommand{\env}{\ensuremath{N_\text{red}}\xspace}
\newcommand{\renorm}{\ensuremath{\lambda}\xspace}
\newcommand{\link}{\ensuremath{k}\xspace}

\definecolor{mymagenta}{RGB}{200, 0, 100}
\definecolor{myblue}{RGB}{45, 48, 146}

\usepackage{ifthen} 
\usepackage{intcalc} 
\usepackage[skins]{tcolorbox}
\usepackage{pgfplots} 
\usepackage{grffile}  

\usetikzlibrary{external}

\usetikzlibrary{arrows,backgrounds,fit,calc,patterns,shapes}

\usetikzlibrary{decorations.pathmorphing} 
\usetikzlibrary{intersections} 
\usetikzlibrary{positioning}	

\usetikzlibrary{plotmarks}
\pgfplotsset{compat=newest}
\usetikzlibrary{plotmarks}
\usetikzlibrary{arrows.meta}
\usepgfplotslibrary{patchplots}

\tikzset{every node/.style={sloped,allow upside down},baseline={([yshift=+0ex]current bounding box.center)},inner sep=.3mm,x=1cm,y=1cm}

\def\pepsWidth{4mm}

\def\xDist{13mm} 
\def\yDist{10mm}
\def\xExt{\dimexpr \xDist-\pepsWidth /2 \relax} 
\def\yExt{\dimexpr \yDist-\pepsWidth /2 \relax}

\def\singularValueWidth{3mm}
\def\swapWidth{1.5mm}
\def\swapHeight{\swapWidth}
\def\gateWidth{\dimexpr \xDist+\pepsWidth \relax}
\def\gateHeight{\pepsWidth}

\def\xWiggly{\dimexpr \xDist/2 \relax} 
\def\physicalLength{\dimexpr \xDist/2-\gateHeight/2-\pepsWidth/2 \relax}

\tikzset{every picture/.style={node distance=\yDist and \xDist}}

\definecolor{tensorblue}{rgb}{0.8,0.8,1}
\definecolor{tensorred}{rgb}{1,0.5,0.5}
\definecolor{tensorpurple}{rgb}{1,0.5,1}

\tikzset{tenop/.style={fill=orange!30}}
\tikzset{tenk/.style={fill=green!50!black!50}}
\tikzset{tenleft/.style={fill=yellow!50}}
\tikzset{tenright/.style={fill=violet!50}}
\tikzset{tenlr/.style={fill=tensorpurple}}
\tikzset{teneff/.style={fill=tensorpurple}}
\tikzset{tentrans/.style={fill=black!20}}
\tikzset{tens/.style={fill=black!30}}

\tikzset{tenred/.style={fill=tensorred}}
\tikzset{tengreen/.style={fill=green!50!black!50}}
\tikzset{tenpurple/.style={fill=tensorpurple}}
\tikzset{tengrey/.style={fill=black!20}}
\tikzset{tenwhite/.style={fill=black!0}}
\tikzset{tenorange/.style={fill=orange!30}}
\tikzset{u/.style={fill=blue!20,draw=black}}
\tikzset{w/.style={fill=green!50!black!80,draw=black}}

\tikzset{ket/.style={circle,draw=black,thick,fill=tensorblue,minimum width=\pepsWidth}}
\tikzset{bra/.style={ket,tens,label=center:{$*$}}}
\tikzset{ketNew/.style={ket,tengreen}}
\tikzset{braNew/.style={bra,tengreen}}
\tikzset{nodeInvisible/.style={circle,draw=none,thick,minimum width=\pepsWidth}}
\tikzset{bMPS/.style={ket,tengrey}}
\tikzset{singularValue/.style={midway,circle,draw=black,thick,tenorange,minimum width=\singularValueWidth}}
\tikzset{swap/.style={diamond,draw=black,thick,fill=black,minimum width=\swapWidth,minimum height=\swapHeight}}
\tikzset{gate/.style={rectangle,draw=black,thick,fill=tensorpurple,minimum width=\gateWidth,minimum height=\gateHeight}}
\tikzset{env/.style={bMPS,rectangle,draw=black,thick,minimum height=\gateHeight}}

\tikzset{wiggly/.style={decorate, decoration={snake, segment length=\pepsWidth/2, amplitude=\pepsWidth/8}}}
\tikzset{physical/.style={densely dashed}}
\tikzset{parity/.style={dotted,very thick}}
\tikzset{boundary/.style={very thick}}

\newcommand{\rightOf}[1]{([shift=({\xDist,0})]#1.center)}

\newcommand{\belowOf}[1]{([shift=({0,-\yDist})]#1.center)}

\newcommand{\distL}[1]{([shift=({-\xExt,0})]#1.center)}
\newcommand{\distR}[1]{([shift=({\xExt,0})]#1.center)}
\newcommand{\distU}[1]{([shift=({0,\yExt})]#1.center)}
\newcommand{\distD}[1]{([shift=({0,-\yExt})]#1.center)}

\newcommand{\distLD}[1]{([shift=({-\xExt,-\yExt})]#1.center)}
\newcommand{\distRU}[1]{([shift=({\xExt,\yExt})]#1.center)}
\newcommand{\distRD}[1]{([shift=({\xExt,-\yExt})]#1.center)}

\newcommand{\lineL}[1]{(#1) -- \distL{#1}}
\newcommand{\lineR}[1]{(#1) -- \distR{#1}}
\newcommand{\lineU}[1]{(#1) -- \distU{#1}}
\newcommand{\lineD}[1]{(#1) -- \distD{#1}}

\newcommand{\lineLD}[1]{(#1) -- \distLD{#1}}

\newcommand{\lineRD}[1]{(#1) -- \distRD{#1}}

\newcommand{\connectD}[2]{(#1) -- (#1|-#2.north)}
\newcommand{\connectU}[2]{(#1) -- (#1|-#2.south)}

\newcommand{\gateLD}[1]{([shift=({-0.5*\xDist,-\yExt})]#1.center) -- ([shift=({-0.5*\xDist,0})]#1.center |- #1.south)}
\newcommand{\gateRD}[1]{([shift=({0.5*\xDist,-\yExt})]#1.center) -- ([shift=({0.5*\xDist,0})]#1.center |- #1.south)}

\newcommand{\connectBezier}[3]{let \p1 = (#1), \p2 = (#2) in (\p1) ..controls (\x1+#3,\y1) and (\x1+#3,\y2)  .. (\x1,\y2) -- (\p2)}

\begin{document}
\title{Simulating both parity sectors of the Hubbard Model with Tensor Networks}

\author{Manuel Schneider}
\affiliation{NIC, DESY Zeuthen, Platanenallee 6, 15738 Zeuthen, Germany}
\affiliation{Institut für Physik, Humboldt-Universität zu Berlin, Newtonstraße 15, 12489 Berlin, Germany}

\author{Johann Ostmeyer}
\affiliation{Helmholtz-Institut f\"ur Strahlen- und Kernphysik, University of Bonn, Nussallee 14-16, 53115 Bonn, Germany}
\affiliation{Bethe Center for Theoretical Physics, University of Bonn, Nussallee 12, 53115 Bonn, Germany}

\author{Karl Jansen}
\affiliation{NIC, DESY Zeuthen, Platanenallee 6, 15738 Zeuthen, Germany}

\author{Thomas Luu}
\affiliation{Helmholtz-Institut f\"ur Strahlen- und Kernphysik, University of Bonn, Nussallee 14-16, 53115 Bonn, Germany}
\affiliation{Institute for Advanced Simulation, Forschungszentrum J\"{u}lich, 54245 J\"{u}lich Germany}
\affiliation{Institut f\"{u}r Kernphysik, Forschungszentrum J\"{u}lich, 54245 J\"{u}lich Germany}

\author{Carsten Urbach}
\affiliation{Helmholtz-Institut f\"ur Strahlen- und Kernphysik, University of Bonn, Nussallee 14-16, 53115 Bonn, Germany}
\affiliation{Bethe Center for Theoretical Physics, University of Bonn, Nussallee 12, 53115 Bonn, Germany}


\date{\today}

\begin{abstract}
	Tensor networks are a powerful tool to simulate a variety of different physical models, including those that suffer from the sign problem in Monte Carlo simulations. The Hubbard model on the honeycomb lattice with non-zero chemical potential is one such problem. Our method is based on projected entangled pair states (PEPS) using imaginary time evolution. We demonstrate that it provides accurate estimators for the ground state of the model, including cases where Monte Carlo simulations fail miserably. In particular it shows near to optimal, that is linear, scaling in lattice size.
	We also present a novel approach to directly simulate the subspace with an odd number of fermions. It allows to independently determine the ground state in both sectors. Without a chemical potential this corresponds to half filling and the lowest energy state with one additional electron or hole. We identify several stability issues, such as degenerate ground states and large single particle gaps, and provide possible fixes.

\end{abstract}

\maketitle

\section{Introduction}
Projected Entangled Pair States (PEPS) are a higher dimensional generalization of Matrix Product States (MPS) \cite{tensor_intro} and have successfully been used to approximate various physical systems. 
Examples include various 2D classical and quantum spin models \cite{Verstraete:2006mgt,Poilblanc:2017dfm}, spin liquids \cite{Mambrini:2016oxl,Chen:2018kfj}, as well as select lattice non-abelian gauge theories \cite{PhysRevX.4.041024, ZOHAR201684}.  
Though originally developed for bosonic systems, calculations nowadays readily use fermionic PEPS \cite{tensor_fermions} to investigate systems with fermionic degrees of freedom.

As the method is variational and deterministic, calculations using tensor networks such as PEPS do not suffer from numerical issues seen in stochastic simulations when dealing with systems with non-zero chemical potential $\mu$.  
In such cases an induced numerical sign problem can completely preclude a stochastic simulation, while tensor networks instead are completely immune and provide results equally precise as compared to cases when $\mu=0$.
Furthermore, tensor network calculations of global ground state energies of various systems are achieving remarkable accuracy~\cite{vanhecke2021entanglement}.
With continual advancements in algorithmic efficiency,  tensor networks are fast becoming the `got-to' method for extracting the global ground state properties of low-dimensional systems.

The situation is less clear, however, when dealing with excited states, or ground state energies in sectors of opposite parity to that of the global ground state.  
To date tensor networks could only access a few low lying energy states by calculating the ground state first and projecting to an orthogonal subspace in the following iterations~\cite{schwinger_spectrum}.
This is unfortunate, as an understanding of the spectrum of states \emph{relative} to the global ground state provides information on novel forms of collective behavior of the system, such as spin and antiferromagnetic correlations \cite{Parsons1253}, or induced Mott transitions \cite{RevModPhys.78.17}, to name a few examples. 
A quintessential example in this regard is the Hubbard model in two-dimensions~\cite{Tasaki_1998,arovas2021hubbard}.
On an infinite honeycomb, or hexagonal lattice, for example, the Hubbard model exhibits a quantum phase transition from a semi-metal to insulator state at a particular critical coupling~\cite{Assaad:2013xua,Buividovich:2018crq}.  
Such a transition can be ascertained by comparing the ground state energy of the odd parity sector to that of the even parity sector, as was done in~\cite{Ostmeyer:2020uov}.
Numerical investigations of the Hubbard model have mostly been confined to chemical potential $\mu=0$ so as to avoid the sign problem, which means that much of its phase diagram is not known.
Thus, were tensor networks able to obtain information of odd parity states with equal precision as obtained with the global ground state and with arbitrary $\mu$, a new era in simulations of non-perturbative phenomena would occur for low-dimensional systems.

In this paper we build off the formalism presented in~\cite{tensor_fermions} and apply it to the odd parity sector of the hexagonal Hubbard model as a test case. Though tensor networks have been applied to the Hubbard model using infinite PEPS (iPEPS)~\cite{iPEPS_hubbard_2016}, such investigations analyzed the global ground state of the Hubbard model on a the square lattice.  In our case, in addition to considering the hexagonal lattice which has important applications in the study of graphene and its derivatives~\cite{Novoselov2007,CastroNeto:2009zz}, we extract the lowest energy state in the odd parity sector of this system, giving us information about the single-particle gap.  We perform these calculations for select values of onsite Hubbard term $U$ and chemical potential $\mu$, benchmarking our results to exact results where possible.  For systems too large for comparison with exact calculations, our results represent predictions.

Our method for simulating in opposite parity sectors is not a complete panacea for tensor network calculations of excited states.  Indeed, we encounter various numerical difficulties and limitations which we carefully document and where possible, address.  Still, our findings provide proof of principle that excited states can be readily extracted using tensor networks, and we anticipate continued development along the lines of our research.

Our paper is organized as follows: in the following section we give a cursory overview of the PEPS formalism and its connection to imaginary time evolution.  We present our algorithm for fermions that allows us to simulate in opposite parity sectors.  We discuss how we implement various update schemes and demonstrate with hard examples the expected scaling of our algorithm with system size, bond dimension, and boundary MPS dimension.  We also introduce the Hubbard Hamiltonian in this section and explain how it is adapted to our method.  In \cref{sect:improvements} we enumerate our various improvements to our algorithm to accelerate convergence as well as increase precision.  We take pains to enumerate the limitations of our procedure so as to present to the reader a clear and honest assessment of our method's efficacy.  Where possible we provide explanations for the sources of these limitations and potential resolutions.  We then show our results for the Hubbard model in \cref{sect:results}.  Our results demonstrably show the ability of our method to simulate both parity sectors of the problem, obtaining precise results for both ground states in both cases and therefore gaining information about the single-particle gap.  Finally we recapitulate in \cref{sect:conclusions}.

%
\section{Formalism}

\subsection{PEPS}
A Projected Entangled Pair State (PEPS) \cite{PEPS_original_bMPS,PEPS_original_2} is an ansatz for a physical state. It is a versatile tool to incorporate different entanglement structures of two and higher dimensional systems and especially applicable for ground state simulations.

We want to study a system of $N$ lattice sites on a $(L_x \times L_y)$ grid\footnote{This corresponds to $(L_x/2 \times L_y)$ unit cells of the honeycomb lattice, up to boundary effects.}, where the local dimension of the Hilbert Space is $d$. Representing a state would therefore require $d^N$ complex numbers, which is not feasible to store on a computer system for larger lattices. Therefore, a truncation of the Hilbert Space is needed. A PEPS reduces the number of parameters drastically while still giving a good approximation of specific states. The truncation corresponds to a cutoff in the entanglement entropy. It was shown that ground states of local Hamiltonians with an energy gap fulfill an area law and can therefore be well approximated by tensor networks \cite{areaLaw_original}. This holds in one spacial dimension, but tensor networks and PEPS in particular were also applied successfully in higher dimensions \cite{areaLaw_overview}.

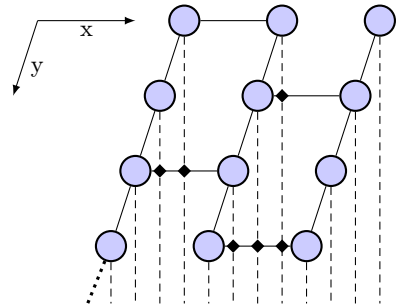
\begin{figure}[ht]
	\centering
	\begin{tikzpicture}
	\def\Lx{3}
	\def\Ly{4}
	\def\xStep{\xDist}
	\def\yStep{-\yDist}
	\def\physLength{-\yExt}
	\def\xShift{\xStep / \Ly} 
	\node (origin) at (- \xShift*3, \yStep) {};
	\draw[-latex] (origin.center) -- +(\xStep, 0) node[midway,label=below:x] {};
	\draw[-latex] (origin.center) -- +(-\xShift, \yStep) node[midway,label=above:y] {};
	\foreach \x in {1,...,\Lx}
		\foreach \y in {1,...,\Ly}
			{
				\node[ket] (ten_\x_\y) at (\xStep*\x - \xShift * \y, \yStep*\y) {};
				\node (physEnd_\x_\y) at (\xStep*\x - \xShift * \y, {\yStep*\Ly + \physLength}) {};
			}
		\draw[parity] (ten_1_\Ly) -- (-\xShift,{\yStep*\Ly + \physLength});
		\foreach \x in {1,...,\Lx}
			\foreach \y in {1,...,\Ly}
			{
				\draw[physical,name path={phys_\x_\y}] (ten_\x_\y) -- (physEnd_\x_\y);
			}
		\foreach \x in {2,...,\Lx}
			\foreach \y in {1,...,\Ly}
			{
				\ifthenelse{\intcalcMod{\x+\y}{2}=0}{
				}{
					\pgfmathtruncatemacro\lastx{\x-1}
					\pgfmathtruncatemacro\lasty{\y-1}
					\draw[name path={xlink_\lastx_\y}] (ten_\lastx_\y) -- (ten_\x_\y);
					\ifthenelse{\y>1}{ 
						\foreach \physy in {1,...,\lasty}
						{
							\path [name intersections={of={xlink_\lastx_\y} and {phys_\lastx_\physy},by={swap_\lastx_\y_\physy}}];
							\node[swap]  at ({swap_\lastx_\y_\physy}) {};
						}
					}
				}
			}
		\foreach \x in {1,...,\Lx}
			\foreach \y in {2,...,\Ly}
			{
				\pgfmathtruncatemacro\lasty{\y-1}
				\draw (ten_\x_\lasty) -- (ten_\x_\y);
			}
\end{tikzpicture}
	\caption{Ket state of a PEPS for a 3x4 fermionic honeycomb lattice.  Description of symbology (see text for more details) -- circles: PEPS tensors; dashed lines: physical indices; solid lines: internal indices; dotted line: parity index; diamonds: swap gates.}
	\label{fig:peps}
\end{figure}

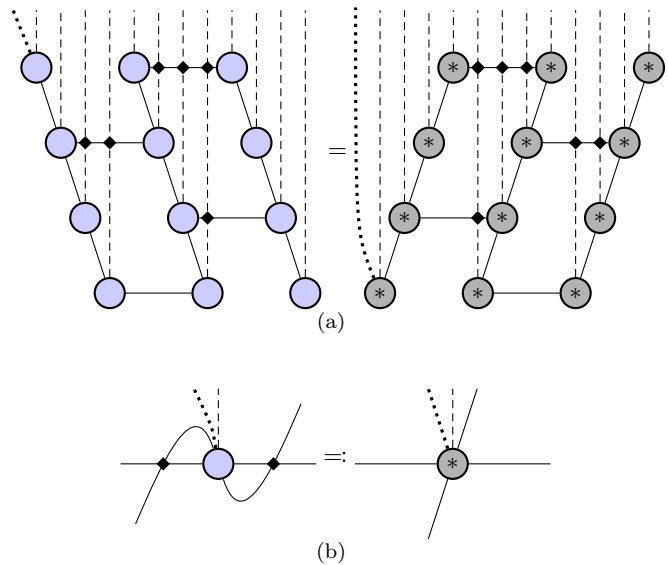
\begin{figure}[ht]
	\centering
	\subfigure[\label{fig:peps_bra_network}]{
		\begin{minipage}[c]{\columnwidth}
			\begin{equation*}
	\begin{tikzpicture}
		\def\Lx{3}
		\def\Ly{4}
		\def\xStep{\xDist}
		\def\yStep{\yDist}
		\def\physLength{\yExt}
		\def\xShift{\xStep / \Ly} 
		\foreach \x in {1,...,\Lx}
			\foreach \y in {1,...,\Ly}
				{
					\node[ket] (ten_\x_\y) at (\xStep*\x - \xShift * \y, \yStep*\y) {};
					\node (physEnd_\x_\y) at (\xStep*\x - \xShift * \y, {\yStep*\Ly + \physLength}) {};
				}
			\draw[parity] (ten_1_\Ly) -- (-\xShift,{\yStep*\Ly + \physLength});
			\foreach \x in {1,...,\Lx}
				\foreach \y in {1,...,\Ly}
				{
					\draw[physical,name path={phys_\x_\y}] (ten_\x_\y) -- (physEnd_\x_\y);
				}
			\foreach \x in {2,...,\Lx}
				\foreach \y in {1,...,\Ly}
				{
					\ifthenelse{\intcalcMod{\x+\y}{2}=0}{
					}{
						\pgfmathtruncatemacro\lastx{\x-1}
						\pgfmathtruncatemacro\lasty{\y-1}
						\draw[name path={xlink_\lastx_\y}] (ten_\lastx_\y) -- (ten_\x_\y);
						\ifthenelse{\y>1}{ 
							\foreach \physy in {1,...,\lasty}
							{
								\path [name intersections={of={xlink_\lastx_\y} and {phys_\lastx_\physy},by={swap_\lastx_\y_\physy}}];
								\node[swap]  at ({swap_\lastx_\y_\physy}) {};
							}
						}
					}
				}
			\foreach \x in {1,...,\Lx}
				\foreach \y in {2,...,\Ly}
				{
					\pgfmathtruncatemacro\lasty{\y-1}
					\draw (ten_\x_\lasty) -- (ten_\x_\y);
				}
	\end{tikzpicture}
	=
	\begin{tikzpicture}
		\def\Lx{3}
		\def\Ly{4}
		\def\xStep{\xDist}
		\def\yStep{-\yDist}
		\def\physLength{\yExt}
		\def\xShift{\xStep / \Ly} 
		\foreach \x in {1,...,\Lx}
		\foreach \y in {1,...,\Ly}
		{
			\node[bra] (ten_\x_\y) at (\xStep*\x - \xShift * \y, \yStep*\y) {};
			\node (physEnd_\x_\y) at (\xStep*\x - \xShift * \y, {\yStep*1 + \physLength}) {};
		}
		\draw[parity] (ten_1_\Ly) .. controls (-\xShift,{\yStep*\Ly + \physLength}) .. (-\xShift,{\yStep*1 + \physLength});
		\foreach \x in {1,...,\Lx}
		\foreach \y in {1,...,\Ly}
		{
			\draw[physical,name path={phys_\x_\y}] (ten_\x_\y) -- (physEnd_\x_\y);
		}
		\foreach \x in {2,...,\Lx}
		\foreach \y in {1,...,\Ly}
		{
			\ifthenelse{\intcalcMod{\x+\y}{2}=0}{
			}{
				\pgfmathtruncatemacro\lastx{\x-1}
				\pgfmathtruncatemacro\lasty{\y-1}
				\draw[name path={xlink_\lastx_\y}] (ten_\lastx_\y) -- (ten_\x_\y);
				\ifthenelse{\y<\Ly}{ 
					\foreach \physy in {2,...,\Ly}
					{
						\path [name intersections={of={xlink_\lastx_\y} and {phys_\x_\physy},by={swap_\x_\y_\physy}}];
						\node[swap]  at ({swap_\x_\y_\physy}) {};
					}
				}
			}
		}
		\foreach \x in {1,...,\Lx}
		\foreach \y in {2,...,\Ly}
		{
			\pgfmathtruncatemacro\lasty{\y-1}
			\draw (ten_\x_\lasty) -- (ten_\x_\y);
		}
	\end{tikzpicture}
\end{equation*}
		\end{minipage}
	}
	\subfigure[	\label{fig:bra_def}]{
		\begin{minipage}[c]{\columnwidth}
			\begin{equation*}
	\begin{tikzpicture}
		\def\Ly{4}
		\def\xStep{\xDist}
		\def\yStep{-\yDist}
		\def\physLength{\yExt}
		\def\xShift{\xStep / \Ly} 
		\node[ket] (ten) at (0, 0) {};
		\draw[name path=linkR] (ten) -- ( \xStep,0) node[midway] (midwayR) {};
		\draw[name path=linkL] (ten) -- (-\xStep,0) node[midway] (midwayL) {};
		\draw[name path=linkD] (ten) .. controls (-\xShift,-\yStep) and (midwayL) .. \distLD{ten};
		\path [name intersections={of=linkL and linkD,by=swapLD}];
		\node[swap]  at (swapLD) {};
		\draw[name path=linkU] (ten) .. controls (\xShift,\yStep) and (midwayR) .. \distRU{ten};
		\path [name intersections={of=linkR and linkU,by=swapRU}];
		\node[swap]  at (swapRU) {};
		\draw[physical] (ten) -- (0,-\yStep);
		\draw[parity] (ten) .. controls (-\xShift/4,-\yStep/2) .. (-\xShift,-\yStep);
	\end{tikzpicture}
	\eqqcolon
	\begin{tikzpicture}
		\def\Ly{4}
		\def\xStep{\xDist}
		\def\yStep{-\yDist}
		\def\physLength{\yExt}
		\def\xShift{\xStep / \Ly} 
		\node[bra] (ten) at (0, 0) {};
		\draw (ten) -- ( \xStep,0);
		\draw (ten) -- (-\xStep,0);
		\draw (ten) -- ( \xShift,-\yStep);
		\draw (ten) -- (-\xShift,\yStep);
		\draw[physical] (ten) -- (0,-\yStep);
		\draw[parity] (ten) -- (-\xShift,-\yStep);
	\end{tikzpicture}
\end{equation*}
		\end{minipage}
	}
	\caption{Bra state of a fermionic PEPS. Symbology -- circles without stars: conjugated PEPS tensors; dashed lines: physical indices; solid lines: internal indices; dotted line: parity index; diamonds: swap gates.
	\subref{fig:peps_bra_network} Bra tensors for a 3x4 honeycomb lattice. \subref{fig:bra_def} Definition of starred tensors with four internal indices. Some indices and the corresponding swap gates can be removed to obtain the honeycomb tensors.}
	\label{fig:peps_bra}
\end{figure}

To construct a PEPS, one places a tensor on each lattice site, pictured as circles in \cref{fig:peps}. Each tensor carries a physical index running from 1 to $d$ (dashed lines).  Additional indices connect the tensor to its nearest neighbors with indices (links) running from 1 to $D_k$, where $k$ enumerates the link between the two nearest neighbors (solid lines). Lines connecting two tensors require summation (contraction) over the corresponding index, while open legs denote the open indices of the resulting tensor after contraction of the network. \Cref{fig:peps_bra} shows different projections of the bra state corresponding to the ket state in \cref{fig:peps}. The swap gates and the parity index in \cref{fig:peps,fig:peps_bra} will be explained in \cref{sec:fermions}.

In the limit of large $D_k$ this ansatz becomes exact. For larger systems, a truncation bond dimension $D$ is chosen and the indices are bound by $D_k \le D\; \forall k$. The accuracy of the ansatz can be improved by increasing $D$.

\subsection{Imaginary time evolution\label{sec:imaginaryTimeEvolution}}
Expressing the state in the form of a PEPS allows us to evolve the system in imaginary time to the ground state $\Ket{\Psi_0}$ of a given Hamiltonian $H$. We start with a (random) initial state $\Ket{\Psi(0)}$ and evolve it to find the ground state:
\begin{align}
	\Ket{\Psi_0} &= \lim\limits_{t\rightarrow\infty}\frac{\Ket{\Psi(t)}}{\sqrt{\Braket{\Psi(t)|\Psi(t)}}}\:,\quad\Ket{\Psi(t)} = \eto{-Ht}\Ket{\Psi(0)}\,.\label{eq:imag_time_evol}
\end{align}
The time evolution operator
\begin{equation}
	U(t)\coloneqq\eto{-Ht}=\left(\eto{-H\delta t}\right)^m
	\label{eq:U_trotterized}
\end{equation}
is decomposed into $m$ time slices of length $\delta t=t/m$. Every single time slice is again split into a product of local terms. We focus on a Hamiltonian that consists only of nearest neighbor interaction terms $H_i$:
\begin{align}
	H &= \sum_{i=1}^{\# H_i} H_i \label{eq:H_local}\\
	U(\delta t) &= \prod_i\eto{-H_i\delta t}+\ordnung{\delta t^2} \label{eq:trotter_error} \\
	U(\delta t) &= \prod_{i=1 \dots \# H_i}\eto{-H_i \frac{\delta t}{2}} \prod_{j=\# H_i \dots 1}\eto{-H_j \frac{\delta t}{2}} + \ordnung{\delta t^3}\,. \label{eq:trotter2}
\end{align}
Here we used the Suzuki-Trotter expansion \cite{SuzukiTrotter_original,SuzukiTrotter_MPS} of first order in \cref{eq:trotter_error} and of second order in \cref{eq:trotter2}. For our calculations we applied the second order expression, where the ordering of terms in the second product is inverse to the ordering in the first.

This way the time evolution operator $U(t)$, which acts \emph{globally}, can be decomposed into a sequence of local terms $\eto{-H_i\delta t}$. These can be efficiently applied to the PEPS, which we discuss in \cref{sec:truncation}.

\subsection{Modifications for fermions\label{sec:fermions}}
If we want to consider fermions in our model, we have to take into account their anti-commuting nature. We need to order the fermionic degrees of freedom and make sure that the sign flips once we exchange two fermionic creation or annihilation operators. In \cref{fig:peps} the two dimensional network is arranged in a way that we can number the physical indices from left to right. The state would be created by applying creation operators in the order of this numbering to the vacuum state. We can apply a nearest neighbor operator in the y-direction directly because the corresponding links have consecutive numbers. Since any observable has an even number of creation and annihilation operators, all sign flips come in pairs and cancel each other. If the creation and annihilation operators of an observable sit on sites with non-consecutive numbers however, we would have to take into account all indices with numbers in between. This happens for the application of a gate in x-direction in our numbering scheme. This would turn nearest-neighbor operators into non-local ones.  Fortunately we can avoid this issue by using the scheme developed in~\cite{tensor_fermions_derivation,tensor_fermions}, which we now discuss.

\subsubsection{Fermion swap gate}
We implement the fermionic anti-commutation relations at the PEPS level rather than at the operator level only. In order to do this we first introduce the concept of fermion number parity. An even number of fermions comes with parity $p=1$, an odd number with parity $p=-1$, i.e.\@ $p\left(\ket 0\right)=p\left(\ket {\uparrow\downarrow}\right)=1$ and $p\left(\ket \uparrow\right)=p\left(\ket\downarrow\right)=-1$. In addition we assign the parity
\begin{align}
	p=(\underbrace{1,\dots,1}_{D_e},\underbrace{-1,\dots,-1}_{D_o})
\end{align}
with $D_e+D_o=D_k$ to every internal tensor index $k$. Parity conservation requires fermions to be created and annihilated in pairs. We can ensure this on the level of the \tn by the rule
\begin{align}
	T_{i_1,i_2,\dots,i_r}&=0\qquad\text{if}\quad p\left(i_1\right)p\left(i_2\right)\cdots p\left(i_r\right)=-1\,.\label{eq:conserve_parity}
\end{align}
The crucial difference to bosonic {\tn}s however is the aforementioned anti-commutation of the creation operators. An exchange of two odd numbers of fermionic operators yields a minus sign. In the PEPS formalism this means that every crossing of two lines introduces additional signs. A minus sign occurs at every crossing for the indices with twice negative parity. In~\cite{tensor_fermions_derivation} this is achieved by introducing fermionic swap gates
\begin{align}
	X^{i_1i_2}_{j_1j_2} &= \delta_{i_1j_1}\delta_{i_2j_2}\,S\left(i_1,i_2\right)\,,\\
	S\left(i_1,i_2\right) &= \begin{cases}
	-1 & p\left(i_1\right)=p\left(i_1\right)=-1\\
	\phantom{-}1 & \text{else}
	\end{cases}
\end{align}
with the two incoming indices $i_1,i_2$ and the two outgoing indices $j_1,j_2$.

In our graphical notation, we place a swap gate on every line crossing. Swap gates are pictured as diamonds in our notation as used in \cref{fig:peps,fig:peps_bra}. We can move the lines freely and create new intersections with swap gates due to $S$ being self inverse. Because of \cref{eq:conserve_parity}, we can also move lines across nodes. When we want to apply nearest-neighbor gates, we can move the physical links to be next to each other. This introduces new swap gates in the network but allows the operators to be applied locally.

The swap gates were used in \cref{fig:peps_bra} to write a bra state in an alternative manner. Originally, the bra state can be obtained by reflecting the ket state on a line in the vertical direction and subsequently conjugating all tensors. This is shown as the left side of \cref{fig:peps_bra_network}. The right side arranges the ket state in the same way as the bra state. This can be realized by appropriately moving the tensors and lines, and introducing new swap gates. Some of the swap gates are included in the definition of the starred tensors according to \cref{fig:bra_def}.

In this manner we never have to contract the swap gate with a tensor explicitly. Instead we can always find a way to write the \tn so that the gate is contracted with two indices of the same tensor. This way the swap gate can be applied element wise to the tensor. More details can be found in \cref{app:swap}. The computational cost scales only with the number of elements of the corresponding tensor. Since this is negligible compared to the most demanding steps we describe later in \cref{sec:resource_scaling}, the fermionic algorithm has the same leading costs as the bosonic one.

\subsubsection{Parity index}
The previously constructed fermionic PEPS requires that all tensors have an overall even parity due to \cref{eq:conserve_parity}. This way, the tensor network can only represent states with even parity. To include the odd parity sector as well, one introduces an additional parity index to one of the tensors. This index can take two values corresponding to even or odd parity. We represent it by dotted lines (see e.g.\ \cref{fig:peps}). When calculating expectation values, the parity index of a bra state gets contracted with the corresponding one in the ket state.

We can also crop the parity index to only have one value with either even or odd parity. This way we simulate the even and odd parity sectors individually. For the Hubbard model, the state of half filling is of even parity. By choosing an odd parity, we can find the state with an additional electron (or hole).


\subsection{Truncation and expectation values\label{sec:truncation}}
To apply the imaginary time evolution described in \cref{sec:imaginaryTimeEvolution}, the nearest neighbor operators $\eto{-H_i\delta t}$ have to be applied to the PEPS sequentially. Since every application of a local operator increases the bond dimension on the corresponding link, a truncation has to be done in order to keep the bond dimension small. We use two different schemes for the truncation:  Simple Update \cite{SimpleUpdate_original,Lubasch_clusterUpdate,iPEPS_introduction,tensor_fermions} and Full Update \cite{iPEPS_original_2008,tensor_fermions,Lubasch_algorithms,iPEPS_introduction}. We limit ourselves to open boundary conditions (OBC) here, though we note that the Simple Update scheme can easily be applied to periodic boundary conditions as well. All results presented in this paper are obtained with Simple Update where not stated otherwise.

\subsubsection{Simple Update\label{sec:SimpleUpdate}}
Simple Update is a truncation method that only takes local properties into account. We use the simplified update, where triads (reduced tensors) are split from the PEPS tensors to reduce the numerical costs \cite{tensor_fermions,Lubasch_algorithms,iPEPS_introduction}.

\begin{figure}[ht]
	\centering
	\def\svgwidth{17\columnwidth}
	\subfigure[]{\begin{tikzpicture}
	\node[ket] (ten) at (0, 0) {$L$};
	\draw \lineL{ten} node[singularValue, label=45:$S_2$] {};
	\draw \lineR{ten};
	\draw \lineU{ten} node[singularValue, label=45:$S_1$] {};
	\draw \lineD{ten} node[singularValue, label=45:$S_3$] {};
	\draw[parity] \lineLD{ten};
	\draw[physical] \lineRD{ten};
\end{tikzpicture}
\begin{tikzpicture}
	\draw[-latex] (0,0) -- (5mm,0);
\end{tikzpicture}
\begin{tikzpicture}
	\node[ket,label=125:{$env_L$}] (envp) at (0, 0) {};
	\node[ket,label=45:{$triad_L$}] (triadp) at \rightOf{envp} {};
	\draw[wiggly] (envp) -- (triadp);
	\draw \lineL{envp};
	\draw \lineR{triadp};
	\draw \lineU{envp};
	\draw \lineD{envp};
	\draw[parity] \lineLD{envp};
	\draw[physical] \lineRD{triadp};
\end{tikzpicture}}
	\subfigure[]{\begin{tikzpicture}
	\node[ket] (ten) at (0, 0) {$R$};
	\draw \lineL{ten};
	\draw \lineR{ten} node[singularValue, label=45:$S_5$] {};
	\draw \lineU{ten} node[singularValue, label=-45:$S_4$] {};
	\draw \lineD{ten} node[singularValue, label=-90:$S_6$] (SD) {};
	\draw[name path=tenD] \lineD{SD} -- (SD) node[midway] (tenDanchor) {};
	\draw[physical,name path=tenP] (ten) .. controls \distRD{ten} and (tenDanchor) .. \distLD{SD};
	\path [name intersections={of=tenD and tenP,by=diamond}];
	\node[swap]  at (diamond) {};
\end{tikzpicture}
\begin{tikzpicture}
	\draw[-latex] (0,0) -- (5mm,0);
\end{tikzpicture}
\begin{tikzpicture}
	\node[ket,label=125:{$triad_R$}] (triadm) at (0, 0) {};
	\node[ket,label=45:{$env_R$}] (envm) at \rightOf{envp} {};
	\draw[wiggly] (triadm) -- (envm);
	\draw \lineL{triadm};
	\draw \lineR{envm};
	\draw \lineU{envm};
	\draw \lineD{envm};
	\draw[physical] \lineLD{triadm};
\end{tikzpicture}}
	\caption{Splitting tensors into triads $triad$ and the environments $env$ for an update in x-direction. Symbology -- small orange circles on indices: singular values; diamond: swap gate for fermionic PEPS.}
	\label{fig:triadSplittingX}
\end{figure}
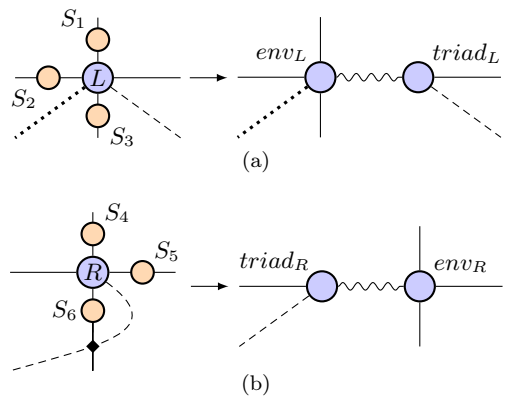

In a first step, each tensor on the updated link is split into a triad and an environment tensor. This is depicted in \cref{fig:triadSplittingX} for the x-direction, where the y-direction gets split similarly. The link to be updated gets combined with the physical index, while all other indices get combined to a second index. The resulting matrix is split in a QR-decomposition, such that the triangular matrix carries the physical index and the link to be updated. The newly emerged index connecting the two matrices of the QR-decomposition is denoted with wiggly lines in our graphical notation. We limit ourselves to updating the triads only as this is numerically cheaper. For fermionic PEPS we apply the swap gates to the initial tensors to put the physical indices next to each other. The two-site operator can then be directly applied to these adjacent indices.

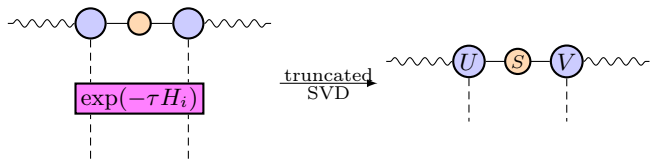
\begin{figure}[ht]
	\centering
	\def\svgwidth{17\columnwidth}
	\begin{tikzpicture}
	\node[ket] (triadp) at (0, 0) {};
	\node[ket] (triadm) at \rightOf{triadp} {};
	\draw (triadp) -- (triadm) node[singularValue] (S) {};
	\draw[wiggly] \lineL{triadp};
	\draw[wiggly] \lineR{triadm};
	\node[gate] (expGate) at \belowOf{S} {$\exp(-\tau H_i)$};
	\draw[physical] \connectD{triadp}{expGate};
	\draw[physical] \connectD{triadm}{expGate};
	\draw[physical] \gateLD{expGate};
	\draw[physical] \gateRD{expGate};
\end{tikzpicture}
\begin{tikzpicture}
	\draw[-latex] (0,0) -- (1.3cm,0) node[midway,above] {\scriptsize truncated} node[midway,below] {\scriptsize SVD};
\end{tikzpicture}
\begin{tikzpicture}
	\node[ket] (triadp) at (0, 0) {$U$};
	\node[ket] (triadm) at \rightOf{triadp} {$V$};
	\draw (triadp) -- (triadm) node[singularValue] (S) {\scriptsize $S$};
	\draw[wiggly] \lineL{triadp};
	\draw[wiggly] \lineR{triadm};
	\draw[physical] \lineD{triadp};
	\draw[physical] \lineD{triadm};
\end{tikzpicture}
	\caption{Simple Update truncation.}
	\label{fig:simpleUpdate}
\end{figure}

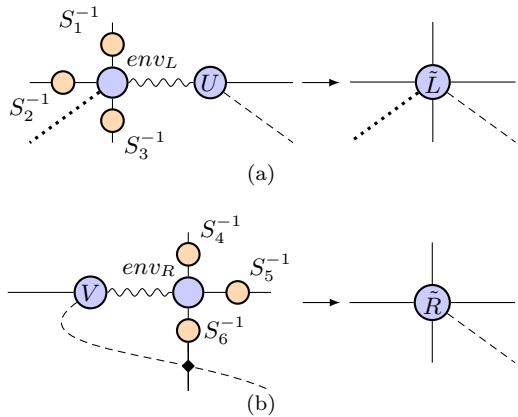
\begin{figure}[ht]
	\centering
	\def\svgwidth{17\columnwidth}
	\subfigure[]{\begin{tikzpicture}
	\node[ket,label=45:{$env_L$}] (envp) at (0, 0) {};
	\node[ket] (triadp) at \rightOf{envp} {$U$};
	\draw[wiggly] (envp) -- (triadp);
	\draw \lineL{envp} node[singularValue, label=45:$S_2^{-1}$] {};
	\draw \lineR{triadp};
	\draw \lineU{envp} node[singularValue, label=45:$S_1^{-1}$] {};
	\draw \lineD{envp} node[singularValue, label=45:$S_3^{-1}$] {};
	\draw[parity] \lineLD{envp};
	\draw[physical] \lineRD{triadp};
\end{tikzpicture}
\begin{tikzpicture}
	\draw[-latex] (0,0) -- (5mm,0);
\end{tikzpicture}
\begin{tikzpicture}
	\node[ket] (ten) at (0, 0) {$\tilde{L}$};
	\draw \lineL{ten};
	\draw \lineR{ten};
	\draw \lineU{ten};
	\draw \lineD{ten};
	\draw[parity] \lineLD{ten};
	\draw[physical] \lineRD{ten};
\end{tikzpicture}}
	\subfigure[]{\begin{tikzpicture}
	\node[ket,label] (triadm) at (0, 0) {$V$};
	\node[ket,label=125:{$env_R$}] (envm) at \rightOf{envp} {};
	\draw[wiggly] (triadm) -- (envm);
	\draw \lineL{triadm};
	\draw \lineR{envm} node[singularValue, label=45:$S_5^{-1}$] {};;
	\draw \lineU{envm} node[singularValue, label=-45:$S_4^{-1}$] {};;
	\draw \lineD{envm} node[singularValue, label=90:$S_6^{-1}$] (SD) {};;
	\draw[name path=envmD] \lineD{SD} -- (SD) node[pos=0.8] (envmDanchor) {};
	\draw[physical,name path=triadmP] (triadm) ..controls \distLD{triadm} and (envmDanchor) .. \distRD{SD};
	\path[name intersections={of=envmD and triadmP,by=diamond}];
	\node[swap]  at (diamond) {};
\end{tikzpicture}
\begin{tikzpicture}
	\draw[-latex] (0,0) -- (5mm,0);
\end{tikzpicture}
\begin{tikzpicture}
	\node[ket] (ten) at (0, 0) {$\tilde{R}$};
	\draw \lineL{ten};
	\draw \lineR{ten};
	\draw \lineU{ten};
	\draw \lineD{ten};
	\draw[physical] \lineRD{ten};
\end{tikzpicture}}
	\caption{Combining triads $U$, $V$ and the environments $env$ to the updated PEPS tensors $\tilde{L}$ and $\tilde{R}$.}
	\label{fig:triadCombiningX}
\end{figure}

The update is performed as depicted in \cref{fig:simpleUpdate}. We apply the local gate to the physical indices of the triads. The resulting network gets contracted and is then split by a Singular Value Decomposition (SVD) into two tensors $U$ and $V$ connected by a singular value matrix $S$. The singular values are truncated in order to keep the bond dimension bound by $D$. Finally, the square roots of the singular values can be multiplied to the left- and right-unitary tensors $U$ and $V$ to form the new triads. We keep the singular values explicitly on the links however, as shown in \cref{fig:triadSplittingX,fig:triadCombiningX} as small orange circles on the links. To increase the precision, the singular values of the surrounding links of a tensors are included in the initial QR-decomposition to calculate the triads and the environment, see \cref{fig:triadSplittingX}. This way more information about the rest of the tensor network is included during the update.

Finally, we update the PEPS tensors by recombining the new triads and the old environment tensors. This is shown in \cref{fig:triadCombiningX}.

For fermionic PEPS we split the tensor on the left side of \cref{fig:simpleUpdate} into two parts. Each one contributes to only one parity sector on the link to be updated. Then, two individual SVDs are performed on these parts. The truncation combines and orders the singular values and keeps the $D$ largest ones. This allows to change the splitting between even and odd parity due to their local contributions to the tensor network.

With this update procedure we also change the norm of the state. In order to keep the norm in a reasonable order, we normalized the new singular values:
\begin{align}
	S &\rightarrow \frac{S}{\renorm} &
	\renorm &= \sqrt{\sum_i S_i^2}\,.
	\label{eq:renormalization}
\end{align}
The renormalization factors \renorm can also be used to estimate the energy, see \cref{sec:energy_estimator}.

\subsubsection{Full Update, Boundary Matrix Product State and Expectation values}
Full Update algorithms take the whole tensor network into account for the truncation. Contracting the complete network scales exponentially in the size of one dimension of the system. To avoid this, an approximate contraction has to be used. We use the boundary Matrix Product State (boundary MPS) method for 2-dimensional systems with OBC \cite{PEPS_original_bMPS,iPEPS_introduction}.

\begin{figure}[ht]
	\centering
	\begin{tikzpicture}
	\def\Lx{5}
	\def\xShift{\xDist/2}
	\def\yShift{-\xShift}
	\def\bMPSyShift{\yShift/5}
	\foreach \x in {1,...,\Lx}
		{
			\node[ket] (ket_\x) at (\xDist*\x, 0) {};
			\node[bra] (bra_\x) at ([shift=({\xShift,\yShift})]ket_\x.center) {};
			\draw[physical] (ket_\x) -- (bra_\x);
			\node[bMPS] (bMPS_\x) at (\xDist*\x + \xShift/2, -\yShift) {};
			\draw (ket_\x) .. controls ([shift=({-\xShift/2,\bMPSyShift})]bMPS_\x.center) .. (bMPS_\x);
			\draw[name path={bMPSlink_\x}] (bra_\x) .. controls ([shift=({\xShift/2,\bMPSyShift})]bMPS_\x.center) .. (bMPS_\x);
			\node[bMPS] (cbMPS_\x) at (\xDist*\x + \xShift/2, \yShift*2) {};
			\draw[name path={cbMPSlink_\x}] (ket_\x) .. controls ([shift=({-\xShift/2,-\bMPSyShift})]cbMPS_\x.center) .. (cbMPS_\x);
			\draw (bra_\x) .. controls ([shift=({\xShift/2,-\bMPSyShift})]cbMPS_\x.center) .. (cbMPS_\x);
			\ifthenelse{\x > 1}
			{
				\pgfmathtruncatemacro\lastx{\x-1}
				\draw[name path={ketlink_\lastx}] (ket_\lastx) -- (ket_\x);
				\path[name intersections={of={bMPSlink_\lastx} and {ketlink_\lastx},by={ketswap_\x}}];
				\node[swap]  at (ketswap_\x) {};
				\draw[boundary] (bMPS_\lastx) -- (bMPS_\x);
				\draw[name path={bralink_\lastx}] (bra_\lastx) -- (bra_\x);
				\path[name intersections={of={cbMPSlink_\x} and {bralink_\lastx},by={braswap_\x}}];
				\node[swap]  at (braswap_\x) {};
				\draw[boundary] (cbMPS_\lastx) -- (cbMPS_\x);
			}
		}
\end{tikzpicture}
	\caption{Overlap of the old and the new boundary MPS.  The 
		upper layer corresponds to the old boundary MPS, 
		the middle layer represents bra and ket PEPS tensors, and the
		lower layer is the conjugate of the new boundary MPS.}
	\label{fig:bMPS_optimization}
\end{figure}
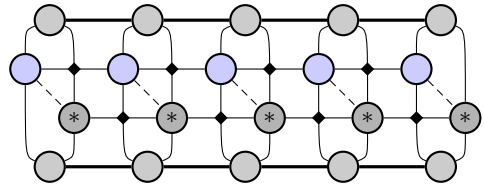


A boundary MPS is a string of tensors each connected to its neighbors. It represents an approximation of the contraction of the bra- and ket-layer of all tensors in the network from the boundary up to a certain $x$ or $y$ value. The upper layer of \cref{fig:bMPS_optimization} represents a boundary MPS in $x$-direction. Each tensor is connected to its neighbor by an index (thick lines) of size up to $\chi$, the boundary MPS dimension.
Additionally, each tensor carries two indices which represent the external indices of the approximated network.  Each is of size up to $D$.

Initially, the boundary MPS is set to a trivial product state of length $L_x$ with all tensors set to rank 0, i.e.\ scalar. To go from one row to the next, we initialize the boundary MPS including one more row with random numbers. Then, a sweep is done through all tensors of the boundary MPS to optimize one tensor at a time, while the others are kept fixed. If $\ket{b}$ is the contraction of the old boundary MPS with the PEPS tensors to be included in the next layer, and $\ket{b'}$ is the new boundary MPS, we minimize
\begin{align}
	\norm{ \ket{b} - \ket{b'} }^2 = \braket{b|b} + \braket{b'|b'}
	- \braket{b|b'} - \braket{b'|b}\,.
\end{align}
The overlap $\braket{b|b'}$ is shown in \cref{fig:bMPS_optimization}. The optimization can be done using alternating least squares (ALS) optimization \cite{Verstraete_introduction,Lubasch_algorithms,iPEPS_introduction}.

\begin{figure}[ht]
	\centering
	\begin{align*}
	\begin{tikzpicture}
		\def\Lx{5}
		\def\xDistLocal{\xDist*2/3}
		\def\xShift{\xDist/2} 
		\def\yShift{-\xShift} 
		\def\yShiftBra{-\xShift*2} 
		\def\bMPSyShift{\yShift/5}
		\node[ket] (ketL) at (0, 0) {};
		\node[bra] (braL) at ([shift=({\xShift,\yShiftBra})]ketL.center) {};
		\node[bMPS] (bMPSxLU) at ([shift=({\xShift/2,-\yShift})]ketL.center) {};
		\draw (ketL) .. controls ([shift=({-\xShift/2,\bMPSyShift})]bMPSxLU.center) .. (bMPSxLU);
		\draw[name path={bMPSxlinkLU}] (braL) .. controls ([shift=({\xShift/2,\bMPSyShift})]bMPSxLU.center) .. (bMPSxLU);
		\node[bMPS] (bMPSxLD) at ([shift=({-\xShift/2,\yShift})]braL.center) {};
		\draw[name path={bMPSxlinkLD}] (ketL) .. controls ([shift=({-\xShift/2,-\bMPSyShift})]bMPSxLD.center) .. (bMPSxLD);
		\draw (braL) .. controls ([shift=({\xShift/2,-\bMPSyShift})]bMPSxLD.center) .. (bMPSxLD);
		\node[bMPS] (bMPSyL) at ([shift=({\yShift,\yShiftBra/2})]ketL.center) {};
		\draw (ketL) .. controls ([shift=({0,-\yShiftBra/2})]bMPSyL.center) and ([shift=({-\bMPSyShift,\xShift/2})]bMPSyL.center) .. (bMPSyL);
		\draw[name path={bMPSylinkL}] (braL) .. controls ([shift=({0,\yShiftBra/2})]bMPSyL.center) and ([shift=({-\bMPSyShift,-\xShift/2})]bMPSyL.center) .. (bMPSyL);
		\path[name intersections={of={bMPSxlinkLD} and {bMPSylinkL},by={swapLD}}];
		\node[swap]  at (swapLD) {};
		\node[ket] (triadLU) at ([shift=({\xDistLocal+\xShift,0})]ketL.center) {};
		\draw[wiggly,name path={wigglyLU}] (ketL) -- (triadLU);
		\path[name intersections={of={bMPSxlinkLU} and {wigglyLU},by={swapLU}}];
		\node[swap]  at (swapLU) {};
		\node[bra] (triadLD) at ([shift=({0,\yShiftBra})]triadLU.center) {};
		\draw[wiggly] (braL) -- (triadLD);
		\node[ket] (triadRU) at ([shift=({\xDistLocal,0})]triadLU.center) {};
		\draw (triadLU) -- (triadRU) node[midway] (middle) {};
		\node[bra] (triadRD) at ([shift=({\xDistLocal,0})]triadLD.center) {};
		\draw (triadLD) -- (triadRD);
		\node[ket] (ketR) at ([shift=({\xDistLocal,0})]triadRU.center) {};
		\draw[wiggly] (triadRU) -- (ketR);
		\node[bra] (braR) at ([shift=({\xShift,\yShiftBra})]ketR.center) {};
		\draw[wiggly,name path={wigglyRD}] (triadRD) -- (braR);
		\node[bMPS] (bMPSxRU) at ([shift=({\xShift/2,-\yShift})]ketR.center) {};
		\draw (ketR) .. controls ([shift=({-\xShift/2,\bMPSyShift})]bMPSxRU.center) .. (bMPSxRU);
		\draw[name path={bMPSxlinkRU}] (braR) .. controls ([shift=({\xShift/2,\bMPSyShift})]bMPSxRU.center) .. (bMPSxRU);
		\node[bMPS] (bMPSxRD) at ([shift=({-\xShift/2,\yShift})]braR.center) {};
		\draw[name path={bMPSxlinkRD}] (ketR) .. controls ([shift=({-\xShift/2,-\bMPSyShift})]bMPSxRD.center) .. (bMPSxRD);
		\draw (braR) .. controls ([shift=({\xShift/2,-\bMPSyShift})]bMPSxRD.center) .. (bMPSxRD);
		\path[name intersections={of={wigglyRD} and {bMPSxlinkRD},by={swapRD}}];
		\node[swap]  at (swapRD) {};
		\node[bMPS] (bMPSyR) at ([shift=({\xShift,-\yShiftBra/2})]braR.center) {};
		\draw[name path={bMPSylinkR}] (ketR) .. controls ([shift=({0,-\yShiftBra/2})]bMPSyR.center) and ([shift=({\bMPSyShift,\xShift/2})]bMPSyR.center) .. (bMPSyR);
		\draw (braR) .. controls ([shift=({0,\yShiftBra/2})]bMPSyR.center) and ([shift=({\bMPSyShift,-\xShift/2})]bMPSyR.center) .. (bMPSyR);
		\path[name intersections={of={bMPSxlinkRU} and {bMPSylinkR},by={swapRU}}];
		\node[swap]  at (swapRU) {};
		\draw[boundary] (bMPSxLU) -- (bMPSxRU);
		\draw[boundary] (bMPSxRU) .. controls ([shift=({\xShift+\xShift/2,0})]bMPSxRU.center) .. (bMPSyR);
		\draw[boundary] (bMPSxRD) .. controls ([shift=({\xShift+\xShift/2,0})]bMPSxRD.center) .. (bMPSyR);
		\draw[boundary] (bMPSxLD) -- (bMPSxRD);
		\draw[boundary] (bMPSxLD) .. controls ([shift=({-\xShift-\xShift/2,0})]bMPSxLD.center) .. (bMPSyL);
		\draw[boundary] (bMPSxLU) .. controls ([shift=({-\xShift-\xShift/2,0})]bMPSxLU.center) .. (bMPSyL);
		\node[gate, minimum width = \xDistLocal+\pepsWidth] (operator) at ([shift=({0,\yShiftBra/2})]middle) {$\hat{o}$};
		\draw[physical] \connectD{triadLU}{operator};
		\draw[physical] \connectD{triadRU}{operator};
		\draw[physical] \connectU{triadLD}{operator};
		\draw[physical] \connectU{triadRD}{operator};
		\draw[dashdotted,thick] ([shift=({-\xDistLocal/2,-\yShift/2})]triadLU.center) rectangle ([shift=({\xDistLocal/2,\yShift/2})]triadRD.center);
	\end{tikzpicture}
	&\eqqcolon
	\begin{tikzpicture}
		\def\Lx{5}
		\def\xDistLocal{\xDist*2/3}
		\def\xShift{\xDist/2} 
		\def\yShift{-\xShift} 
		\def\yShiftBra{-\xShift*2} 
		\def\bMPSyShift{\yShift/5}
		\node[ket] (triadLU) at (0,0) {};
		\node[bra] (triadLD) at ([shift=({0,\yShiftBra})]triadLU.center) {};
		\node[ket] (triadRU) at ([shift=({\xDistLocal,0})]triadLU.center) {};
		\draw (triadLU) -- (triadRU) node[midway] (middleUp) {};
		\node[bra] (triadRD) at ([shift=({\xDistLocal,0})]triadLD.center) {};
		\draw (triadLD) -- (triadRD) node[midway] (middleDown) {};;
		\node[gate, minimum width = \xDistLocal+\pepsWidth] (operator) at ([shift=({0,\yShiftBra/2})]middleUp) {$\hat{o}$};
		\draw[physical] \connectD{triadLU}{operator};
		\draw[physical] \connectD{triadRU}{operator};
		\draw[physical] \connectU{triadLD}{operator};
		\draw[physical] \connectU{triadRD}{operator};
		\node[env] (env) at ([shift=({0,-\yShift})]middleUp) {\env};
		\draw[wiggly] (triadLU.west) arc(270:90:-\yShift/2-\gateHeight/4) -- (env.south west);
		\draw[wiggly] (triadLD.west) arc(270:180:-\yShift) -- ([shift=({\yShift,\gateHeight/2})]triadLU.west) arc(180:90:-\yShift) -- (env.north west);
		\draw[wiggly] (triadRU.east) arc(-90:90:-\yShift/2-\gateHeight/4) -- (env.south east);
		\draw[wiggly] (triadRD.east) arc(-90:0:-\yShift) -- ([shift=({-\yShift,\gateHeight/2})]triadRU.east) arc(0:90:-\yShift) -- (env.north east);
		\node[env,opacity=0] (envPhant) at ([shift=({0,\yShift})]middleDown) {\env};
		\draw[wiggly,opacity=0] (triadRU.east) arc(90:0:-\yShift) -- ([shift=({-\yShift,-\gateHeight/2})]triadRD.east) arc(0:-90:-\yShift) -- (envPhant.south east);
	\end{tikzpicture}
\end{align*}
	\caption{Calculating the expectation value of a two site operator $\hat{o}$ in x-direction with a fermionic tensor network using boundary MPS. Wiggly lines denote a splitting of triads and environment tensors according to \cref{fig:triadCombiningX} but without singular values $S$.
	Contraction of all tensors outside the dash dotted rectangle define the environment tensor \env.}
	\label{fig:expectation_value}
\end{figure}
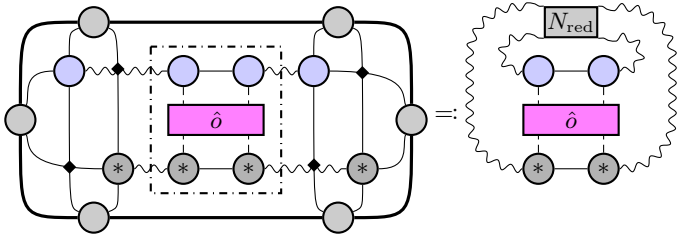

This procedure is done from all boundaries and allows us to calculate expectation values as depicted in \cref{fig:expectation_value}. In the update step, the same surrounding can be used. We split the nodes we want to update into triads and the rest as described in \cref{sec:SimpleUpdate}. The whole network except for the triads is then contracted to form an environment tensor \env as defined in \cref{fig:expectation_value}.

For the truncation step, we make \env hermitian and semipositive before we gauge it for better conditioning of the following steps. This preparation is explained in detail in \cite{Lubasch_algorithms}. For the update itself there are two common procedures. The first is called the Second Renormalization Group (SRG) \cite{SRG_original}. The idea here is to generalize Simple Update and do a singular value decomposition that includes the environment \env. The truncation is done by dropping the smallest singular values of this decomposition. Since every singular value corresponds to either even or odd parity on the link to be updated, the splitting between the two parity sectors is done dynamically according to the sizes of the singular values in the sectors.  The second update procedure optimizes the sites iteratively while keeping the remaining sites fixed, using the same ALS optimization techniques as before for the boundary MPS \cite{Verstraete_introduction, Lubasch_algorithms,iPEPS_introduction}. We applied this algorithm to fermionic PEPS, so we also had to sweep through the parity sectors. The details of the optimization can be found in \cref{app:FU_optimization}.

Despite its benefit of being able to dynamically track and split the parity sectors, we found that truncation errors of the SRG update process were rather large for the systems we tested and this precluded its use as a standalone updating process. On the other hand, the ALS procedure provided much smaller truncation errors, but the parity sectors have to be kept at a constant size during the updating process. This prevented us from dividing the parity sectors on a link according to their contribution to the state. Our solution to addressing the shortfall of each update procedure was to combine both procedures during the updating step, as follows:  We first used SRG to make an initial guess for the new triads and to choose the size of the  different parity sectors. The relatively large truncation errors at this point were subsequently reduced by application of the ALS updating procedure with a fixed size of the sectors. By combining both methods, we were able to achieve our desired accuracy goals with full accounting of both parity sectors.

After each update step we divided the new state $\ket{\Psi'}$ by $\renorm = \sqrt{\frac{\braket{\Psi'|\Psi'}}{\braket{\Psi|\Psi}}}$ to normalize it.


\subsection{Resource scaling}\label{sec:resource_scaling}
Now that we know which operations have to be performed, we can analyze the required resources in terms of runtime and memory. In both cases the leading contribution is made by the calculation of the environment \env as shown in \cref{fig:expectation_value}. The overall runtime $T_\text{CPU}$ then asymptotically scales as
\begin{align}
	T_\text{CPU}^\square &\propto N_\text{in}\left(2 \cdot \chi^3 D^4 d^2 + 2 \cdot \chi^2 D^6 d^2 + \chi^2 D^4 d^4\right)
\end{align}
for a 2D square lattice where $N_\text{in}=(L_x-2)(L_y-2)$ denotes the number of inner lattice points. We do not need the full number of points here because the contraction at the boundary proceeds much faster due to the lower number of links on the tensors.


In the case of the honeycomb lattice some of the links are removed. This can be interpreted (and implemented) as setting $D=1$ for these links. The runtime reduces to
\begin{align}
	T_\text{CPU}^{\hexago} &\propto N_\text{in}\left(2 \cdot \chi^3 D^4 d^2 + 2 \cdot \chi^2 D^5 d^2 + \chi^2 D^4 d^4\right)
\end{align}

It is common to choose $\chi \propto D^2$ which leads to the well known and feared scaling in $\ordnung{D^{10}}$. Our findings in \cref{sec:boundaryMPSbondDimension} however suggest that choosing $\chi\propto D$ (usually $\chi=2D$ or $\chi=3D$) already yields very good results in many cases. In particular, the error on observables introduced by the choice of $\chi$ is much smaller than that introduced by the choice of $D$. With $\chi\propto D$ the overall scaling of our algorithm reduces to $\ordnung{D^{8}}$ for the square and $\ordnung{D^{7}}$ for the honeycomb lattice.

For practical considerations such as job submissions, an understanding of the required memory $M$ is often even more important than the scaling of the runtime. We found that we can predict the memory usage very accurately (up to $\ordnung{\SI{100}{\mega\byte}}$) with the formula
\begin{align}\label{eqn:memory}
M &= \SI{700}{\mega\byte} + 2\chi^2D^4d^2 M_\text{num} + 2N\chi^2D^2d M_\text{num}
\end{align}
where $M_\text{num}$ is the memory required for a single number, in our case $M_\text{num}=\SI{16}{\byte}$ for complex doubles.  Notice that the constant offset in \cref{eqn:memory} is system dependent and in our case represents the approximate memory usage of  \texttt{MATLAB}~\cite{MATLAB} when idle.

Our simulations were single-threaded and accomplished within 48 hours.  They were performed on \texttt{MATLAB}  with heavy reliance on \texttt{ncon}~\cite{pfeifer2015ncon}.



\subsection{The Hubbard model}
The Hamiltonian of the Hubbard model can be decomposed into sums of operators acting on a pair of neighboring sites only. It reads
\begin{align}
	H&=-\kappa \sum_{\erwartung{x,y},s}c^\dagger_{x,s}c^\pdagger_{y,s}+\frac{U}{2}\sum_{x}q_x^2
	+\mu \sum_{x,s}\left(c_{x,s}^\dagger c_{x,s}^\pdagger-\frac12\right)
	\label{eq:hubbard_hamilton}
\end{align}
with $\erwartung{x,y}$ denoting nearest neighbors, whereas $s\in\left\{\uparrow,\,\downarrow\right\}$ denotes the two possibilities for the spin of an electron. $c^\dagger_{x,s}$ ($c^\pdagger_{x,s}$) is a creation (annihilation) operator of an electron at position $x$ with spin $s$ and follows the usual anti-commutation relations
\begin{equation}
	\left\{c^\dagger_i,c^\dagger_j\right\}=\left\{c^\pdagger_i,c^\pdagger_j\right\}=0\:,\quad\left\{c^\pdagger_i,c^\dagger_j\right\}=\delta_{ij}\,.
\end{equation}
The onsite coupling $U$ gives the interaction strength of two electrons with combined charge $q_x=c^\dagger_{x,\uparrow}c^\pdagger_{x,\uparrow}+c^\dagger_{x,\downarrow}c^\pdagger_{x,\downarrow}-1$ on the same lattice point. The hopping parameter is denoted by $\kappa$ and $\mu$ is the chemical potential.
The form of \cref{eq:hubbard_hamilton} is such that at $\mu=0$ the system is at half-filling.

The Hubbard model is studied on a honeycomb lattice. The connectivity of nearest neighbors $\erwartung{x,y}$ is the same as the connectivity of the \tn, as can be seen in \cref{fig:peps}. For larger systems in x- (y-) direction, the figure is extended to the right (bottom) in our convention. The parity link is always attached to the tensor in the left bottom corner.

We rewrite the Hamiltonian \cref{eq:hubbard_hamilton} purely as a sum of two site terms $h_{xy}$ as follows
\begin{align}
	H&=\sum_{\erwartung{x,y}}h_{xy}\,,\label{eq:hamilton_as_sum_over_gates}\\
	h_{xy}&=
	\begin{aligned}[t]
		&-\kappa\sum_s\left(c^\dagger_{x,s}c^\pdagger_{y,s}+c^\dagger_{y,s}c^\pdagger_{x,s}\right)
		+\frac U2\left(\frac{1}{n_x}q_x^2+\frac{1}{n_y}q_y^2\right)\\
		&+\mu\sum_s
		\begin{aligned}[t]
			&\left( \frac{1}{n_x}\left(c_{x,s}^\dagger c_{x,s}^\pdagger-\frac12\right) \right.\\
			&\left. +\frac{1}{n_y} \left(c_{y,s}^\dagger c_{y,s}^\pdagger-\frac12\right) \right)
			\,.
		\end{aligned}
	\end{aligned}
\end{align}

Here $n_x$ is the number of nearest neighbors of site $x$.
$h_{xy}$ is chosen hermitian and therefore every link $\erwartung{x,y}$ is counted only once, i.e.\ $\erwartung{y,x}$ is ignored.

For the imaginary time evolution we then use the Hamiltonian gate $\eto{-h_{xy}\delta t}$.

\subsubsection{Two site basis}\label{sec_two_site_basis}
We now define our basis for the two site states. For this we fix the order of creation operators acting on the empty state $\ket{0}$ as $c^\dagger_{x,\uparrow}$, $c^\dagger_{x,\downarrow}$, $c^\dagger_{y,\uparrow}$, $c^\dagger_{y,\downarrow}$ and receive a $4\otimes4=16$-dimensional basis, where the one-site basis consisting of $\ket0$, $\ket{\uparrow\downarrow}$, $\ket\uparrow$ and $\ket \downarrow$ is $d\equiv4$-dimensional. We write
\begin{align}
	\ket k&\coloneqq\ket{k_x}_x\otimes\ket{k_y}_y\,,\\
	k&=4k_x+k_y
\end{align}
and the $k_{x,y}\in\left\{0,\dots,3\right\}$ is the respective state index, so for example $\ket{7}=\ket{1}_x\otimes\ket{3}_y\equiv\ket{\uparrow\downarrow}_x\otimes\ket{\downarrow}_y$.

In this basis we can rewrite the gate operator $h_{xy}$ using the defining properties
\begin{align}
	\Braket{k_2|c^\dagger_{x,\uparrow}c^\pdagger_{y,\uparrow}|k_1}&=\begin{cases}
		\phantom{-}1&\left(k_2,k_1\right)\in\left\{(8,2),(11,1)\right\}\\
		-1&\left(k_2,k_1\right)\in\left\{(4,14),(7,13)\right\}\\
		\phantom{-}0&\text{else}
	\end{cases},\\
	\Braket{k_2|c^\dagger_{x,\downarrow}c^\pdagger_{y,\downarrow}|k_1}&=\begin{cases}
		\phantom{-}1&\left(k_2,k_1\right)\in\left\{(12,3),(4,11)\right\}\\
		-1&\left(k_2,k_1\right)\in\left\{(14,1),(6,9)\right\}\\
		\phantom{-}0&\text{else}
	\end{cases},
\end{align}
and
\begin{align}
	q_x^2&=\matr{rrrr}{1&0&0&0\\0&1&0&0\\0&0&0&0\\0&0&0&0}\otimes\mathbb{1}_4\,,\\
	c_{x,\uparrow}^\dagger c_{x,\uparrow}^\pdagger+c_{x,\downarrow}^\dagger c_{x,\downarrow}^\pdagger-1 &= \matr{rrrr}{-1&0&0&0\\0&1&0&0\\0&0&0&0\\0&0&0&0}\otimes\mathbb{1}_4\,.\label{eq:local_particle_number}
\end{align}

\subsubsection{Operators}
Any local operator, i.e.\ involving only onsite and nearest neighbor interactions, can be expressed in terms of gates in the same way as the Hamiltonian in \cref{eq:hamilton_as_sum_over_gates}. We derive this form explicitly for the magnetization operator because it becomes important for breaking degeneracies later on. The operator reads
\begin{align}	
	M &= \sum_{x}\left(c_{x,\uparrow}^\dagger c_{x,\uparrow}^\pdagger-c_{x,\downarrow}^\dagger c_{x,\downarrow}^\pdagger\right)\\	
	&= \sum_{\erwartung{x,y}}
	\begin{aligned}[t]
		&\left(\frac{1}{n_x}\left(c_{x,\uparrow}^\dagger c_{x,\uparrow}^\pdagger-c_{x,\downarrow}^\dagger c_{x,\downarrow}^\pdagger\right)\right.\nonumber\\	
		&\left.+\frac{1}{n_y}\left(c_{y,\uparrow}^\dagger c_{y,\uparrow}^\pdagger-c_{y,\downarrow}^\dagger c_{y,\downarrow}^\pdagger\right)\right)
	\end{aligned}
	\\	
	&\eqqcolon \sum_{\erwartung{x,y}} m_{xy}
	\label{eq:magnetization_definition}
\end{align}
with the matrix form	
\begin{multline}	
	m_{xy} = \left(\frac{1}{n_x}\matr{rrrr}{0&0&0&0\\0&0&0&0\\0&0&1&0\\0&0&0&-1}\otimes\mathbb{1}_4\right.\\	
	\left.+\frac{1}{n_y}\mathbb{1}_4\otimes\matr{rrrr}{0&0&0&0\\0&0&0&0\\0&0&1&0\\0&0&0&-1}\right)\,.\label{eqn_magnetization_gate}	
\end{multline}
The integers $n_x$ and $n_y$ denote the number of nearest neighbors at sites $x$ and $y$.

The particle number
\begin{align}
	n &= \sum_{x}\left(c_{x,\uparrow}^\dagger c_{x,\uparrow}^\pdagger+c_{x,\downarrow}^\dagger c_{x,\downarrow}^\pdagger\right)
	\label{eq:particle_number_definition}
\end{align}
is another important operator as it directly couples to the chemical potential. Its matrix form follows immediately from \cref{eq:local_particle_number}.
\section{Improvements\label{sect:improvements}}
\subsection{Boundary MPS bond dimension dependence\label{sec:boundaryMPSbondDimension}}
Since our observables consist of nearest neighbor operators only, one can use the boundary MPS environment for each set of nearest neighbors and calculate the corresponding expectation value as explained in \cref{sec:truncation} and \cref{fig:expectation_value}. The sum of these terms leads to the desired global observables.

In the literature a typical value of $\chi \ge D^2$ is standard \cite{Lubasch_clusterUpdate,Lubasch_algorithms,bMPS,PEPS_original_bMPS,tensor_fermions,iPEPS_original_2008}. This gives an exact representation of the boundaries itself and a good approximation of the rest of the network. We considered the dependence of expectation values on the boundary MPS bond dimension more carefully and found that a much smaller $\chi$ leads to a good precision already.

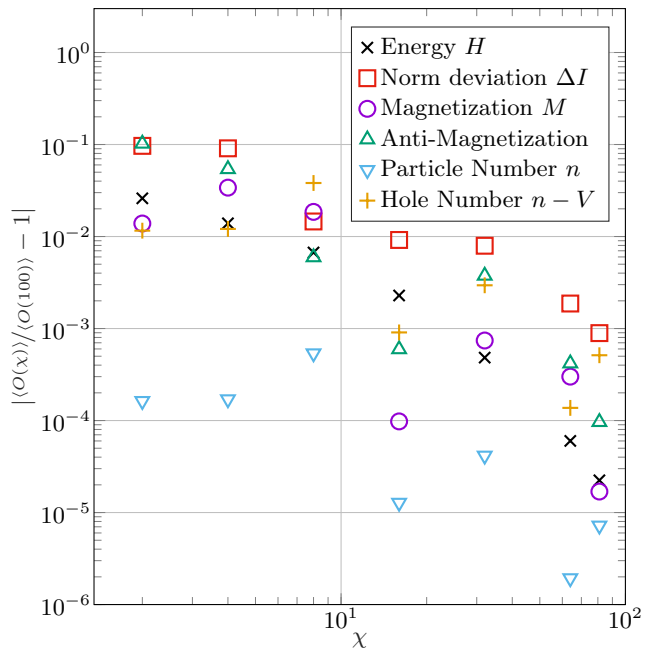
\begin{figure}[ht]
	\centering
%
%
\definecolor{mycolor1}{rgb}{0.898,0.122,0.059}%
\definecolor{mycolor2}{rgb}{0.58,0,0.831}%
\definecolor{mycolor3}{rgb}{0.00000,0.62,0.451}%
\definecolor{mycolor4}{rgb}{0.341,0.71,0.91}%
\definecolor{mycolor5}{rgb}{0.898,0.62,0}%
\begin{tikzpicture}

\begin{axis}[%
width=\linewidth,
height=1.1\linewidth,
unbounded coords=jump,
xmin=0,
xmax=100,
xlabel style={font=\color{white!15!black}},
xlabel={$\chi$},
xmode=log,
ymode=log,
ymin=1e-06,
ymax=3,
yminorticks=true,
ylabel style={font=\color{white!15!black},{rotate=-90}},
ylabel={$\left|\nicefrac{\Braket{O\left(\chi\right)}}{\Braket{O\left(100\right)}}-1 \right|$},
axis background/.style={fill=white},
title style={font=\bfseries},
xmajorgrids,
ymajorgrids,
legend style={legend cell align=left, align=left, draw=white!15!black, 
	legend pos=north east}
]
\addplot [color=black, only marks, mark=x, mark options={solid, black, scale=1.5, thick}]
  table[row sep=crcr]{%
2	0.0260069192947757\\
4	0.0139208000983909\\
8	0.00670613395298232\\
16	0.00228523723092908\\
32	0.000482984546681127\\
64	6.00624143457044e-05\\
81	2.24047932330323e-05\\
};
\addlegendentry{Energy $H$}

\addplot [color=mycolor1, only marks, mark=square, mark options={solid, mycolor1, scale=1.5, thick}]
table[row sep=crcr]{%
	2	0.096678357998241\\
	4	0.0909057863263887\\
	8	0.0145723957507468\\
	16	0.00917393406855137\\
	32	0.00793899922872473\\
	64	0.00186960981674376\\
	81	0.000891719282645178\\
};
\addlegendentry{Norm deviation $\Delta I$}

\addplot [color=mycolor2, only marks, mark=o, mark options={solid, mycolor2, scale=1.5, thick}]
  table[row sep=crcr]{%
2	0.0138474308563559\\
4	0.0339913338759446\\
8	0.0185252877052031\\
16	9.786140862874e-05\\
32	0.000741071632818017\\
64	0.000300076933798946\\
81	1.69141248368413e-05\\
};
\addlegendentry{Magnetization $M$}

\addplot [color=mycolor3, only marks, mark=triangle, mark options={solid, mycolor3, scale=1.5, thick}]
  table[row sep=crcr]{%
2	0.10215346611069\\
4	0.0540001546162627\\
8	0.0059501487046594\\
16	0.000595619536942853\\
32	0.00375281289439241\\
64	0.000417644105515267\\
81	9.63358389874289e-05\\
};
\addlegendentry{Anti-Magnetization}

\addplot [color=mycolor4, only marks, mark=triangle, mark options={rotate=180, solid, mycolor4, scale=1.5, thick}]
  table[row sep=crcr]{%
2	0.00016304109633034\\
4	0.000170167308784585\\
8	0.000537324077486438\\
16	1.27690627588389e-05\\
32	4.1544419683653e-05\\
64	1.93252344431274e-06\\
81	7.21477258247843e-06\\
};
\addlegendentry{Particle Number $n$}

\addplot [color=mycolor5, only marks, mark=+, mark options={solid, mycolor5, scale=1.5, thick}]
  table[row sep=crcr]{%
2	0.0115816895380439\\
4	0.0120879028921297\\
8	0.038169030918162\\
16	0.000907055484127961\\
32	0.00295112448115164\\
64	0.000137277576395053\\
81	0.000512504258242693\\
};
\addlegendentry{Hole Number $n-V$}

\end{axis}
\end{tikzpicture}%
	\caption{Dependence of different operators $\Braket{O}$ on the boundary MPS bond dimension $\chi$ for the ground state in the odd parity sector. Shown are the relative errors compared to results at $\chi = 100$. Honeycomb Hubbard model with $D=12$, $L=12\times6$, $\kappa=1$, $U=2$, $\mu=0.1$, $B=0.01$.}
	\label{fig:boundaryMPSbondDimension}
\end{figure}

For example, \Cref{fig:boundaryMPSbondDimension} shows the relative error of different observables as a function of $\chi$. In this case the ground state of the odd parity sector of the Hubbard model was obtained with Simple Update and a bond dimension of $D=12$. Expectation values were then calculated with different boundary MPS dimensions $\chi$ and compared to the result using $\chi=100$. We found that for $\chi \ge 16$ all errors are on a sub-percent level. This is much smaller than the usual $D^2 = 144$. Thus if a sub-percent precision is sufficient, as it was in our studies, the boundary MPS bond dimension can be decreased significantly.
We found similar results for different parameters $U$ and $\mu$, bond dimensions and both parity sectors of the Hubbard model.

Let us stress however that these results only apply to the boundary MPS contraction after a time evolution with Simple Update. They do not apply to Full Update. On the contrary, for Full Update to be stable we found that in some cases $\chi\gtrsim D^2$ is indeed required.

Obviously, choosing a smaller $\chi$ improves the computational costs drastically. This in turn can be used to increase $D$ which then improves the overall precision.

Note that we need to normalize our state or calculate the norm explicitly for expectation values:
\begin{align}
	\Braket{\hat{O}} &= \frac{\Braket{\Psi|\hat{O}|\Psi}}{\Braket{\Psi|\Psi}}\,.
\end{align}
Instead of normalizing the state once, we found it helpful to calculate the norm $\braket{\Psi|\Psi}_\link$ for each nearest-neighbor pair \link by calculating it the same way as the expectation values, but without inserting any operators. This way, the local expectation values can be divided by the norm with respect to the same environment \env and therefore the same truncation of the \tn. This increases the precision of expectation values.
Moreover, the standard deviation $\Delta I$ of the local norm evaluations can be used as an estimator for the error introduced by the boundary MPS truncation:
\begin{equation}
	\Delta I = \frac{\sqrt{\sum_\link \left( \braket{\Psi|\Psi}_\link - \overline{\braket{\Psi|\Psi}} \right)^2}}{\overline{\braket{\Psi|\Psi}} \cdot \sqrt{|\link|-1}}\,.\label{eq:norm_std_deviation}
\end{equation}
The term $|\link|$ denotes the number of nearest-neighbor pairs in the system and $\overline{\braket{\Psi|\Psi}} = \frac{\sum_\link \braket{\Psi|\Psi}_\link}{|\link|} $ is the mean value of the norm.
The error estimation by $\Delta I$ can be seen in \cref{fig:boundaryMPSbondDimension} as red squares. In all cases we studied, $\Delta I$ was of the order of the relative errors of any observable considered. This allowed us to estimate the effects of the boundary MPS truncation on expectation values without needing to do a full scan in $\chi$.

We observed that a good precision can be reached with $\chi=3D$. The error introduced by a finite $D$ is then larger than that caused by a finite $\chi$. A similar behavior was observed for infinite PEPS in \cite{iPEPS_hubbard_2018}. We checked that our results do not crucially depend on $\chi$ by repeating most of our simulations with $\chi=2D$.

\subsection{Convergence}
The imaginary time evolution as described in \cref{eq:imag_time_evol,eq:U_trotterized,eq:H_local,eq:trotter_error,eq:trotter2} has to be tuned carefully. It is crucial to choose an appropriate value for $\delta t$ at every step. To understand the details better we first have to differentiate two kinds of errors. Firstly, we have the pollution of the ground state\footnote{Here we mean the ground state of the trotterized time evolution operator.} by excited states. This error arises from \cref{eq:imag_time_evol}, decays exponentially for any given $\delta t$, and we therefore call it \textit{exponential error}. Secondly, the discretization into a scheme of order $r-1$ introduces an error decreasing as $\delta t^r$ (see \cref{eq:trotter_error}). This error remains for any given $\delta t$ even at $t\rightarrow\infty$. We designate this as the \textit{Trotter error}.

A well tuned convergence scheme should use as few steps as possible to achieve the smallest total error. Naively this can be achieved by choosing several, successively decreasing, values of $\delta t$ and performing the complete imaginary time evolution for every one of them. This however turns out to be highly inefficient. Instead, we first perform the imaginary time evolution with a coarse step size and then proceed from the resulting state with a smaller value of $\delta t$. As the ground state energies of both discretizations are (asymptotically) close to each other, we only have a small number of steps to perform with the finer step size. Once the new discretization shows no significant exponential error any more, the step size is reduced again and the procedure is iterated.

\begin{figure}[H]
	\centering
	\subfigure[Too rapid reduction of $\delta t$.\label{dt_slow}]{\input{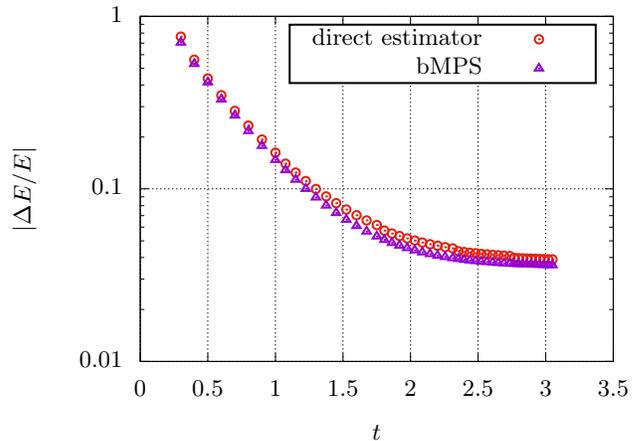}}
	\subfigure[Efficient reduction of $\delta t$.\label{dt_optimal}]{\input{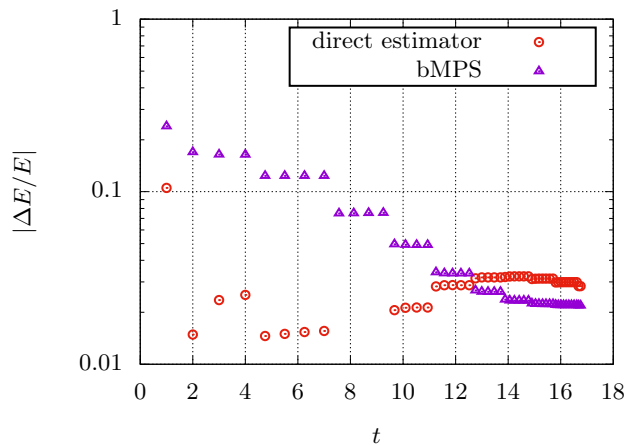}}
	\subfigure[Too slow reduction of $\delta t$.\label{dt_rapid}]{\input{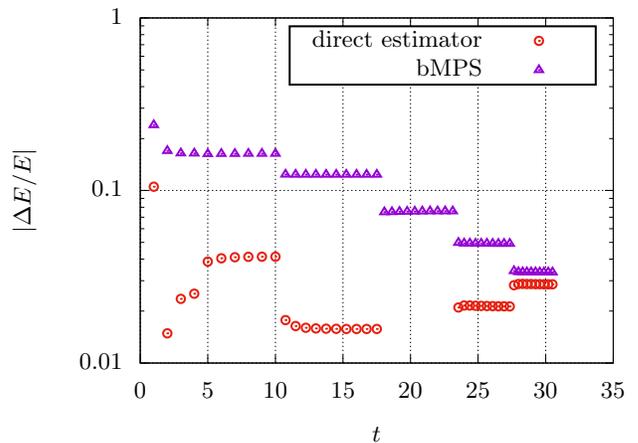}}
	\caption{Relative deviations of the energies during the imaginary time evolution with $\kappa=1$, $U=4$ and $\mu=B=0$ on the $2\times 2$-lattice with $D=6$. In all three cases exactly 50 steps have been performed. Energies calculated using boundary MPS (bMPS) and the direct estimator derived in \cref{sec:energy_estimator}.}
	\label{fig:stair}
\end{figure}

Thus, the overall convergence behavior resembles a stair as can be seen in \cref{fig:stair}. If the `steps' of this stair have a significant slope at the time a smaller $\delta t$ is chosen, this indicates a remnant exponential error and thus a loss of accuracy (see \cref{dt_slow}). Therefore, $\delta t$ has to be decreased slower. If on the other hand there are long plateaus or the energy difference between adjacent `steps' is very small, no accuracy is lost, but the algorithms efficiency could be increased by a more rapid decrease of $\delta t$ (see \cref{dt_rapid}).

\subsubsection{Extrapolation}
For energy values $E(\delta t)$ with pure (or at least strongly dominating) Trotter error we make the ansatz
\begin{align}
	E(\delta t) &= E_0 + \gamma \delta t^r + \ordnung{\delta t^{r+1}}
\end{align}
where $E_0$ is the exact ground state energy and $\gamma$ is some coefficient independent of $\delta t$. We then obtain an extrapolated $E_0$ by choosing two values of $\delta t$, preferably the smallest two available, and calculate
\begin{align}
	E_0 &= \left(\frac{1}{\delta t_1^r}-\frac{1}{\delta t_2^r}\right)^{-1}\left(\frac{E(\delta t_1)}{\delta t_1^r}-\frac{E(\delta t_2)}{\delta t_2^r}\right) + \ordnung{\delta t^{r+1}}\,.
\end{align}

\subsubsection{$\delta t$-reduction scheme}
The energy including the exponential error can be approximated by
\begin{align}
	E(t) &= E_0\left(1+\alpha\eto{-\Delta_\text{tr} t}\right)
\end{align}
where $\alpha$ is some coefficient independent of $t$ (and at least asymptotically of $\delta t$) and $\Delta_\text{tr}$ is the current gap to the first excited state. This approximation is accurate up to exponentially suppressed higher state contributions and therefore very precise at large enough times. At large $t$ and small $\delta t$ the gap approaches the difference in Trotter errors between the last $\delta t$ and the current one. We therefore write
\begin{align}
	\Delta_\text{tr}\,\xrightarrow{t\rightarrow\infty,\,\delta t \rightarrow 0}\, \beta \Delta_\text{ex}\delta t^r
\end{align}
with another coefficient $\beta$ constant in $t$ and $\delta t$. The physical gap $\Delta_\text{ex}$ to the first excited state\footnote{Only the gap in the simulated sector is relevant here. If e.g.\ only the even parity sector is simulated, then $\Delta_\text{ex}$ denotes the gap between even ground state and first excited even state.} has been included to set a comparable energy scale.

We consider the local relative energy change
\begin{align}
	E'(t) &\coloneqq \frac{E(t+\delta t)-E(t)}{E(t)}\\
	&= \frac{\alpha \eto{-\Delta_\text{tr} t}\left(\eto{-\Delta_\text{tr} \delta t}-1\right)}{1+\alpha\eto{-\Delta_\text{tr} t}}\\
	&= \alpha\eto{-\Delta_\text{tr} t}\left(-\Delta_\text{tr} \delta t + \ordnung{\left(\Delta_\text{tr}\delta t\right)^2}\right) + \ordnung{\eto{-2\Delta_\text{tr} t}}\\
	&\rightarrow \underbrace{-\alpha \beta \eto{-\Delta_\text{tr} t}}_{\varepsilon} \Delta_\text{ex}\delta t^{r+1}
\end{align}
and find that at any given discretization the exponential error can be considered small enough once some threshold
\begin{align}
	\left|\varepsilon\right| = \left|\frac{E'(t)}{\Delta_\text{ex}}\right|\delta t^{-(r+1)} < \varepsilon_\text{tol}
\end{align}
is reached. The target precision of the exponential error $\varepsilon_\text{tol}$ therefore can be chosen independently of $t$ and $\delta t$.

The remaining problem is that a priori the factors $\beta$ and especially $\Delta_\text{ex}$ are not known. We observe that $\beta$ is usually of the order of one and therefore does not greatly influence the performance of our algorithm.
$\Delta_\text{ex}$ on the other hand can vary over many orders of magnitude if the parameters of the model are changed and a good initial guess can drastically improve the performance of the algorithm.

As explained earlier, we reduce $\delta t$ once the exponential error decreases below the threshold set by $\varepsilon_\text{tol}$. We chose to perform this reduction geometrically, i.e.\ $\delta t \mapsto\xi\delta t$ with some factor $\xi\in\left(0,1\right)$. It is important to choose $\xi$ carefully. If it is chosen too small, unnecessarily many steps are performed with the small step size where a few larger steps would have sufficed. On the other hand too large values of $\xi$ lead to a slow decrease of $\delta t$ implying a slow continuous time convergence. We find that usually the results are best for $\xi\in\left[0.7,0.9\right]$.

\subsubsection{Convergence criteria}
Our termination criterion for the evolution is the point where the exponential error is diminished as explained before and $\delta t$ has dropped below some threshold. This threshold usually can be chosen surprisingly high because the extrapolation of the values without exponential error reduces the Trotter error to $\ordnung{\delta t^{r+1}}$.

\subsection{Energy estimator\label{sec:energy_estimator}}
The energy can be measured as described before by calculating the expectation values of the local terms of the Hamiltonian as in \cref{fig:expectation_value}. This requires the calculation \env, which is the leading cost of the whole algorithm, for every site. If we use a Full Update algorithm, \env has to be calculated in each update step as well. Therefore, calculating the energy is as expensive as a step of the imaginary time evolution. It can be even cheaper if parts of the previously calculated boundary MPS are reused.

The situation looks different, however, when we use Simple Update. Each update step scales at most as $\ordnung{D^4}$ in the bond dimension for the honeycomb lattice. This is much cheaper than the calculation of an expectation value. Hence, we used an estimator for the energy to avoid calculating \env after each update step.

From \cref{eq:U_trotterized} follows (up to discretization errors)
\begin{align}
	H &= - \frac{1}{\delta t} \ln{U(\delta t)} \\
	\Rightarrow\,E &= - \frac{1}{\delta t} \frac{\bra{\Psi} \ln{U(\delta t)} \ket{\Psi}}{\braket{\Psi|\Psi}}\,.
\end{align}
If $\ket{\Psi}$ is the ground state of the system, we can approximate this as
\begin{align}
	E &\approx - \frac{1}{\delta t} \ln{\sqrt{\frac{\braket{\Psi | U(\delta t)^2 |\Psi}}{\braket{\Psi|\Psi}}}}\\
	&= - \frac{1}{\delta t} \ln{\sqrt{\frac{\braket{\Psi'|\Psi'}}{\braket{\Psi|\Psi}}}}
	= - \frac{1}{\delta t} \ln{\prod \lambda}\,.
\end{align}
In the last line we named the new state after an imaginary time step $\ket{\Psi'}$. The renormalization factor \renorm corresponds to the change of the norm of a state when an imaginary time evolution operator is applied locally, see \cref{eq:renormalization}. In the product $\prod \renorm$  we combine the individual renormalization factors  from all two-site updates involved in a global update. With this procedure we get an estimator of the energy from the Simple Update algorithm without significant additional costs.

The estimator is only exact up to the Trotter error. It also assumes that we converged to the ground state. In some cases the estimator reproduces the energy quite reliably even if convergence or a small $\delta t$ are not given. In other cases the estimator was far off the real energy. See \cref{fig:stair,fig:degenerate,fig:no_coupling_jump} for examples.

We found the estimator useful to check for convergence within a given trotter step size $\delta t$. If the energy estimator does not change anymore after an imaginary time step with a given step size, the state converged to the ground state of the trotterized hamiltonian. Therefore, we can reduce the step size. This convergence check is much cheaper than the calculation of an expectation value after each imaginary time step.

\subsection{Instabilities caused by degenerate states\label{sec:instabilities_degeneracies}}
\begin{figure*}[ht]
	\centering
	\subfigure[$\mu=B=0$\label{fig:degenerate_unstable}]{\input{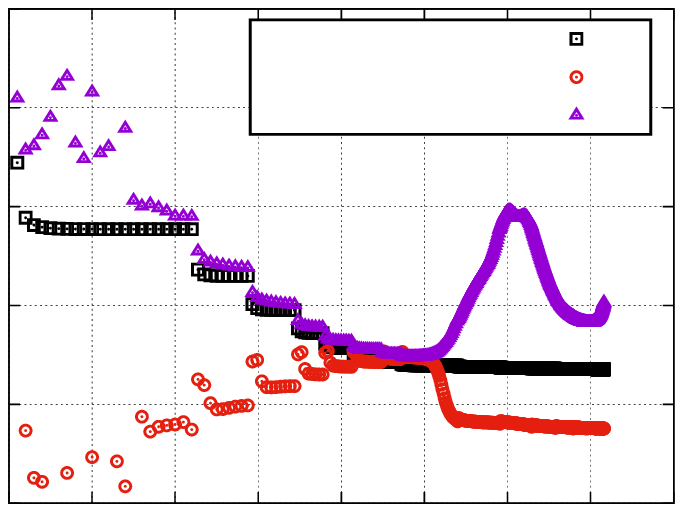}}
	\subfigure[$\mu=B=0.1$\label{fig:degenerate_stable}]{\input{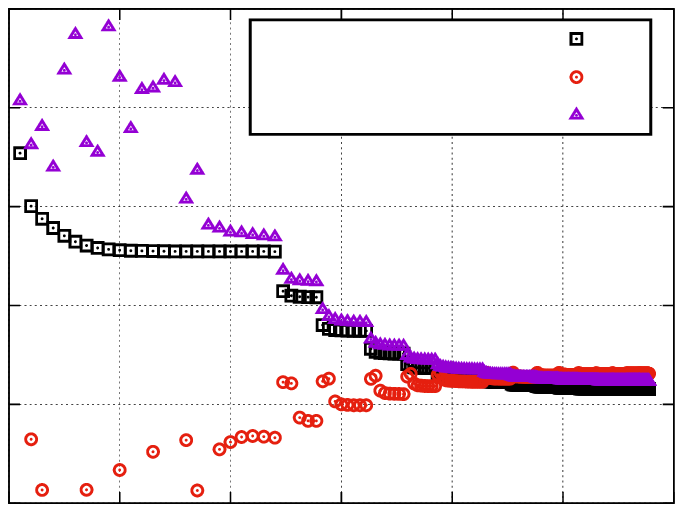}}
	\caption{Imaginary time evolution with $\kappa=1$, $U=4$ on the $2\times 4$-lattice with $D=8$ and odd parity.
		Energies calculated using boundary MPS (bMPS) and the direct estimator derived in \cref{sec:energy_estimator}. As a reference we provide the exact imaginary time evolution of the state vector obtained via full contraction of the initial PEPS.}
	\label{fig:degenerate}
\end{figure*}

Even if all the criteria described above are fulfilled (including a good estimation of the gap $\Delta_\text{ex}$), the imaginary time evolution is still not guaranteed to converge to the correct value. As a matter of fact it might not converge at all. One such example is shown in \cref{fig:degenerate_unstable}, where the results first approach the correct values. At some point however the energy estimator does not decrease monotonously any more. It begins to rise and to fluctuate. Again much later the results become numerically unstable and chaotic.

Fortunately, the onset of this problem has various indicators which we can track. Not only does the full contraction energy estimator fail to decrease further, in our implementation it also gains a significant imaginary part as compared to machine precision. Under normal circumstances the imaginary part is on the order of $10^{-13}$, but when the time evolution starts to fail it increases drastically. It becomes clear that we are facing numerical instabilities when taking into account that even the norm estimator obtains an imaginary part. One further, though less reliable, indicator is a step-like decrease of the Simple Update energy estimator. The step does not always occur and it sometimes stabilises at a new level, but other times it continues decreasing without bounds.

We find that such instabilities occur only if the ground state of the simulated system is degenerate. It is plausible that degeneracies lead to numerical instabilities because the ambiguity of the unitary matrices in an SVD is dramatically increased once the principal values are no longer pairwise distinct. This is similar to simulations without gauge fixing which are known to be unstable. We find that breaking the degeneracy, if ever slightly, resolves the problem. In \cref{fig:degenerate_unstable} the simulated system has a four-fold degenerate ground state and we clearly observe the described instability. With all symmetries broken (see next section for details) and a non-degenerate ground state the evolution converges stably as can be seen in \cref{fig:degenerate_stable}.

\subsubsection{Degeneracies in the Hubbard model}
The ground state of the Hubbard model at half filling (i.e.\ $\mu$=0) lies in the even-parity sector and is non-degenerate~\cite{arovas2021hubbard}. Therefore, simulations of the full space and the even-parity subspace do not feature the instabilities described above. One has to be careful nevertheless because influences like a strong chemical potential can reintroduce degeneracies. In the case of large $\mu$ the ground state falls into a sector with non-zero\footnote{In our convention half filling has exactly zero particles though of course the electron number has to equal the number of lattice sites. It is a simple shift by a constant.} particle number which might be degenerate.

The ground state of the odd parity sector on the other hand a priori comes with a four-fold degeneracy. It can have one spin up or down particle less or more than the ground state, all leading to the same energy at $\mu=0$ with no external magnetic field $B$. This degeneracy is easily broken by introducing a small but non-zero chemical potential and a magnetic field of similar magnitude. The magnitude has to be chosen carefully as it has to be large enough to ensure numerical stability but small enough to keep the ground state in the desired sector. In practice $\mu\sim 10^{-2} \kappa$ usually yields good results and the quality of the simulation is not very sensitive to the exact value of $\mu$.

The ground state energy $E_0$ without chemical potential and magnetic field can be easily regained after a simulation from the measured one:
\begin{align}
	E_0(\mu,B) &= E_0 - |n\mu| - |MB|\,.
	\label{eq:degenerecy_correction}
\end{align}
We exploit the fact here that particle number $n$ and magnetization $M$ are both good quantum numbers of the Hubbard model, i.e.\  the corresponding operators commute with the Hamiltonian. We use the absolute values in \cref{eq:degenerecy_correction} for clarity as the ground state of the broken system always has lower energy than the ground state of the unbroken one.

Finally, we point out that the lattice geometry itself can exhibit an additional, more subtle, degeneracy that is not as easily broken and regained after the simulation. This degeneracy comes from a high spatial symmetry of the lattice or, put differently, is a degeneracy in momentum space. It occurs only if the symmetry group of the lattice has at least one irreducible representation of dimension larger than one. In this case the degeneracy is protected by symmetry and cannot be broken by any translationally invariant Hamiltonian. Any ring with an even number $N$ of sites exhibits this feature as it falls into the dihedral symmetry group $D_N$. In our case of rectangular pieces of the hexagonal lattice with OBC it turns out that the 6-ring is the only lattice with such a high symmetry. All the other lattices can only have reflection symmetries and thus only one-dimensional irreducible representations.
We can therefore easily avoid this degeneracy by choosing the geometry accordingly from here on.

\subsection{Instabilities caused by the single particle gap}
A different problem arises in the case of a very large single particle gap $\Delta_\text{sp}\gg\Delta_\text{ex}$. The single particle gap is the difference between even and odd parity ground state energies $E_0^e$ and $E_0^o$ respectively. It turns out that in this regime the simulation in the even parity sector works fine, but the odd parity simulation becomes unreliable. During the time evolution the energy in the odd parity sector can drop to that of the even parity sector for example. To demonstrate this type of instability explicitly, we simulate in the regime where $\kappa=0$, thereby enforcing $\Delta_\text{sp}\gg\Delta_\text{ex}$.
\begin{figure}[ht]
	\centering
	\input{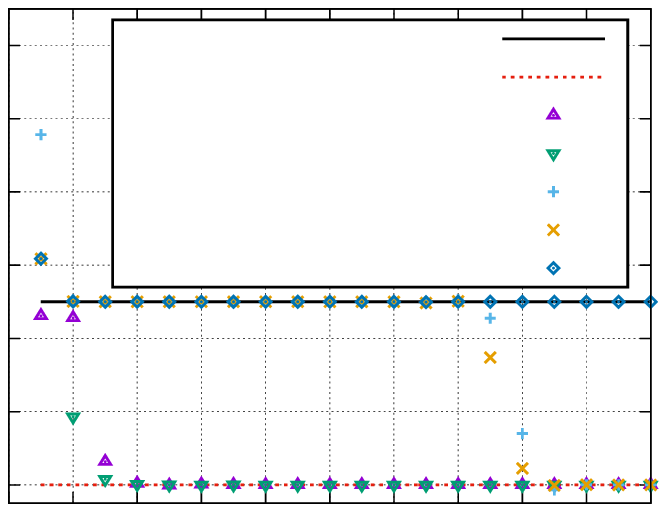}
	\caption{Energy during the imaginary time evolution with $U=1$ and $\kappa=\mu=B=0$ on the $3\times 4$-lattice with odd parity. The energies have been calculated using boundary MPS (bMPS) and the direct estimator derived in \cref{sec:energy_estimator} for Simple Update. Full Update simulation with bMPS energy calculation.}
	\label{fig:no_coupling_jump}
\end{figure}

Such a case is shown in \cref{fig:no_coupling_jump} where an odd parity simulation has been attempted. The energy estimator from the change of the norm during the Simple Update goes down to $E_0^e=0$ immediately. The energy estimator from the boundary MPS contraction originally converges correctly against $E_0^o=\frac12 U$, but then jumps down to $E_0^e$ after some time $t\approx 35$. Only the simulation using Full Update remains stably at the correct energy.

The chosen bond dimension $D=2$ suffices to solve the problem at hand exactly because with no hopping the Hamiltonian decomposes into its local contributions. On every lattice point there is a twofold degenerate ground state (one particle up or down) with energy 0 and a twofold degenerate excited state (none or two particles) with energy $U/2$. That means that only two independent dimensions have to be considered. Thus the instability cannot be alleviated with larger $D$.

\subsubsection{Tensor structure}
To understand how this can happen, let us revisit the structure we imposed onto the tensors by the condition that only parity-conserving entries can be non-zero. We consider the simplest possible example, namely a two site lattice with OBC and $d=D=2$. Then both tensors can be written as $2\times2$-matrices. We add a parity-index to the left tensor and fix it to be odd. \Cref{eq:conserve_parity} now guarantees that the full tensor network takes the form
\begin{align}
	M_1M_2 &\coloneqq \matr{ll}{0&a_1\\b_1&0}\matr{ll}{a_2&0\\0&b_2}\\
	&= \matr{ll}{0&a_1b_2\\a_2b_1&0}\,,\label{eq:two_site_tn_odd}
\end{align}
which is equivalent to the state vector $v\coloneqq\left(0,a_1b_2,a_2b_1,0\right)$ encoding the two-dimensional odd subspace of the $d^2=4$-dimensional state space.
Next, we apply a time evolution operator $U$ to both sites independently. Again to keep things simple, we choose $U=\mathrm{diag}\left(1,\epsilon\right)$ where we are interested in the case of $\epsilon\ll1$. The evolved network reads
\begin{align}
	\left(UM_1\right)\left(M_2U\right) &= \matr{ll}{0&a_1\\\epsilon b_1&0}\matr{ll}{a_2&0\\0&\epsilon b_2}\label{eq:local_time_evol}\\
	&= \matr{ll}{0&\epsilon a_1b_2\\\epsilon a_2b_1&0}\,.
\end{align}
This means that ultimately the eigenstate $v$ is multiplied by $\epsilon$. The tensors $M_{1,2}$ however are not `eigentensors' of the time evolution.

\subsubsection{Suppression of a subspace}
\Cref{eq:local_time_evol} shows why the application of this energy estimator of \cref{sec:energy_estimator} using the local norm change is problematic. Repeated applications of $U$ lead to a suppression of the $b$ terms. Once they are much smaller than the $a$ terms, the matrix norm no longer changes significantly with further applications of $U$. The energy estimator thus becomes zero, or more generally $E_0^e$.

With this in hand, we can also estimate when the jump, i.e.\ the more important problem, will occur. We expect that the odd parity sector will not be represented sufficiently well any more once the $b$ terms reach the order of machine precision relative to the $a$ terms. We call the threshold below which stability cannot be guaranteed any more $\varepsilon_\text{stable}$.  This leads to the stability condition
\begin{align}
	\eto{-\Delta_\text{sp} t} &> \varepsilon_\text{stable} \\
	\Rightarrow\,t &< \frac{\ln\left(1/ \varepsilon_\text{stable}\right)}{\Delta_\text{sp}}\,.\label{eq:stable_time}
\end{align}
In the example of \cref{fig:no_coupling_jump} we find $\varepsilon_\text{stable}\approx10^{-8}$.

One would expect $\varepsilon_\text{stable}$ to depend on $D$ because more numerical errors can accumulate with a larger number of entries. This is not always the case however, as can be clearly seen in \cref{fig:no_coupling_jump}. Unfortunately, this property is not generalizable and there are cases where larger $D$ can worsen the problem.
Such examples are visualized in \cref{fig:deviations,fig:even_odd_energy_3x4,fig:smallest_unstable_D} and explained in detail in \cref{sec:results_ground_states}.

As the stability condition takes the same form as the convergence condition for the exponential error due to the gap $\Delta_\text{ex}$, we can easily extract the ratio between the time $t_\text{stable}$ for which a stable evolution is possible and the time $t_\text{conv}$ needed to reach the precision $\varepsilon_\text{conv}$, giving
\begin{align}
	\frac{t_\text{stable}}{t_\text{conv}} &=\frac{\Delta_\text{ex}}{\Delta_\text{sp}} \frac{\ln \varepsilon_\text{stable}}{\ln \varepsilon_\text{conv}}\,.
\end{align}
The ratio of the logarithms can vary depending on the desired precision and the dimensionality of the system, but in all realistic scenarios it is of order 1 or slightly larger. Therefore, the relevant factor is the ratio between the different gaps. One finds that a reliable simulation is possible if and only if $\Delta_\text{ex}\gtrsim\Delta_\text{sp}$.

Note that \cref{eq:stable_time} provides a lower bound for $t_\text{stable}$, i.e. the time in which stability is guaranteed. It is nevertheless possible that simulations proceed stably for a much longer time. We observe that this is in fact usually the case. A possible explanation is that interactions involving more than one site average out the effect of suppressed subspaces on a larger part of the lattice and thus mitigates the problem. The example above has no such interactions and hence accumulates the factors locally.

\subsubsection{Adding the even parity sector}
The example above \cref{eq:two_site_tn_odd} assumes an odd external parity index at $M_1$. In the more general case we would allow this external index to take both values and therefore obtain two copies of the matrix $M_1^o$ and $M_1^e$ where the odd version $M_1^o$ corresponds to $M_1$ in \cref{eq:two_site_tn_odd}. In this case an imaginary time step evolutes the network as
\begin{align}
	M_1^oM_2 &\sim \matr{ll}{0&1\\\epsilon&0}\matr{ll}{1&0\\0&\epsilon}\\
	&= \matr{ll}{0&\epsilon\\\epsilon&0}
\end{align}
in the odd parity part and as
\begin{align}
	M_1^eM_2 &\sim \matr{ll}{1&0\\0&\epsilon}\matr{ll}{1&0\\0&\epsilon}\\
	&= \matr{ll}{1&0\\0&\epsilon^2}
\end{align}
in the even parity sector.

\subsubsection{Full Update as a possible fix}
So far we are not aware of any solution to this problem as long as Simple Update is used. Full Update on the other hand provides a solution as can be seen in \cref{fig:no_coupling_jump}, be it a very costly one. In future hybrid solutions might be considered performing several steps with Simple Update and then a stabilizing step with Full Update. We believe that Full Update does not suffer instabilities from this problem because the full contraction of the network and the following normalization prevents the scale separation of individual blocks as in \cref{eq:local_time_evol}. More research however is required and ongoing with respect to Full Update's stability and possible alternative fixes.
\section{Results\label{sect:results}}
We now apply the algorithms explained above to specific test models. We provide most of our results for the $3\times 4$ honeycomb lattice with OBC because this is the largest system for which results from exact diagonalization are feasible for a comparison. Some tests on larger lattices as a proof of principle are presented as well.

We find that the common choice of $\chi=D^2$ is much larger than required, see \cref{sec:boundaryMPSbondDimension}. A good convergence is observed already at $\chi=3D$. From now on all our simulations are performed with both $\chi=2D$ and $\chi=3D$. Usually the deviation between these two points is negligible compared to the differences between results at different $D$. We use Simple Update in the imaginary time steps.

\subsection{Ground state and first excited state of the Hubbard Model\label{sec:results_ground_states}}
\begin{figure*}[ht]
	\centering
	\subfigure[Even parity\label{fig:even_energy_3x4}]{\input{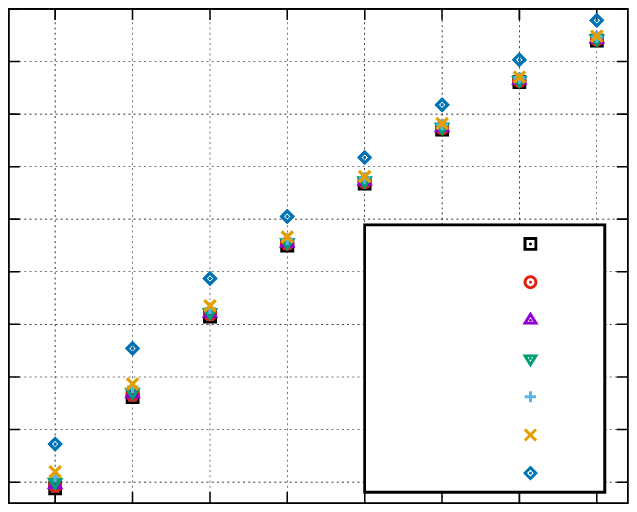}}
	\subfigure[Odd parity\label{fig:odd_energy_3x4}]{\input{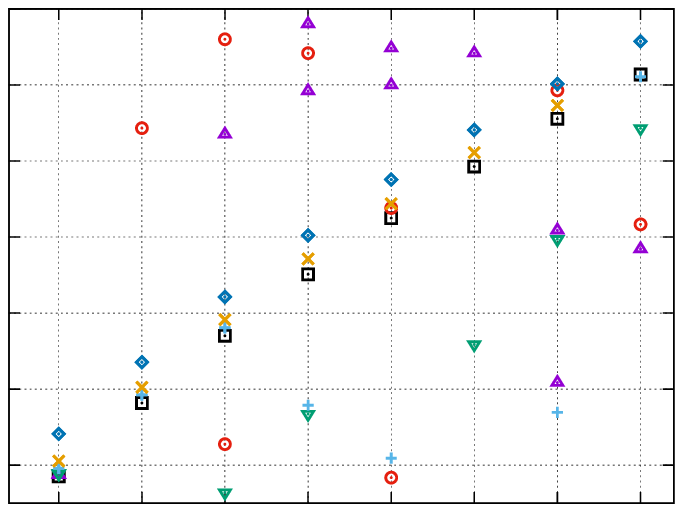}}
	\caption{Energies of the $3\times 4$ hexagonal lattice with $\kappa=1$ and $\mu=B=0$ at different values of the coupling $U$. Duplicate points correspond to $\chi=2D$ and $\chi=3D$. The legend in the left plot applies to both.}\label{fig:even_odd_energy_3x4}
\end{figure*}

\begin{figure}[ht]
	\centering
	\input{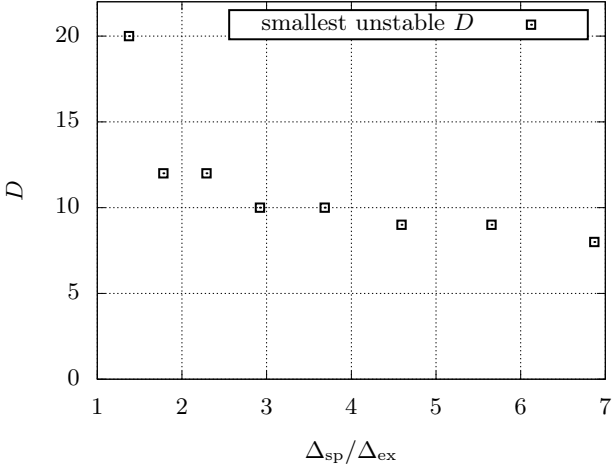}
	\caption{Smallest bond dimension $D$ for which the simulation of the $3\times 4$ hexagonal lattice with $\kappa=1$ and $\mu=B=0$ in the odd parity sector becomes unstable against the ratio of single particle and excitation gaps. Values of $U$ as in \cref{fig:even_odd_energy_3x4} resulting in different ratios $\Delta_\text{sp}/\Delta_\text{ex}$.}\label{fig:smallest_unstable_D}
\end{figure}

We first demonstrate the capabilities and limitations of our algorithm with a scan of different on-site couplings $U$ which provides a broad range of single particle gaps $\Delta_\text{sp}$. The results for both even and odd parity sectors are depicted in \cref{fig:even_odd_energy_3x4}. While the even parity calculation yields reliably good results (with errors in the sub-percent level for $D\ge 10$) for all chosen couplings, this is not the case in the odd parity sector. There, the results are similarly good for $U\lesssim 2$ corresponding to the case of $\Delta_\text{sp} < \Delta_\text{ex}$. Above this threshold, however, the stability decreases. At $U=2.5$ only simulations with bond dimensions $D\ge 20$ produce wrong results. The ever smaller bond dimensions at which the instabilities occur, are plotted in \cref{fig:smallest_unstable_D}.

\begin{figure}[ht]
	\centering
	\input{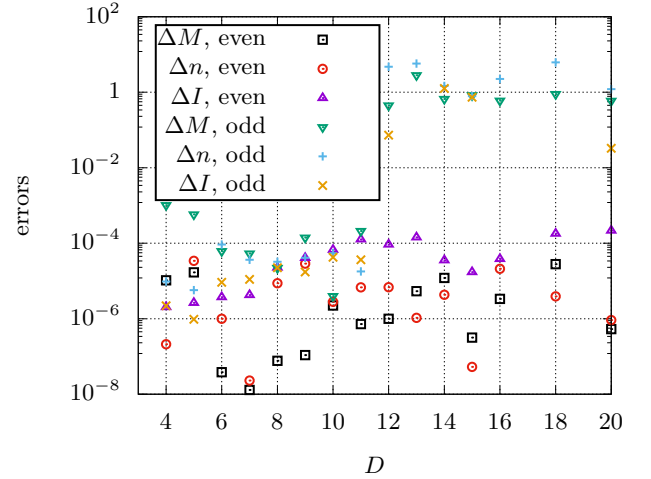}
	\caption{Standard deviation of the norm $\Delta I$ (see \cref{eq:norm_std_deviation}) and deviations of the magnetization $M$ (see \cref{eq:magnetization_definition}) and the particle number $n$ (see \cref{eq:particle_number_definition}) from the exact value.  $3\times 4$ hexagonal lattice with $\kappa=1$, $U=3$ and $\mu=B=0$ for different bond dimensions $D$ using $\chi=3D$.}\label{fig:deviations}
\end{figure}

Fortunately, the breakdown in the odd parity simulations is easily identifiable. Not only the energy but also particle number $n$ and magnetization $M$ yield completely wrong results. Since the latter are known to be integer values in a non-degenerate ground state, they can be used for a simple consistency check. A large standard deviation of the norm $\Delta I$ provides an even more reliable indicator for untrustworthy results (see \cref{eq:norm_std_deviation,fig:boundaryMPSbondDimension}). We visualize all three observables in \cref{fig:deviations} for a conditionally stable simulation. The region of $D\ge12$, where the odd parity sector features stability issues, is clearly distinct from the stable regions for all three indicators. There are two orders of magnitude separating the best unstable result from the worst stable one.

This observation implies that, even though stability problems can occur, we can still trust the results from odd parity simulations as long as they have been tested for the stability indicators, in particular $\Delta I$.

\subsection{Finite chemical potential}
\begin{figure}[ht]
	\centering
	\includegraphics[width=\columnwidth]{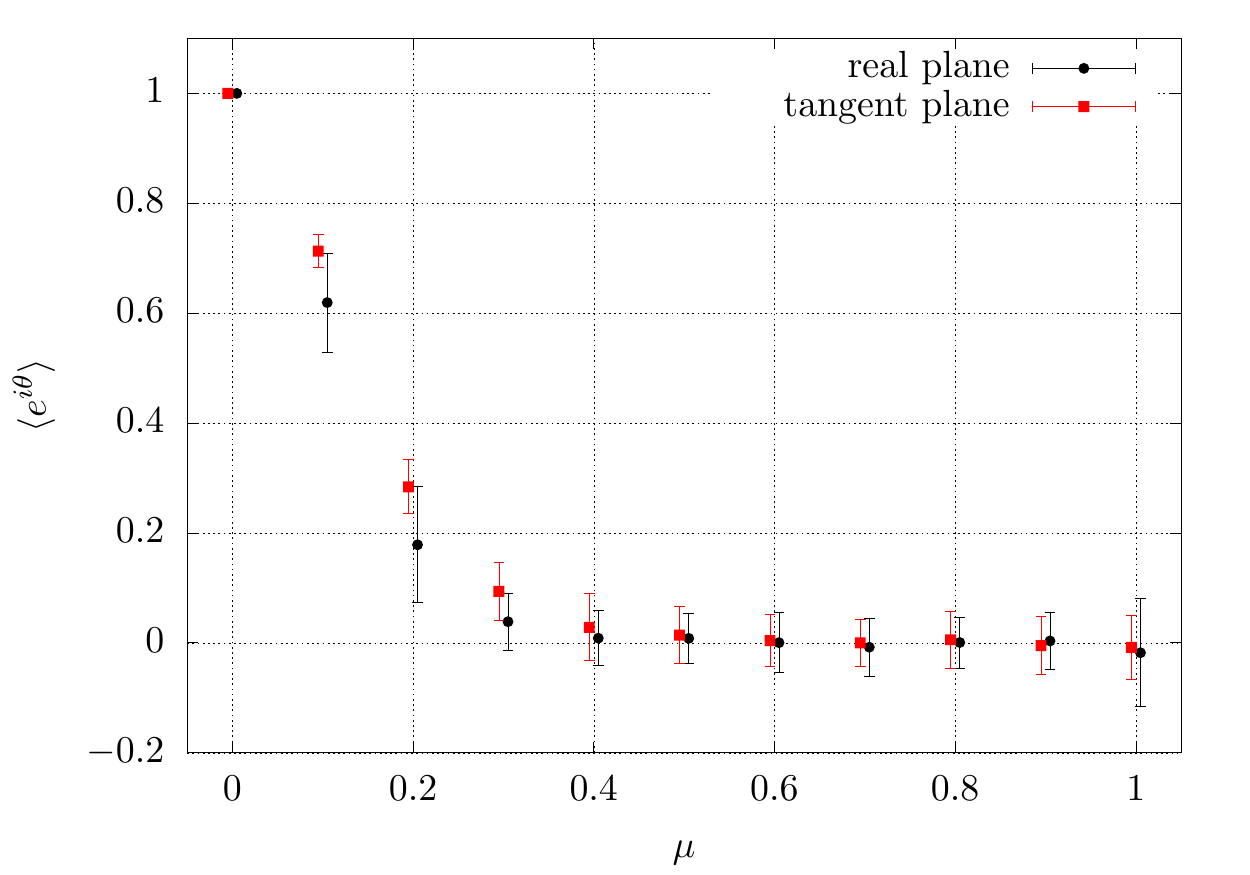}
	\caption{The expectation value of the exponential of the complex phase $\theta$ of the action for the $3\times 4$ honeycomb lattice as a function of chemical potential $\mu$.  Results are shown for integration along the real plane (black points) and integration pushed into the complex plane along the tangent plane (red squares) as described in the text.  For $\mu=0$ there is no sign problem and the expectation value is equal to 1 for both calculations.}
	\label{fig:statPower}
\end{figure}

To emphasize the efficacy of our tensor method when dealing with non-zero chemical potential $\mu$, we compare our results to those obtained from hybrid (or Hamiltonian) Monte Carlo (HMC) for the $3\times 4$ lattice with open boundaries.  The non-zero value of $\mu$ induces a complex phase in the action, which we label $\theta$ for convenience\footnote{The phase $\theta$ is itself a function of the field configurations.}. Standard HMC calculations must therefore resort to reweighting techniques when calculating some observable $O$,
\begin{equation}\label{eq:reweighting}
	\langle O\rangle = \frac{\langle O e^{i\theta}\rangle}{\langle e^{i\theta}\rangle}\,.
\end{equation}
The expectation values on the right-hand side of the equation are evaluated with the real part of the action only. The reliability of such reweighting calculations depends strongly on the fluctuations of $e^{i\theta}$ and this in turn influences its mean value $\langle e^{i\theta}\rangle$, or statistical power.  In the limit of no complex phase (i.e.\ no sign problem), the statistical power goes to one and we recover the original HMC algorithm.  On the other hand, in the limit of large fluctuations the statistical power goes to zero and the right-hand side of \cref{eq:reweighting} becomes ill-defined, precluding any reliable estimate of $\langle O\rangle$.  In \cref{fig:statPower} we show the statistical power related to reweighting (black points) for this system as a function of $\mu$.  It is clear that the sign-problem quickly becomes prohibitive for moderate $\mu\sim 0.3$ and larger.  One can attempt to alleviate the sign problem by deforming the contour of integration into the complex plane \cite{Cristoforetti:2013wha,Cristoforetti:2014gsa,Ulybyshev:2019fte}.  The simplest contour deformation is that of a constant imaginary shift of the fields to a ``tangent'' plane that intersects the action's main critical point \cite{Alexandru:2015sua,Alexandru:2015xva}.  Examples have been found where such a simple transformation of the fields greatly alleviate the sign problem \cite{leveragingML}.  We have done similar calculations with the honeycomb Hubbard system, as shown in \cref{fig:statPower} (red squares).  Though there is slight improvement in the statistical power for $\mu\le 0.3$, the sign problem still quickly overwhelms the calculation as $\mu$ increases.  

\begin{figure}[H]
	\centering
	\subfigure[Even parity]{\input{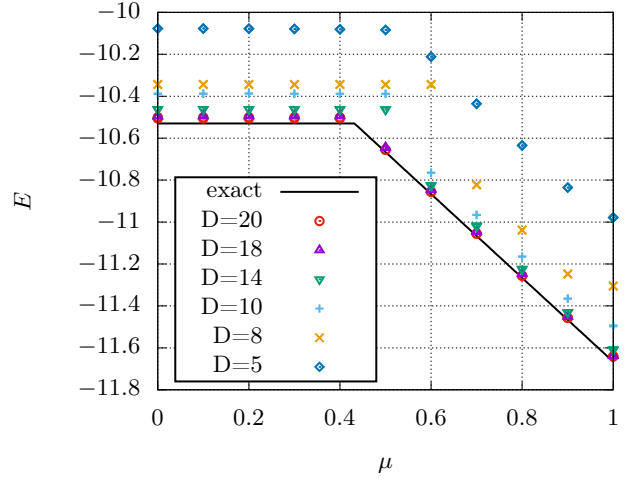}}
	\subfigure[Odd parity]{\input{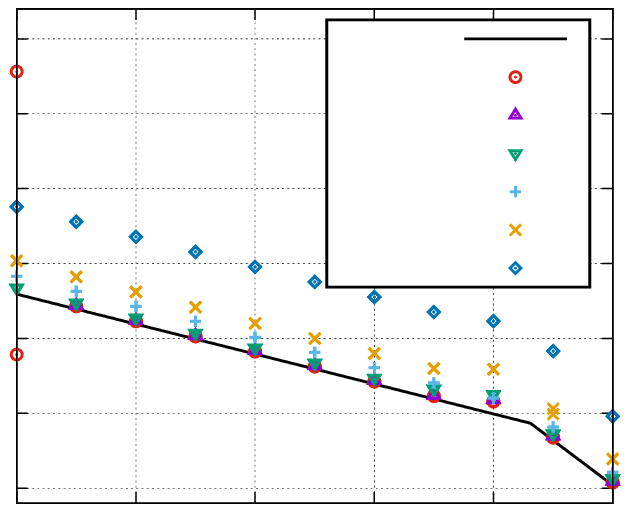}}
	\subfigure[Energy gap between even and odd parity sectors]{\input{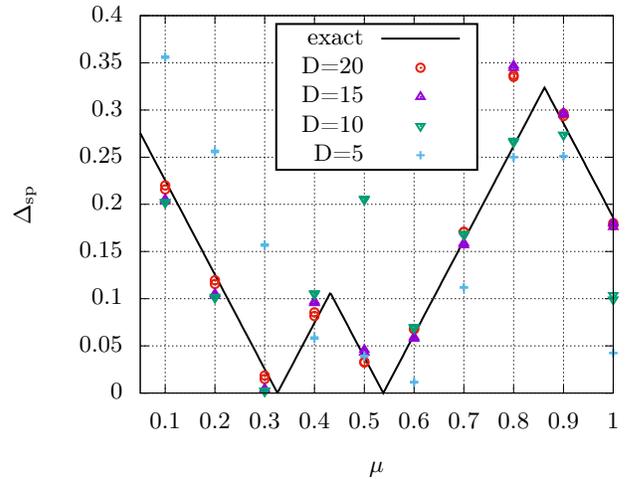}}
	\caption{Energies of the $3\times 4$ hexagonal lattice with $\kappa=1$, $U=2$ and $B=0$ at different values of $\mu$. Duplicate points correspond to $\chi=2D$ and $\chi=3D$.}\label{fig:energies_of_mu}
\end{figure}

For \tn\ simulations this problem is completely absent. \Cref{fig:energies_of_mu} visualizes the energy obtained with different bond dimensions $D$ in the even and odd parity sectors as well as the single particle gap as the difference of these energies. The same system and the same range of chemical potentials as in \cref{fig:statPower} were used.
The simulations have a lower precision in the regions where the ground state changes, i.e.\ near the kinks. This can be explained by the near degeneracy due to the small excitation gap $\Delta_\text{ex}$. We also encounter severe problems in the odd parity sector at $\mu=0$ which is to be expected as the ground state in this sector is degenerate and we encounter the problem described in \autoref{sec:instabilities_degeneracies}.
Otherwise, the results are very precise, in particular for a broad range of values $\mu>0.3$ where the HMC completely fails.

\subsection{Scalability with Volume}
\begin{figure}[ht]
	\centering
	\input{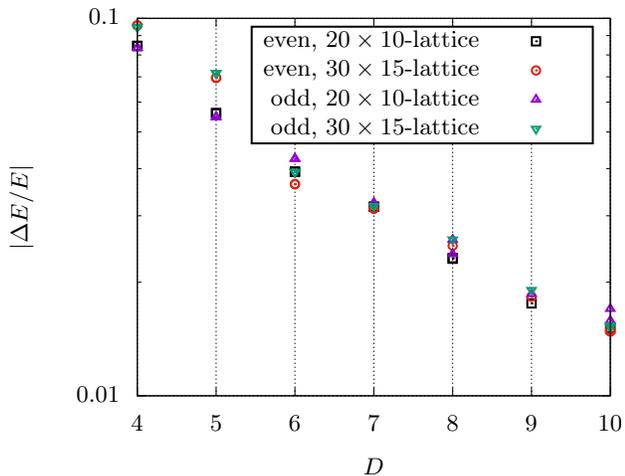}
	\caption{Relative deviations of the energies in the non-interacting limit with finite chemical potential ($\kappa=1$, $U=0$, $\mu=0.5$, $B=0$) from the exact values for different lattice sizes and bond dimensions $D$. Duplicate points correspond to $\chi=2D$ and $\chi=3D$.}\label{fig:different_large_sizes}
\end{figure}

The ability to simulate away from half filling by itself is a great advantage of the \tn\ ansatz over stochastic methods. Yet, the method is only beneficial if it demonstrates a feasible scaling in spatial volume. We know from \cref{sec:resource_scaling} that the runtime depends linearly on the lattice size and the volume's influence on memory is of subleading order. It is however not clear a priori that the same bond dimension $D$ is enough to describe larger lattices with similar precision. Therefore, the implicit effect on runtime and memory might, in principle, be significantly larger.
We find that this is not the case. As can be seen in \cref{fig:different_large_sizes}, the relative error for any specific bond dimension $D$ is virtually independent of the lattice size. This is a strong indicator that the area law is fulfilled for this system.


There is, however, another hidden cost of larger volumes. As the lattice approaches the thermodynamic limit, the excitation gap $\Delta_\text{ex}$ usually goes to zero. Thus, a longer imaginary time evolution is required to reach convergence of the true ground state. In addition, it turns out that the time steps $\delta t$ have to be chosen smaller from the very beginning for larger lattices. Too large time steps lead to numerical instabilities. The combination of these two constraints results in a significantly increased number of steps required for the imaginary time evolution.

\begin{figure*}[ht]
	\centering
	\subfigure[$U=0$]{\input{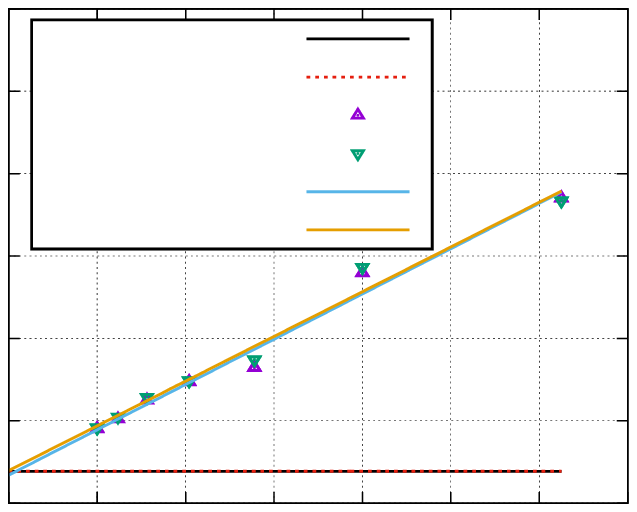}}
	\subfigure[$U=2$]{\input{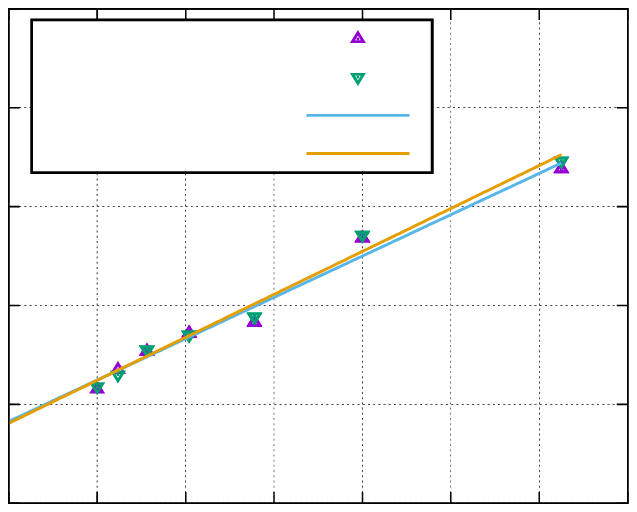}}
	\caption{Energies with finite chemical potential ($\kappa=1$, $\mu=0.5$, $B=0$) for the $30\times 15$-lattice against the inverse squared bond dimension. Duplicate points correspond to $\chi=2D$ and $\chi=3D$.}
	\label{fig:large_sizes_of inv_D2}
\end{figure*}

We present the results of a $30\times 15$-lattice\footnote{This corresponds, up to boundary conditions, to the bipartite $15\times15$ honeycomb lattice with two sites per unit cell.} simulation at non-zero chemical potential both in the non-interacting limit and at finite coupling $U=2$ in \cref{fig:large_sizes_of inv_D2}. The first serves as a proof of principle where we can compare to results obtained from exact diagonalization of the hopping matrix. The latter simulation is the first of its kind and it predicts energies that could not have been obtained otherwise to date.

The results in \cref{fig:large_sizes_of inv_D2} are depicted as a function of $D^{-2}$. We find empirically that this dependence approximates the convergence behavior very well. It should be possible to further improve the extrapolation by using the truncation error or similar quantities instead of the bond dimension~\cite{iPEPS_hubbard_2016,precise_extrapolation}.

Though we can calculate the energy well below sub-percentage precision in this way, we cannot give a reliable estimator of the single particle gap $\Delta_\text{sp}$ because for this particular choice of parameters the gap is too small to be resolved, i.e.\ even and odd energy extrapolations are compatible within uncertainties.
The results of the extrapolation in the non-interacting case are $E_\text{even} = \num{-693.2(17)}$ and $E_\text{odd} = \num{-692.1(18)}$ which in both sectors are fully compatible with the exact value $E_\text{exact} \approx \num{-692.3}$. The exact energy is the same for both parity sectors on our level of precision.
In the case of $U=2$ we find $E_\text{even} = \num{-483.5(14)}$ and $E_\text{odd} = \num{-483.8(12)}$.
The fit has been done using the Levenberg–Marquardt algorithm~\cite{levenberg,marquardt}. Ref.~\cite{lm_explained_gsl} provides a comprehensible explanation of the uncertainty estimation.
\section{Conclusion\label{sect:conclusions}}
We presented an algorithm for direct simulations of fermionic even and odd parity sectors and demonstrate its efficacy using the example of the Hubbard model. For this, we use fermionic PEPS with an additional parity link confined to one of the two subspaces. This allows us to probe explicitly for specific excited states and, among others, calculate the single particle gap. Our ability to study physical systems in both parity sectors with tensor networks is a great novelty and stands in contrast to infinite PEPS~\cite{PEPS_original_bMPS,iPEPS_original_2008,iPEPS_hubbard_2016,iPEPS_hubbard_2018} or other renormalization based algorithms~\cite{SRG_original,HOTRG_original,triad_renormalization} which often reach higher precisions for ground state calculations.

The ground state energies of the respective parity sectors can be determined with high accuracy. Using the boundary Matrix Product State method with a truncation dimension $\chi$ that scales linear in the bond dimension $D$ allowed us to use large values of $D$ to increase the precision. Extrapolations in squared inverse bond dimension $D^{-2}$ allow for uncertainties in the order of $10^{-3}$ with relatively low computational costs. Nevertheless, a higher accuracy is required before physically meaningful interpretations of the single particle gap on large lattices are possible.
The single particle gap is, of course, only one out of many different order parameters of the Hubbard model on the honeycomb lattice~\cite{more_observables} and it might be worthwhile investigating additional observables like the staggered magnetization or correlator functions.

Our simulations work reliably and efficiently for large lattices and away from half filling, but not necessarily with large onsite interactions. In the latter case Simple Update becomes increasingly unstable. We also find that degenerate ground states can significantly decrease the stability of the simulations. While the only feasible solution to the first problem seems to be the usage of Full Update, the latter problem can usually be alleviated by explicitly breaking the degeneracy.

\begin{acknowledgments}
  We thank Stefan Kühn, Román Orús and Mari Carmen Bañuls for helpful
  discussions and remarks. We acknowledge the Computer Center at DESY
  Zeuthen for the compute time. This work was funded, in part, through
  financial support by the Helmholtz Einstein International Berlin
  Research School in Data Science (HEIBRiDS),
  by the Deutsche Forschungsgemeinschaft (DFG,
  German Research Foundation) and by the 
  NSFC through the funds provided to the Sino-German
  Collaborative Research Center CRC 110 “Symmetries
  and the Emergence of Structure in QCD” (DFG Project-ID 196253076 - TRR 110, NSFC Grant No. 12070131001).
\end{acknowledgments}

\appendix
\section{Fermionic swap gates\label{app:swap}}
We introduced the fermionic swap gates in \cref{sec:fermions} as additional tensors in the \tn. In our case we never have to contract the swap gate with a tensor explicitly. Instead, we can always find a way to write the \tn such that the gate is contracted with two indices of a tensor $T_{i_1i_2}$, where any additional indices of $T$ are ignored at the moment because they are irrelevant for the consideration. With two contracted indices we can write
\begin{align}
T_{i_1i_2}X^{i_1i_2}_{j_1j_2} &= T_{i_1i_2}\delta_{i_1j_1}\delta_{i_2j_2}\,S\left(i_1,i_2\right)\\
&= T_{j_1j_2}\,S\left(j_1,j_2\right)\,.\label{eq:elementwise_prod_swap}
\end{align}
Note that the indices $j_1$ and $j_2$ in the last term \cref{eq:elementwise_prod_swap} are not contracted. This can be realized by an element wise multiplication of $T$ with the matrix $S$. Any additional indices of $T$ simply yield a multiplicity of the same operation.

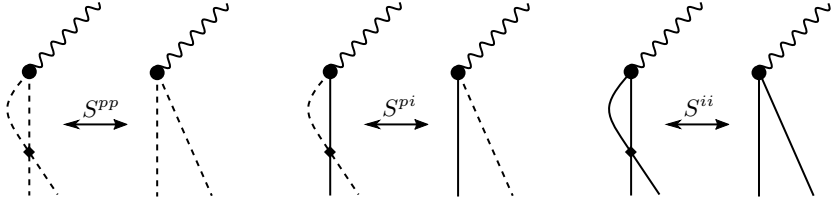
\begin{figure*}[ht]
	\centering
\begingroup%
  \makeatletter%
  \providecommand\color[2][]{%
    \errmessage{(Inkscape) Color is used for the text in Inkscape, but the package 'color.sty' is not loaded}%
    \renewcommand\color[2][]{}%
  }%
  \providecommand\transparent[1]{%
    \errmessage{(Inkscape) Transparency is used (non-zero) for the text in Inkscape, but the package 'transparent.sty' is not loaded}%
    \renewcommand\transparent[1]{}%
  }%
  \providecommand\rotatebox[2]{#2}%
  \newcommand*\fsize{\dimexpr\f@size pt\relax}%
  \newcommand*\lineheight[1]{\fontsize{\fsize}{#1\fsize}\selectfont}%
  \ifx\svgwidth\undefined%
    \setlength{\unitlength}{310.6411422bp}%
    \ifx\svgscale\undefined%
      \relax%
    \else%
      \setlength{\unitlength}{\unitlength * \real{\svgscale}}%
    \fi%
  \else%
    \setlength{\unitlength}{\svgwidth}%
  \fi%
  \global\let\svgwidth\undefined%
  \global\let\svgscale\undefined%
  \makeatother%
  \begin{picture}(1,0.23607628)%
    \lineheight{1}%
    \setlength\tabcolsep{0pt}%
    \put(0,0){\includegraphics[width=\unitlength,page=1]{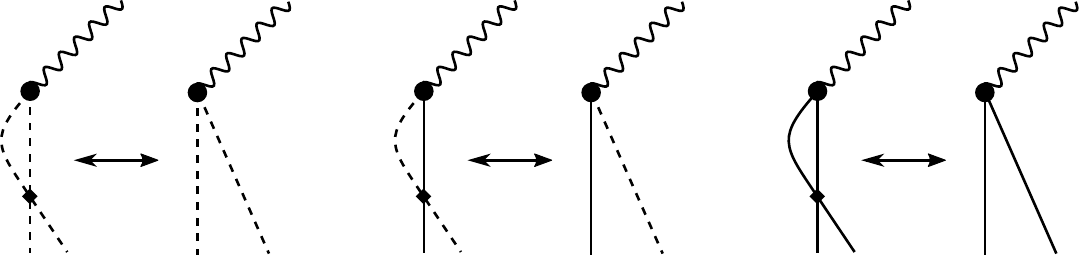}}%
    \put(0.09187633,0.09685034){\color[rgb]{0,0,0}\makebox(0,0)[lt]{\lineheight{1.25}\smash{\begin{tabular}[t]{l}$S^{pp}$\end{tabular}}}}%
    \put(0.45617084,0.09749692){\color[rgb]{0,0,0}\makebox(0,0)[lt]{\lineheight{1.25}\smash{\begin{tabular}[t]{l}$S^{pi}$\end{tabular}}}}%
    \put(0.82151404,0.09749693){\color[rgb]{0,0,0}\makebox(0,0)[lt]{\lineheight{1.25}\smash{\begin{tabular}[t]{l}$S^{ii}$\end{tabular}}}}%
  \end{picture}%
\endgroup%

	\caption{Contraction of a single swap gate with a tensor.	The left figure shows the two physical indices, 
		the middle represents a physical and an internal index, and the right the two internal indices.}
	\label{fig:swap_gate}
\end{figure*}

\Cref{fig:swap_gate} shows the different possible contractions of a swap gate. Wiggly lines summarize any number of indices that are not explicitly relevant for a given contraction but cannot be neglected nevertheless. Arrows between diagrams show the transformation under contraction.

The transformations in \cref{fig:swap_gate} are the different contractions of a swap gate as written in \cref{eq:elementwise_prod_swap}. It becomes clear that a swap gate does precisely what one would expect by interchanging two lines (best to be seen in the central panel). The swap gate is self inverse, so the transformation works the same way by a single contraction in both directions. The figure seems very redundant, but the swapping of different line types comes with different $S$-matrices. Thus the first panel interchanging two physical indices translates to
\begin{align}
	S^{pp} &= \matr{rrrr}{1&1&1&1\\1&1&1&1\\1&1&-1&-1\\1&1&-1&-1}\,,\label{eq:S_mat_phys_phys}
\end{align}
the second physical-internal panel yields
\begin{align}
	S^{pi} &= \matr{rcrrcr}{1&\dots&1&1&\dots&1\\1&\dots&1&1&\dots&1\\1&\dots&1&-1&\dots&-1\\1&\dots&1&-1&\dots&-1}\label{eq:S_mat_phys_intern}
\end{align}
and the twice internal one reads
\begin{align}
	S^{ii} &= \matr{rcrrcr}{1&\dots&1&1&\dots&1\\\vdots&\ddots&\vdots&\vdots&\ddots&\vdots\\1&\dots&1&1&\dots&1\\1&\dots&1&-1&\dots&-1\\\vdots&\ddots&\vdots&\vdots&\ddots&\vdots\\1&\dots&1&-1&\dots&-1}\,.\label{eq:S_mat_intern_intern}
\end{align}
\section{Full Update truncation for fermions\label{app:FU_optimization}}
\begin{figure}[ht]
	\centering
	\begin{equation*}
	\ket{\Psi} =
	\begin{tikzpicture}
		\def\xDistLocal{\xDist*2/3}
		\def\yShiftBra{-\xDist} 
		\def\wigglyLength{\xDistLocal/2}
		\def\physicalLength{\xDistLocal/2}
		\node[ket] (triadLU) at (0,0) {};
		\node (triadLD) at ([shift=({0,\yShiftBra/2-\gateHeight/2-\physicalLength})]triadLU.center) {};
		\node[ket] (triadRU) at ([shift=({\xDistLocal,0})]triadLU.center) {};
		\draw (triadLU) -- (triadRU) node[midway] (middleUp) {};
		\node (triadRD) at ([shift=({\xDistLocal,0})]triadLD.center) {};
		\draw[wiggly] (triadLU) -- ([shift=({-\wigglyLength,0})]triadLU.west);
		\draw[wiggly] (triadRU) -- ([shift=({\wigglyLength,0})]triadRU.east);
		\node[gate, minimum width = \xDistLocal+\pepsWidth] (operator) at ([shift=({0,\yShiftBra/2})]middleUp) {$\scriptstyle\exp(-\tau H_i)$};
		\draw[physical] \connectD{triadLU}{operator};
		\draw[physical] \connectD{triadRU}{operator};
		\draw[physical] \connectU{triadLD}{operator};
		\draw[physical] \connectU{triadRD}{operator};
	\end{tikzpicture}
	\,\rightarrow\,
	\ket{\Psi'} =
	\begin{tikzpicture}
		\def\xDistLocal{\xDist*2/3}
		\def\yShiftBra{-\xDist} 
		\def\wigglyLength{\xDistLocal/2}
		\def\physicalLength{\xDistLocal/2}
		\node[ketNew] (triadLU) at (0,0) {};
		\node (triadLD) at ([shift=({0,\yShiftBra/2-\gateHeight/2-\physicalLength})]triadLU.center) {};
		\node[ketNew] (triadRU) at ([shift=({\xDistLocal,0})]triadLU.center) {};
		\draw (triadLU) -- (triadRU) node[midway] (middleUp) {};
		\node (triadRD) at ([shift=({\xDistLocal,0})]triadLD.center) {};
		\draw[wiggly] (triadLU) -- ([shift=({-\wigglyLength,0})]triadLU.west);
		\draw[wiggly] (triadRU) -- ([shift=({\wigglyLength,0})]triadRU.east);
		\draw[physical] (triadLU) -- ([shift=({0,-\physicalLength})]triadLU.south);
		\draw[physical] (triadRU) -- ([shift=({0,-\physicalLength})]triadRU.south);
	\end{tikzpicture}
\end{equation*}
	\caption{Truncation after the application of an imaginary time evolution operator. Left: original triads with operator applied. Right: new triads with given bond dimension $D$ on the link between them.}
	\label{fig:fu_optimization}
\end{figure}
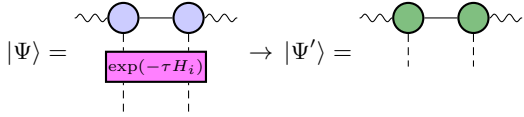

The operator $\exp(-\tau H_i)$ is applied to a pair of nearest neighbors in the imaginary time evolution. This increases the bond dimension and requires a truncation. We want to find new triads with a link of fixed bond dimension $D$ approximating the new state. This is depicted in \cref{fig:fu_optimization}. The bond dimension splits into an even and an odd parity part $D=D_e+D_o$. We keep the size of the parity sectors fixed for this truncation procedure.

The truncation shall be optimal in the sense that the difference
\begin{equation}
	f = \norm{\ket{\Psi} - \ket{\Psi'}}^2 = \braket{\Psi | \Psi} - \braket{\Psi | \Psi'} - \braket{\Psi' | \Psi} + \braket{\Psi' | \Psi'}
\end{equation}
between the old state and the new state is minimal. Scalar products mean summation of the corresponding physical indices between the bra and the ket as well as a contraction of the wiggly lines with the rest of the tensor network \env (see \cref{fig:expectation_value}).

The optimization involves two triads with two parity sectors each. Therefore we have to update four tensors. As an example, we present the update of the even parity part of the left tensor here. The other tensors follow correspondingly. We first calculate the derivative of our loss function $f$ with respect to the tensor to be updated:
\begin{align}
	\pdv{f}{\begin{tikzpicture}[scale=0.6, every node/.style={scale=0.7}]
	\def\xDistLocal{\xDist*2/3}
	\def\yShiftBra{-\xDist} 
	\def\wigglyLength{\xDistLocal/2}
	\def\physicalLength{\xDistLocal/2}
	\node[braNew] (triadLU) at (0,0) {};
	\draw[wiggly] (triadLU) -- ([shift=({-\wigglyLength,0})]triadLU.west);
	\draw[physical] (triadLU) -- ([shift=({0,\physicalLength})]triadLU.north);
	\draw (triadLU) -- ([shift=({\wigglyLength,0})]triadLU.east) node[midway, above] {$e$};
\end{tikzpicture}} =& \braket{\pdv{\Psi'}{\begin{tikzpicture}[scale=0.6, every node/.style={scale=0.7}]
	\def\xDistLocal{\xDist*2/3}
	\def\yShiftBra{-\xDist} 
	\def\wigglyLength{\xDistLocal/2}
	\def\physicalLength{\xDistLocal/2}
	\node[braNew] (triadLU) at (0,0) {};
	\draw[wiggly] (triadLU) -- ([shift=({-\wigglyLength,0})]triadLU.west);
	\draw[physical] (triadLU) -- ([shift=({0,\physicalLength})]triadLU.north);
	\draw (triadLU) -- ([shift=({\wigglyLength,0})]triadLU.east) node[midway, above] {$e$};
\end{tikzpicture}}|\Psi'} - \braket{\pdv{\Psi'}{\begin{tikzpicture}[scale=0.6, every node/.style={scale=0.7}]
	\def\xDistLocal{\xDist*2/3}
	\def\yShiftBra{-\xDist} 
	\def\wigglyLength{\xDistLocal/2}
	\def\physicalLength{\xDistLocal/2}
	\node[braNew] (triadLU) at (0,0) {};
	\draw[wiggly] (triadLU) -- ([shift=({-\wigglyLength,0})]triadLU.west);
	\draw[physical] (triadLU) -- ([shift=({0,\physicalLength})]triadLU.north);
	\draw (triadLU) -- ([shift=({\wigglyLength,0})]triadLU.east) node[midway, above] {$e$};
\end{tikzpicture}}|\Psi}\\
	=& \begin{tikzpicture}
	\def\xDistLocal{\dimexpr \xDist*2/3 \relax}
	\def\xShift{\dimexpr  \xDist/2 \relax} 
	\def\yShift{\dimexpr \xShift \relax} 
	\def\yShiftBra{\dimexpr \xShift*2 \relax} 
	\def\textDist{\xDist/15} 
	\node[ketNew] (triadLU) at (0,0) {};
	\node[nodeInvisible] (triadLD) at ([shift=({0,-\yShiftBra})]triadLU.center) {};
	\node[ketNew] (triadRU) at ([shift=({\xDistLocal,0})]triadLU.center) {};
	\draw (triadLU) -- (triadRU) node[midway,label={below:$e$}] (middleUp) {};
	\node[braNew] (triadRD) at ([shift=({\xDistLocal,0})]triadLD.center) {};
	\draw (triadLD) -- (triadRD) node[midway,label={below:$e$}] {};
	\draw[physical] (triadLU) -- (triadLD);
	\draw[physical] (triadRU) -- (triadRD);
	\node[env] (env) at ([shift=({0,\yShift})]middleUp) {\env};
	\draw[wiggly] \connectBezier{triadLU.west}{node cs:name=env,angle=-160}{-\xWiggly/2};
	\draw[wiggly] \connectBezier{triadLD.west}{node cs:name=env,angle=160}{-\xWiggly};
	\draw[wiggly] \connectBezier{triadRU.east}{node cs:name=env,angle=-20}{\xWiggly/2};
	\draw[wiggly] \connectBezier{triadRD.east}{node cs:name=env,angle=20}{\xWiggly};
\end{tikzpicture} + \begin{tikzpicture}
	\def\xDistLocal{\dimexpr \xDist*2/3 \relax}
	\def\xShift{\dimexpr  \xDist/2 \relax} 
	\def\yShift{\dimexpr \xShift \relax} 
	\def\yShiftBra{\dimexpr \xShift*2 \relax} 
	\def\textDist{\xDist/15} 
	\node[ketNew] (triadLU) at (0,0) {};
	\node[nodeInvisible] (triadLD) at ([shift=({0,-\yShiftBra})]triadLU.center) {};
	\node[ketNew] (triadRU) at ([shift=({\xDistLocal,0})]triadLU.center) {};
	\draw (triadLU) -- (triadRU) node[midway,label={below:$o$}] (middleUp) {};
	\node[braNew] (triadRD) at ([shift=({\xDistLocal,0})]triadLD.center) {};
	\draw (triadLD) -- (triadRD) node[midway,label={below:$e$}] {};
	\draw[physical] (triadLU) -- (triadLD);
	\draw[physical] (triadRU) -- (triadRD);
	\node[env] (env) at ([shift=({0,\yShift})]middleUp) {\env};
	\draw[wiggly] \connectBezier{triadLU.west}{node cs:name=env,angle=-160}{-\xWiggly/2};
	\draw[wiggly] \connectBezier{triadLD.west}{node cs:name=env,angle=160}{-\xWiggly};
	\draw[wiggly] \connectBezier{triadRU.east}{node cs:name=env,angle=-20}{\xWiggly/2};
	\draw[wiggly] \connectBezier{triadRD.east}{node cs:name=env,angle=20}{\xWiggly};
\end{tikzpicture} \nonumber \\
	&- \begin{tikzpicture}
	\def\xDistLocal{\dimexpr \xDist*2/3 \relax}
	\def\xShift{\dimexpr  \xDist/2 \relax} 
	\def\yShift{\dimexpr \xShift \relax} 
	\def\yShiftBra{\dimexpr \xShift*2 \relax} 
	\def\textDist{\xDist/15} 
	\node[ket] (triadLU) at (0,0) {};
	\node[nodeInvisible] (triadLD) at ([shift=({0,-\yShiftBra})]triadLU.center) {};
	\node[ket] (triadRU) at ([shift=({\xDistLocal,0})]triadLU.center) {};
	\draw (triadLU) -- (triadRU) node[midway] (middleUp) {};
	\node[braNew] (triadRD) at ([shift=({\xDistLocal,0})]triadLD.center) {};
	\draw (triadLD) -- (triadRD) node[midway,label={below:$e$}] {};
	\node[gate, minimum width = \xDistLocal+\pepsWidth] (operator) at ([shift=({0,-\yShiftBra/2})]middleUp) {$\scriptstyle\exp(-\tau H_i)$};
	\draw[physical] \connectD{triadLU}{operator};
	\draw[physical] \connectD{triadRU}{operator};
	\draw[physical] \connectU{triadLD}{operator};
	\draw[physical] \connectU{triadRD}{operator};
	\node[env] (env) at ([shift=({0,\yShift})]middleUp) {\env};
	\draw[wiggly] \connectBezier{triadLU.west}{node cs:name=env,angle=-160}{-\xWiggly/2};
	\draw[wiggly] \connectBezier{triadLD.west}{node cs:name=env,angle=160}{-\xWiggly};
	\draw[wiggly] \connectBezier{triadRU.east}{node cs:name=env,angle=-20}{\xWiggly/2};
	\draw[wiggly] \connectBezier{triadRD.east}{node cs:name=env,angle=20}{\xWiggly};
\end{tikzpicture}\,.
\end{align}
To find the minimum of $f$, we set the derivative to zero. This leads to a linear equation for the triad of interest. Its solution is
\begin{align}
	\begin{tikzpicture}[scale=0.6, every node/.style={scale=0.7}]
	\def\xDistLocal{\xDist*2/3}
	\def\yShiftBra{-\xDist} 
	\def\wigglyLength{\xDistLocal/2}
	\def\physicalLength{\xDistLocal/2}
	\node[ketNew] (triadLU) at (0,0) {};
	\draw[wiggly] (triadLU) -- ([shift=({-\wigglyLength,0})]triadLU.west);
	\draw[physical] (triadLU) -- ([shift=({0,-\physicalLength})]triadLU.south);
	\draw (triadLU) -- ([shift=({\wigglyLength,0})]triadLU.east) node[midway, above] {e};
\end{tikzpicture} =&	\begin{tikzpicture}
	\def\xDistLocal{\dimexpr \xDist*2/3 \relax}
	\def\xShift{\dimexpr  \xDist/2 \relax} 
	\def\yShift{\dimexpr \xShift \relax} 
	\def\yShiftBra{\dimexpr \xShift*2 \relax} 
	\def\textDist{\xDist/15} 
	\node[ket] (triadLU) at (0,0) {};
	\node[nodeInvisible] (triadLD) at ([shift=({0,-\yShiftBra})]triadLU.center) {};
	\node[env] (Minv) at ([shift=({0,-\yShiftBra/2})]triadLD.center) {$M^{-1}$};
	\node[ket] (triadRU) at ([shift=({\xDistLocal,0})]triadLU.center) {};
	\draw (triadLU) -- (triadRU) node[midway] (middleUp) {};
	\node[braNew] (triadRD) at ([shift=({\xDistLocal,0})]triadLD.center) {};
	\draw (node cs:name=Minv, angle=45) .. controls +(0,\yShiftBra/2-\gateHeight/2) .. (triadRD) node[midway,label={below:$e$}] {};
	\node[gate, minimum width = \xDistLocal+\pepsWidth] (operator) at ([shift=({0,-\yShiftBra/2})]middleUp) {$\scriptstyle\exp(-\tau H_i)$};
	\draw[physical] \connectD{triadLU}{operator};
	\draw[physical] \connectD{triadRU}{operator};
	\draw[physical] \connectU{triadLD}{operator};
	\draw[physical] \connectU{triadRD}{operator};
	\node[env] (env) at ([shift=({0,\yShift})]middleUp) {\env};
	\draw[wiggly] \connectBezier{triadLU.west}{node cs:name=env,angle=-160}{-\xWiggly/2};
	\draw[wiggly] \connectBezier{triadRU.east}{node cs:name=env,angle=-20}{\xWiggly/2};
	\draw[wiggly] \connectBezier{triadRD.east}{node cs:name=env,angle=20}{\xWiggly};
	\draw[wiggly] let \p1 = (node cs:name=env,angle=160), \p2 = (triadLD.west) in (\p1) -- (\x2-\xWiggly/2,\y1) ..controls (\x2-\xWiggly,\y1) and (\x2-\xWiggly,\y2)  .. (\x2-\xWiggly/2,\y2) ..controls (triadLD.west) .. (node cs:name=Minv, angle=135);
	\draw[wiggly] (node cs:name=Minv, angle=-135) .. controls +(0,-\physicalLength) .. +(-\physicalLength,-\physicalLength);
	\draw (node cs:name=Minv, angle=-45) ..controls +(0,-\physicalLength) .. +(\physicalLength,-\physicalLength) node[midway, above=\textDist] (extEven) {};
	\node at (extEven) {$e$};
\end{tikzpicture} - \begin{tikzpicture}
	\def\xDistLocal{\dimexpr \xDist*2/3 \relax}
	\def\xShift{\dimexpr  \xDist/2 \relax} 
	\def\yShift{\dimexpr \xShift \relax} 
	\def\yShiftBra{\dimexpr \xShift*2 \relax} 
	\def\textDist{\xDist/15} 
	\node[ketNew] (triadLU) at (0,0) {};
	\node[nodeInvisible] (triadLD) at ([shift=({0,-\yShiftBra})]triadLU.center) {};
	\node[env] (Minv) at ([shift=({0,-\yShiftBra/2})]triadLD.center) {$M^{-1}$};
	\node[ketNew] (triadRU) at ([shift=({\xDistLocal,0})]triadLU.center) {};
	\draw (triadLU) -- (triadRU) node[midway,label={below:$o$}] (middleUp) {};
	\node[braNew] (triadRD) at ([shift=({\xDistLocal,0})]triadLD.center) {};
	\draw (node cs:name=Minv, angle=45) .. controls +(0,\yShiftBra/2-\gateHeight/2) .. (triadRD) node[midway,label={below:$e$}] {};
	\draw[physical] (triadLU) -- (triadLD);
	\draw[physical] (triadRU) -- (triadRD);
	\node[env] (env) at ([shift=({0,\yShift})]middleUp) {\env};
	\draw[wiggly] \connectBezier{triadLU.west}{node cs:name=env,angle=-160}{-\xWiggly/2};
	\draw[wiggly] \connectBezier{triadRU.east}{node cs:name=env,angle=-20}{\xWiggly/2};
	\draw[wiggly] \connectBezier{triadRD.east}{node cs:name=env,angle=20}{\xWiggly};
	\draw[wiggly] let \p1 = (node cs:name=env,angle=160), \p2 = (triadLD.west) in (\p1) -- (\x2-\xWiggly/2,\y1) ..controls (\x2-\xWiggly,\y1) and (\x2-\xWiggly,\y2)  .. (\x2-\xWiggly/2,\y2) ..controls (triadLD.west) .. (node cs:name=Minv, angle=135);
	\draw[wiggly] (node cs:name=Minv, angle=-135) .. controls +(0,-\physicalLength) .. +(-\physicalLength,-\physicalLength);
	\draw (node cs:name=Minv, angle=-45) ..controls +(0,-\physicalLength) .. +(\physicalLength,-\physicalLength) node[midway, above=\textDist] (extEven) {};
	\node at (extEven) {$e$};
\end{tikzpicture} \,,\label{eq:optimalTensor}
\end{align}
where $M^{-1}$ is the inverse of $M$:
\begin{align}
	\begin{tikzpicture}
	\def\xDistLocal{\dimexpr \xDist*2/3 \relax}
	\def\xShift{\dimexpr  \xDist/2 \relax} 
	\def\yShift{\dimexpr \xShift \relax} 
	\def\yShiftBra{\dimexpr \xShift*2 \relax} 
	\def\textDist{\xDist/15} 
	\phantom{
		\node[nodeInvisible] (triadLU) at (0,0) {};
		\node[nodeInvisible] (triadLD) at ([shift=({0,-\yShiftBra})]triadLU.center) {};
		\node[ketNew] (triadRU) at ([shift=({\xDistLocal,0})]triadLU.center) {};
		\draw (triadLU) -- (triadRU) node[midway,label={below:e}] (middleUp) {};
		\node[braNew] (triadRD) at ([shift=({\xDistLocal,0})]triadLD.center) {};
		\draw (triadLD) -- (triadRD) node[midway,label={below:e}] {};
		\draw[physical] (triadRU) -- (triadRD);
		\node[env] (env) at ([shift=({0,\yShift})]middleUp) {\env};
		\draw[wiggly] \connectBezier{triadLU.west}{node cs:name=env,angle=-160}{-\xWiggly/2};
		\draw[wiggly] \connectBezier{triadLD.west}{node cs:name=env,angle=160}{-\xWiggly};
		\draw[wiggly] \connectBezier{triadRU.east}{node cs:name=env,angle=-20}{\xWiggly/2};
		\draw[wiggly] \connectBezier{triadRD.east}{node cs:name=env,angle=20}{\xWiggly};
		\node[env] (Minv) at ([shift=({0,-\yShiftBra/2})]middleUp) {$M^{-1}$};
	}
	\tcbsetmacrotowidthofnode{\Mwidth}{Minv}
	\node[env,minimum width=\Mwidth] (M) at ([shift=({0,-\yShiftBra/2})]middleUp) {$M$};
	\draw[wiggly] (node cs:name=M, angle=135) ..controls +(0,\yShift/5) and ([shift=({-\xWiggly,-\yShift/5})]triadLU.west) .. (triadLU.west);
	\draw[wiggly] (node cs:name=M, angle=-135) ..controls +(0,-\yShift/5) and ([shift=({-\xWiggly,\yShift/5})]triadLD.west) .. (triadLD.west);
	\draw (node cs:name=M, angle=45) ..controls +(0,\yShift/5) and ([shift=({\xWiggly,-\yShift/5})]triadRU.east) .. (triadRU.east) node[midway,label=below:$e$] {};
	\draw (node cs:name=M, angle=-45) ..controls +(0,-\yShift/5) and ([shift=({\xWiggly,\yShift/5})]triadRD.east) .. (triadRD.east) node[midway,label=above:$e$] {};
\end{tikzpicture} \coloneqq&	\begin{tikzpicture}
	\def\xDistLocal{\dimexpr \xDist*2/3 \relax}
	\def\xShift{\dimexpr  \xDist/2 \relax} 
	\def\yShift{\dimexpr \xShift \relax} 
	\def\yShiftBra{\dimexpr \xShift*2 \relax} 
	\def\textDist{\xDist/15} 
	\node[nodeInvisible] (triadLU) at (0,0) {};
	\node[nodeInvisible] (triadLD) at ([shift=({0,-\yShiftBra})]triadLU.center) {};
	\node[ketNew] (triadRU) at ([shift=({\xDistLocal,0})]triadLU.center) {};
	\draw (triadLU) -- (triadRU) node[midway,label={below:$e$}] (middleUp) {};
	\node[braNew] (triadRD) at ([shift=({\xDistLocal,0})]triadLD.center) {};
	\draw (triadLD) -- (triadRD) node[midway,label={below:$e$}] {};
	\draw[physical] (triadRU) -- (triadRD);
	\node[env] (env) at ([shift=({0,\yShift})]middleUp) {\env};
	\draw[wiggly] \connectBezier{triadLU.west}{node cs:name=env,angle=-160}{-\xWiggly/2};
	\draw[wiggly] \connectBezier{triadLD.west}{node cs:name=env,angle=160}{-\xWiggly};
	\draw[wiggly] \connectBezier{triadRU.east}{node cs:name=env,angle=-20}{\xWiggly/2};
	\draw[wiggly] \connectBezier{triadRD.east}{node cs:name=env,angle=20}{\xWiggly};
\end{tikzpicture}\,, \\
	\begin{tikzpicture}
	\def\yShift{\dimexpr \xDist*11/20 \relax} 

	\node[env] (Minv) at (0,0) {$M^{-1}$};
	\tcbsetmacrotowidthofnode{\Mwidth}{Minv}
	\node[env,minimum width=\Mwidth] (M) at ([shift=({0,\yShift})]Minv) {$M$};
	\draw[wiggly] (node cs:name=M, angle=-135) -- (node cs:name=Minv, angle=135);
	\draw (node cs:name=M, angle=-45) -- (node cs:name=Minv, angle=45) node[midway,label=above:$e$] {};
	\draw[wiggly] (node cs:name=M, angle=135) -- +(0,\physicalLength);
	\draw (node cs:name=M, angle=45) -- +(0,\physicalLength) node[midway,label=below:$e$] {};
	\draw[wiggly] (node cs:name=Minv, angle=-135) -- +(0,-\physicalLength);
	\draw (node cs:name=Minv, angle=-45) -- +(0,-\physicalLength) node[midway,label=above:$e$] {};
\end{tikzpicture} =&	\begin{tikzpicture}
	\def\yShift{\dimexpr \xDist*11/20 \relax} 
	\phantom{
		\node[env] (Minv) at (0,0) {$M^{-1}$};
		\tcbsetmacrotowidthofnode{\Mwidth}{Minv}
		\node[env,minimum width=\Mwidth] (M) at ([shift=({0,-\yShift})]Minv) {$M$};
	}
	\path (node cs:name=Minv, angle=135) -- +(0,\physicalLength) node (LU) {};
	\path (node cs:name=Minv, angle=45) -- +(0,\physicalLength) node (RU) {};
	\path (node cs:name=M, angle=-135) -- +(0,-\physicalLength) node (LD) {};
	\path (node cs:name=M, angle=-45) -- +(0,-\physicalLength) node (RD) {};
	\draw[wiggly] (LU) -- (LD);
	\draw (RU) -- (RD) node[midway, label=above:$e$] {};
\end{tikzpicture}\,.
\end{align}

We get the optimal new tensor according to \cref{eq:optimalTensor} while keeping the other triads fixed. All four tensors are updated sequentially this way. We then calculate the truncation error $f$ and repeat the procedure iteratively until $f$ converged.

\bibliographystyle{apsrev4-2}
\bibliography{bibliography}

\begin{thebibliography}{51}%
\makeatletter
\providecommand \@ifxundefined [1]{%
 \@ifx{#1\undefined}
}%
\providecommand \@ifnum [1]{%
 \ifnum #1\expandafter \@firstoftwo
 \else \expandafter \@secondoftwo
 \fi
}%
\providecommand \@ifx [1]{%
 \ifx #1\expandafter \@firstoftwo
 \else \expandafter \@secondoftwo
 \fi
}%
\providecommand \natexlab [1]{#1}%
\providecommand \enquote  [1]{``#1''}%
\providecommand \bibnamefont  [1]{#1}%
\providecommand \bibfnamefont [1]{#1}%
\providecommand \citenamefont [1]{#1}%
\providecommand \href@noop [0]{\@secondoftwo}%
\providecommand \href [0]{\begingroup \@sanitize@url \@href}%
\providecommand \@href[1]{\@@startlink{#1}\@@href}%
\providecommand \@@href[1]{\endgroup#1\@@endlink}%
\providecommand \@sanitize@url [0]{\catcode `\\12\catcode `\$12\catcode
  `\&12\catcode `\#12\catcode `\^12\catcode `\_12\catcode `\%12\relax}%
\providecommand \@@startlink[1]{}%
\providecommand \@@endlink[0]{}%
\providecommand \url  [0]{\begingroup\@sanitize@url \@url }%
\providecommand \@url [1]{\endgroup\@href {#1}{\urlprefix }}%
\providecommand \urlprefix  [0]{URL }%
\providecommand \Eprint [0]{\href }%
\providecommand \doibase [0]{https://doi.org/}%
\providecommand \selectlanguage [0]{\@gobble}%
\providecommand \bibinfo  [0]{\@secondoftwo}%
\providecommand \bibfield  [0]{\@secondoftwo}%
\providecommand \translation [1]{[#1]}%
\providecommand \BibitemOpen [0]{}%
\providecommand \bibitemStop [0]{}%
\providecommand \bibitemNoStop [0]{.\EOS\space}%
\providecommand \EOS [0]{\spacefactor3000\relax}%
\providecommand \BibitemShut  [1]{\csname bibitem#1\endcsname}%
\let\auto@bib@innerbib\@empty
\bibitem [{\citenamefont {Orús}(2014)}]{tensor_intro}%
  \BibitemOpen
  \bibfield  {author} {\bibinfo {author} {\bibfnamefont {R.}~\bibnamefont
  {Orús}},\ }\href {https://doi.org/10.1016/j.aop.2014.06.013} {\bibfield
  {journal} {\bibinfo  {journal} {Annals of Physics}\ }\textbf {\bibinfo
  {volume} {349}},\ \bibinfo {pages} {117–158} (\bibinfo {year}
  {2014})}\BibitemShut {NoStop}%
\bibitem [{\citenamefont {Verstraete}\ \emph {et~al.}(2006)\citenamefont
  {Verstraete}, \citenamefont {Wolf}, \citenamefont {Perez-Garcia},\ and\
  \citenamefont {Cirac}}]{Verstraete:2006mgt}%
  \BibitemOpen
  \bibfield  {author} {\bibinfo {author} {\bibfnamefont {F.}~\bibnamefont
  {Verstraete}}, \bibinfo {author} {\bibfnamefont {M.~M.}\ \bibnamefont
  {Wolf}}, \bibinfo {author} {\bibfnamefont {D.}~\bibnamefont {Perez-Garcia}},\
  and\ \bibinfo {author} {\bibfnamefont {J.~I.}\ \bibnamefont {Cirac}},\ }\href
  {https://doi.org/10.1103/PhysRevLett.96.220601} {\bibfield  {journal}
  {\bibinfo  {journal} {Phys. Rev. Lett.}\ }\textbf {\bibinfo {volume} {96}},\
  \bibinfo {pages} {220601} (\bibinfo {year} {2006})},\ \Eprint
  {https://arxiv.org/abs/quant-ph/0601075} {arXiv:quant-ph/0601075}
  \BibitemShut {NoStop}%
\bibitem [{\citenamefont {Poilblanc}(2017)}]{Poilblanc:2017dfm}%
  \BibitemOpen
  \bibfield  {author} {\bibinfo {author} {\bibfnamefont {D.}~\bibnamefont
  {Poilblanc}},\ }\href {https://doi.org/10.1103/PhysRevB.96.121118} {\bibfield
   {journal} {\bibinfo  {journal} {Phys. Rev. B}\ }\textbf {\bibinfo {volume}
  {96}},\ \bibinfo {pages} {121118} (\bibinfo {year} {2017})},\ \Eprint
  {https://arxiv.org/abs/1707.07844} {arXiv:1707.07844 [cond-mat.str-el]}
  \BibitemShut {NoStop}%
\bibitem [{\citenamefont {Mambrini}\ \emph {et~al.}(2016)\citenamefont
  {Mambrini}, \citenamefont {Orus},\ and\ \citenamefont
  {Poilblanc}}]{Mambrini:2016oxl}%
  \BibitemOpen
  \bibfield  {author} {\bibinfo {author} {\bibfnamefont {M.}~\bibnamefont
  {Mambrini}}, \bibinfo {author} {\bibfnamefont {R.}~\bibnamefont {Orus}},\
  and\ \bibinfo {author} {\bibfnamefont {D.}~\bibnamefont {Poilblanc}},\ }\href
  {https://doi.org/10.1103/PhysRevB.94.205124} {\bibfield  {journal} {\bibinfo
  {journal} {Phys. Rev. B}\ }\textbf {\bibinfo {volume} {94}},\ \bibinfo
  {pages} {205124} (\bibinfo {year} {2016})},\ \Eprint
  {https://arxiv.org/abs/1608.06003} {arXiv:1608.06003 [cond-mat.str-el]}
  \BibitemShut {NoStop}%
\bibitem [{\citenamefont {Chen}\ \emph {et~al.}(2018)\citenamefont {Chen},
  \citenamefont {Vanderstraeten}, \citenamefont {Capponi},\ and\ \citenamefont
  {Poilblanc}}]{Chen:2018kfj}%
  \BibitemOpen
  \bibfield  {author} {\bibinfo {author} {\bibfnamefont {J.-Y.}\ \bibnamefont
  {Chen}}, \bibinfo {author} {\bibfnamefont {L.}~\bibnamefont
  {Vanderstraeten}}, \bibinfo {author} {\bibfnamefont {S.}~\bibnamefont
  {Capponi}},\ and\ \bibinfo {author} {\bibfnamefont {D.}~\bibnamefont
  {Poilblanc}},\ }\href {https://doi.org/10.1103/PhysRevB.98.184409} {\bibfield
   {journal} {\bibinfo  {journal} {Phys. Rev. B}\ }\textbf {\bibinfo {volume}
  {98}},\ \bibinfo {pages} {184409} (\bibinfo {year} {2018})},\ \Eprint
  {https://arxiv.org/abs/1807.04385} {arXiv:1807.04385 [cond-mat.str-el]}
  \BibitemShut {NoStop}%
\bibitem [{\citenamefont {Tagliacozzo}\ \emph {et~al.}(2014)\citenamefont
  {Tagliacozzo}, \citenamefont {Celi},\ and\ \citenamefont
  {Lewenstein}}]{PhysRevX.4.041024}%
  \BibitemOpen
  \bibfield  {author} {\bibinfo {author} {\bibfnamefont {L.}~\bibnamefont
  {Tagliacozzo}}, \bibinfo {author} {\bibfnamefont {A.}~\bibnamefont {Celi}},\
  and\ \bibinfo {author} {\bibfnamefont {M.}~\bibnamefont {Lewenstein}},\
  }\href {https://doi.org/10.1103/PhysRevX.4.041024} {\bibfield  {journal}
  {\bibinfo  {journal} {Phys. Rev. X}\ }\textbf {\bibinfo {volume} {4}},\
  \bibinfo {pages} {041024} (\bibinfo {year} {2014})}\BibitemShut {NoStop}%
\bibitem [{\citenamefont {Zohar}\ \emph {et~al.}(2016)\citenamefont {Zohar},
  \citenamefont {Wahl}, \citenamefont {Burrello},\ and\ \citenamefont
  {Cirac}}]{ZOHAR201684}%
  \BibitemOpen
  \bibfield  {author} {\bibinfo {author} {\bibfnamefont {E.}~\bibnamefont
  {Zohar}}, \bibinfo {author} {\bibfnamefont {T.~B.}\ \bibnamefont {Wahl}},
  \bibinfo {author} {\bibfnamefont {M.}~\bibnamefont {Burrello}},\ and\
  \bibinfo {author} {\bibfnamefont {J.~I.}\ \bibnamefont {Cirac}},\ }\href
  {https://doi.org/https://doi.org/10.1016/j.aop.2016.08.008} {\bibfield
  {journal} {\bibinfo  {journal} {Annals of Physics}\ }\textbf {\bibinfo
  {volume} {374}},\ \bibinfo {pages} {84} (\bibinfo {year} {2016})}\BibitemShut
  {NoStop}%
\bibitem [{\citenamefont {Corboz}\ \emph {et~al.}(2010)\citenamefont {Corboz},
  \citenamefont {Or\'us}, \citenamefont {Bauer},\ and\ \citenamefont
  {Vidal}}]{tensor_fermions}%
  \BibitemOpen
  \bibfield  {author} {\bibinfo {author} {\bibfnamefont {P.}~\bibnamefont
  {Corboz}}, \bibinfo {author} {\bibfnamefont {R.}~\bibnamefont {Or\'us}},
  \bibinfo {author} {\bibfnamefont {B.}~\bibnamefont {Bauer}},\ and\ \bibinfo
  {author} {\bibfnamefont {G.}~\bibnamefont {Vidal}},\ }\href
  {https://doi.org/10.1103/PhysRevB.81.165104} {\bibfield  {journal} {\bibinfo
  {journal} {Phys. Rev. B}\ }\textbf {\bibinfo {volume} {81}},\ \bibinfo
  {pages} {165104} (\bibinfo {year} {2010})}\BibitemShut {NoStop}%
\bibitem [{\citenamefont {Vanhecke}\ \emph {et~al.}(2021)\citenamefont
  {Vanhecke}, \citenamefont {Verstraete},\ and\ \citenamefont
  {Acoleyen}}]{vanhecke2021entanglement}%
  \BibitemOpen
  \bibfield  {author} {\bibinfo {author} {\bibfnamefont {B.}~\bibnamefont
  {Vanhecke}}, \bibinfo {author} {\bibfnamefont {F.}~\bibnamefont
  {Verstraete}},\ and\ \bibinfo {author} {\bibfnamefont {K.~V.}\ \bibnamefont
  {Acoleyen}},\ }\href@noop {} {\bibinfo {title} {Entanglement scaling for
  $\lambda\phi_2^4$}} (\bibinfo {year} {2021}),\ \Eprint
  {https://arxiv.org/abs/2104.10564} {arXiv:2104.10564 [hep-lat]} \BibitemShut
  {NoStop}%
\bibitem [{\citenamefont {Bañuls}\ \emph {et~al.}(2013)\citenamefont
  {Bañuls}, \citenamefont {Cichy}, \citenamefont {Cirac},\ and\ \citenamefont
  {Jansen}}]{schwinger_spectrum}%
  \BibitemOpen
  \bibfield  {author} {\bibinfo {author} {\bibfnamefont {M.}~\bibnamefont
  {Bañuls}}, \bibinfo {author} {\bibfnamefont {K.}~\bibnamefont {Cichy}},
  \bibinfo {author} {\bibfnamefont {J.}~\bibnamefont {Cirac}},\ and\ \bibinfo
  {author} {\bibfnamefont {K.}~\bibnamefont {Jansen}},\ }\bibfield  {journal}
  {\bibinfo  {journal} {Journal of High Energy Physics}\ }\textbf {\bibinfo
  {volume} {2013}},\ \href {https://doi.org/10.1007/jhep11(2013)158}
  {10.1007/jhep11(2013)158} (\bibinfo {year} {2013})\BibitemShut {NoStop}%
\bibitem [{\citenamefont {Parsons}\ \emph {et~al.}(2016)\citenamefont
  {Parsons}, \citenamefont {Mazurenko}, \citenamefont {Chiu}, \citenamefont
  {Ji}, \citenamefont {Greif},\ and\ \citenamefont {Greiner}}]{Parsons1253}%
  \BibitemOpen
  \bibfield  {author} {\bibinfo {author} {\bibfnamefont {M.~F.}\ \bibnamefont
  {Parsons}}, \bibinfo {author} {\bibfnamefont {A.}~\bibnamefont {Mazurenko}},
  \bibinfo {author} {\bibfnamefont {C.~S.}\ \bibnamefont {Chiu}}, \bibinfo
  {author} {\bibfnamefont {G.}~\bibnamefont {Ji}}, \bibinfo {author}
  {\bibfnamefont {D.}~\bibnamefont {Greif}},\ and\ \bibinfo {author}
  {\bibfnamefont {M.}~\bibnamefont {Greiner}},\ }\href
  {https://doi.org/10.1126/science.aag1430} {\bibfield  {journal} {\bibinfo
  {journal} {Science}\ }\textbf {\bibinfo {volume} {353}},\ \bibinfo {pages}
  {1253} (\bibinfo {year} {2016})}\BibitemShut {NoStop}%
\bibitem [{\citenamefont {Lee}\ \emph {et~al.}(2006)\citenamefont {Lee},
  \citenamefont {Nagaosa},\ and\ \citenamefont {Wen}}]{RevModPhys.78.17}%
  \BibitemOpen
  \bibfield  {author} {\bibinfo {author} {\bibfnamefont {P.~A.}\ \bibnamefont
  {Lee}}, \bibinfo {author} {\bibfnamefont {N.}~\bibnamefont {Nagaosa}},\ and\
  \bibinfo {author} {\bibfnamefont {X.-G.}\ \bibnamefont {Wen}},\ }\href
  {https://doi.org/10.1103/RevModPhys.78.17} {\bibfield  {journal} {\bibinfo
  {journal} {Rev. Mod. Phys.}\ }\textbf {\bibinfo {volume} {78}},\ \bibinfo
  {pages} {17} (\bibinfo {year} {2006})}\BibitemShut {NoStop}%
\bibitem [{\citenamefont {Tasaki}(1998)}]{Tasaki_1998}%
  \BibitemOpen
  \bibfield  {author} {\bibinfo {author} {\bibfnamefont {H.}~\bibnamefont
  {Tasaki}},\ }\href {https://doi.org/10.1088/0953-8984/10/20/004} {\bibfield
  {journal} {\bibinfo  {journal} {Journal of Physics: Condensed Matter}\
  }\textbf {\bibinfo {volume} {10}},\ \bibinfo {pages} {4353} (\bibinfo {year}
  {1998})}\BibitemShut {NoStop}%
\bibitem [{\citenamefont {Arovas}\ \emph {et~al.}(2021)\citenamefont {Arovas},
  \citenamefont {Berg}, \citenamefont {Kivelson},\ and\ \citenamefont
  {Raghu}}]{arovas2021hubbard}%
  \BibitemOpen
  \bibfield  {author} {\bibinfo {author} {\bibfnamefont {D.~P.}\ \bibnamefont
  {Arovas}}, \bibinfo {author} {\bibfnamefont {E.}~\bibnamefont {Berg}},
  \bibinfo {author} {\bibfnamefont {S.}~\bibnamefont {Kivelson}},\ and\
  \bibinfo {author} {\bibfnamefont {S.}~\bibnamefont {Raghu}},\ }\href@noop {}
  {\bibinfo {title} {The hubbard model}} (\bibinfo {year} {2021}),\ \Eprint
  {https://arxiv.org/abs/2103.12097} {arXiv:2103.12097 [cond-mat.str-el]}
  \BibitemShut {NoStop}%
\bibitem [{\citenamefont {Assaad}\ and\ \citenamefont
  {Herbut}(2013)}]{Assaad:2013xua}%
  \BibitemOpen
  \bibfield  {author} {\bibinfo {author} {\bibfnamefont {F.~F.}\ \bibnamefont
  {Assaad}}\ and\ \bibinfo {author} {\bibfnamefont {I.~F.}\ \bibnamefont
  {Herbut}},\ }\href {https://doi.org/10.1103/PhysRevX.3.031010} {\bibfield
  {journal} {\bibinfo  {journal} {Phys. Rev. X}\ }\textbf {\bibinfo {volume}
  {3}},\ \bibinfo {pages} {031010} (\bibinfo {year} {2013})},\ \Eprint
  {https://arxiv.org/abs/1304.6340} {arXiv:1304.6340 [cond-mat.str-el]}
  \BibitemShut {NoStop}%
\bibitem [{\citenamefont {Buividovich}\ \emph {et~al.}(2019)\citenamefont
  {Buividovich}, \citenamefont {Smith}, \citenamefont {Ulybyshev},\ and\
  \citenamefont {von Smekal}}]{Buividovich:2018crq}%
  \BibitemOpen
  \bibfield  {author} {\bibinfo {author} {\bibfnamefont {P.}~\bibnamefont
  {Buividovich}}, \bibinfo {author} {\bibfnamefont {D.}~\bibnamefont {Smith}},
  \bibinfo {author} {\bibfnamefont {M.}~\bibnamefont {Ulybyshev}},\ and\
  \bibinfo {author} {\bibfnamefont {L.}~\bibnamefont {von Smekal}},\ }\href
  {https://doi.org/10.1103/PhysRevB.99.205434} {\bibfield  {journal} {\bibinfo
  {journal} {Phys. Rev. B}\ }\textbf {\bibinfo {volume} {99}},\ \bibinfo
  {pages} {205434} (\bibinfo {year} {2019})},\ \Eprint
  {https://arxiv.org/abs/1812.06435} {arXiv:1812.06435 [cond-mat.str-el]}
  \BibitemShut {NoStop}%
\bibitem [{\citenamefont {Ostmeyer}\ \emph {et~al.}(2020)\citenamefont
  {Ostmeyer}, \citenamefont {Berkowitz}, \citenamefont {Krieg}, \citenamefont
  {L\"ahde}, \citenamefont {Luu},\ and\ \citenamefont
  {Urbach}}]{Ostmeyer:2020uov}%
  \BibitemOpen
  \bibfield  {author} {\bibinfo {author} {\bibfnamefont {J.}~\bibnamefont
  {Ostmeyer}}, \bibinfo {author} {\bibfnamefont {E.}~\bibnamefont {Berkowitz}},
  \bibinfo {author} {\bibfnamefont {S.}~\bibnamefont {Krieg}}, \bibinfo
  {author} {\bibfnamefont {T.~A.}\ \bibnamefont {L\"ahde}}, \bibinfo {author}
  {\bibfnamefont {T.}~\bibnamefont {Luu}},\ and\ \bibinfo {author}
  {\bibfnamefont {C.}~\bibnamefont {Urbach}},\ }\href
  {https://doi.org/10.1103/PhysRevB.102.245105} {\bibfield  {journal} {\bibinfo
   {journal} {Phys. Rev. B}\ }\textbf {\bibinfo {volume} {102}},\ \bibinfo
  {pages} {245105} (\bibinfo {year} {2020})},\ \Eprint
  {https://arxiv.org/abs/2005.11112} {arXiv:2005.11112 [cond-mat.str-el]}
  \BibitemShut {NoStop}%
\bibitem [{\citenamefont {Corboz}(2016)}]{iPEPS_hubbard_2016}%
  \BibitemOpen
  \bibfield  {author} {\bibinfo {author} {\bibfnamefont {P.}~\bibnamefont
  {Corboz}},\ }\bibfield  {journal} {\bibinfo  {journal} {Physical Review B}\
  }\textbf {\bibinfo {volume} {93}},\ \href
  {https://doi.org/10.1103/physrevb.93.045116} {10.1103/physrevb.93.045116}
  (\bibinfo {year} {2016})\BibitemShut {NoStop}%
\bibitem [{\citenamefont {Geim}\ and\ \citenamefont
  {Novoselov}(2007)}]{Novoselov2007}%
  \BibitemOpen
  \bibfield  {author} {\bibinfo {author} {\bibfnamefont {A.~K.}\ \bibnamefont
  {Geim}}\ and\ \bibinfo {author} {\bibfnamefont {K.~S.}\ \bibnamefont
  {Novoselov}},\ }\href {https://doi.org/10.1038/nmat1849} {\bibfield
  {journal} {\bibinfo  {journal} {Nature Materials}\ }\textbf {\bibinfo
  {volume} {6}},\ \bibinfo {pages} {183} (\bibinfo {year} {2007})}\BibitemShut
  {NoStop}%
\bibitem [{\citenamefont {Castro~Neto}\ \emph {et~al.}(2009)\citenamefont
  {Castro~Neto}, \citenamefont {Guinea}, \citenamefont {Peres}, \citenamefont
  {Novoselov},\ and\ \citenamefont {Geim}}]{CastroNeto:2009zz}%
  \BibitemOpen
  \bibfield  {author} {\bibinfo {author} {\bibfnamefont {A.~H.}\ \bibnamefont
  {Castro~Neto}}, \bibinfo {author} {\bibfnamefont {F.}~\bibnamefont {Guinea}},
  \bibinfo {author} {\bibfnamefont {N.~M.~R.}\ \bibnamefont {Peres}}, \bibinfo
  {author} {\bibfnamefont {K.~S.}\ \bibnamefont {Novoselov}},\ and\ \bibinfo
  {author} {\bibfnamefont {A.~K.}\ \bibnamefont {Geim}},\ }\href
  {https://doi.org/10.1103/RevModPhys.81.109} {\bibfield  {journal} {\bibinfo
  {journal} {Rev. Mod. Phys.}\ }\textbf {\bibinfo {volume} {81}},\ \bibinfo
  {pages} {109} (\bibinfo {year} {2009})},\ \Eprint
  {https://arxiv.org/abs/0709.1163} {arXiv:0709.1163 [cond-mat.other]}
  \BibitemShut {NoStop}%
\bibitem [{\citenamefont {{Verstraete}}\ and\ \citenamefont
  {{Cirac}}(2004{\natexlab{a}})}]{PEPS_original_bMPS}%
  \BibitemOpen
  \bibfield  {author} {\bibinfo {author} {\bibfnamefont {F.}~\bibnamefont
  {{Verstraete}}}\ and\ \bibinfo {author} {\bibfnamefont {J.~I.}\ \bibnamefont
  {{Cirac}}},\ }\href@noop {} {\bibfield  {journal} {\bibinfo  {journal} {arXiv
  e-prints}\ ,\ \bibinfo {eid} {cond-mat/0407066}} (\bibinfo {year}
  {2004}{\natexlab{a}})},\ \Eprint {https://arxiv.org/abs/cond-mat/0407066}
  {arXiv:cond-mat/0407066 [cond-mat.str-el]} \BibitemShut {NoStop}%
\bibitem [{\citenamefont {{Verstraete}}\ and\ \citenamefont
  {{Cirac}}(2004{\natexlab{b}})}]{PEPS_original_2}%
  \BibitemOpen
  \bibfield  {author} {\bibinfo {author} {\bibfnamefont {F.}~\bibnamefont
  {{Verstraete}}}\ and\ \bibinfo {author} {\bibfnamefont {J.~I.}\ \bibnamefont
  {{Cirac}}},\ }\href {https://doi.org/10.1103/PhysRevA.70.060302} {\bibfield
  {journal} {\bibinfo  {journal} {\pra}\ }\textbf {\bibinfo {volume} {70}},\
  \bibinfo {eid} {060302} (\bibinfo {year} {2004}{\natexlab{b}})},\ \Eprint
  {https://arxiv.org/abs/quant-ph/0311130} {arXiv:quant-ph/0311130 [quant-ph]}
  \BibitemShut {NoStop}%
\bibitem [{\citenamefont {{Hastings}}(2007)}]{areaLaw_original}%
  \BibitemOpen
  \bibfield  {author} {\bibinfo {author} {\bibfnamefont {M.~B.}\ \bibnamefont
  {{Hastings}}},\ }\href {https://doi.org/10.1088/1742-5468/2007/08/P08024}
  {\bibfield  {journal} {\bibinfo  {journal} {Journal of Statistical Mechanics:
  Theory and Experiment}\ }\textbf {\bibinfo {volume} {2007}},\ \bibinfo
  {pages} {08024} (\bibinfo {year} {2007})},\ \Eprint
  {https://arxiv.org/abs/0705.2024} {arXiv:0705.2024 [quant-ph]} \BibitemShut
  {NoStop}%
\bibitem [{\citenamefont {{Eisert}}\ \emph {et~al.}(2010)\citenamefont
  {{Eisert}}, \citenamefont {{Cramer}},\ and\ \citenamefont
  {{Plenio}}}]{areaLaw_overview}%
  \BibitemOpen
  \bibfield  {author} {\bibinfo {author} {\bibfnamefont {J.}~\bibnamefont
  {{Eisert}}}, \bibinfo {author} {\bibfnamefont {M.}~\bibnamefont {{Cramer}}},\
  and\ \bibinfo {author} {\bibfnamefont {M.~B.}\ \bibnamefont {{Plenio}}},\
  }\href {https://doi.org/10.1103/RevModPhys.82.277} {\bibfield  {journal}
  {\bibinfo  {journal} {Reviews of Modern Physics}\ }\textbf {\bibinfo {volume}
  {82}},\ \bibinfo {pages} {277} (\bibinfo {year} {2010})},\ \Eprint
  {https://arxiv.org/abs/0808.3773} {arXiv:0808.3773 [quant-ph]} \BibitemShut
  {NoStop}%
\bibitem [{\citenamefont {{Sornborger}}\ and\ \citenamefont
  {{Stewart}}(1998)}]{SuzukiTrotter_original}%
  \BibitemOpen
  \bibfield  {author} {\bibinfo {author} {\bibfnamefont {A.~T.}\ \bibnamefont
  {{Sornborger}}}\ and\ \bibinfo {author} {\bibfnamefont {E.~D.}\ \bibnamefont
  {{Stewart}}},\ }\href@noop {} {\bibfield  {journal} {\bibinfo  {journal}
  {arXiv e-prints}\ ,\ \bibinfo {eid} {quant-ph/9809009}} (\bibinfo {year}
  {1998})},\ \Eprint {https://arxiv.org/abs/quant-ph/9809009}
  {arXiv:quant-ph/9809009 [quant-ph]} \BibitemShut {NoStop}%
\bibitem [{\citenamefont {{Vidal}}(2007)}]{SuzukiTrotter_MPS}%
  \BibitemOpen
  \bibfield  {author} {\bibinfo {author} {\bibfnamefont {G.}~\bibnamefont
  {{Vidal}}},\ }\href {https://doi.org/10.1103/PhysRevLett.98.070201}
  {\bibfield  {journal} {\bibinfo  {journal} {\prl}\ }\textbf {\bibinfo
  {volume} {98}},\ \bibinfo {eid} {070201} (\bibinfo {year} {2007})},\ \Eprint
  {https://arxiv.org/abs/cond-mat/0605597} {arXiv:cond-mat/0605597
  [cond-mat.str-el]} \BibitemShut {NoStop}%
\bibitem [{\citenamefont {Corboz}\ and\ \citenamefont
  {Vidal}(2009)}]{tensor_fermions_derivation}%
  \BibitemOpen
  \bibfield  {author} {\bibinfo {author} {\bibfnamefont {P.}~\bibnamefont
  {Corboz}}\ and\ \bibinfo {author} {\bibfnamefont {G.}~\bibnamefont {Vidal}},\
  }\bibfield  {journal} {\bibinfo  {journal} {Physical Review B}\ }\textbf
  {\bibinfo {volume} {80}},\ \href {https://doi.org/10.1103/physrevb.80.165129}
  {10.1103/physrevb.80.165129} (\bibinfo {year} {2009})\BibitemShut {NoStop}%
\bibitem [{\citenamefont {{Jiang}}\ \emph {et~al.}(2008)\citenamefont
  {{Jiang}}, \citenamefont {{Weng}},\ and\ \citenamefont
  {{Xiang}}}]{SimpleUpdate_original}%
  \BibitemOpen
  \bibfield  {author} {\bibinfo {author} {\bibfnamefont {H.~C.}\ \bibnamefont
  {{Jiang}}}, \bibinfo {author} {\bibfnamefont {Z.~Y.}\ \bibnamefont
  {{Weng}}},\ and\ \bibinfo {author} {\bibfnamefont {T.}~\bibnamefont
  {{Xiang}}},\ }\href {https://doi.org/10.1103/PhysRevLett.101.090603}
  {\bibfield  {journal} {\bibinfo  {journal} {\prl}\ }\textbf {\bibinfo
  {volume} {101}},\ \bibinfo {eid} {090603} (\bibinfo {year} {2008})},\ \Eprint
  {https://arxiv.org/abs/0806.3719} {arXiv:0806.3719 [cond-mat.str-el]}
  \BibitemShut {NoStop}%
\bibitem [{\citenamefont {{Lubasch}}\ \emph
  {et~al.}(2014{\natexlab{a}})\citenamefont {{Lubasch}}, \citenamefont
  {{Cirac}},\ and\ \citenamefont {{Ba{\~n}uls}}}]{Lubasch_clusterUpdate}%
  \BibitemOpen
  \bibfield  {author} {\bibinfo {author} {\bibfnamefont {M.}~\bibnamefont
  {{Lubasch}}}, \bibinfo {author} {\bibfnamefont {J.~I.}\ \bibnamefont
  {{Cirac}}},\ and\ \bibinfo {author} {\bibfnamefont {M.-C.}\ \bibnamefont
  {{Ba{\~n}uls}}},\ }\href {https://doi.org/10.1088/1367-2630/16/3/033014}
  {\bibfield  {journal} {\bibinfo  {journal} {New Journal of Physics}\ }\textbf
  {\bibinfo {volume} {16}},\ \bibinfo {eid} {033014} (\bibinfo {year}
  {2014}{\natexlab{a}})},\ \Eprint {https://arxiv.org/abs/1311.6696}
  {arXiv:1311.6696 [quant-ph]} \BibitemShut {NoStop}%
\bibitem [{\citenamefont {{Bruognolo}}\ \emph {et~al.}(2020)\citenamefont
  {{Bruognolo}}, \citenamefont {{Li}}, \citenamefont {{von Delft}},\ and\
  \citenamefont {{Weichselbaum}}}]{iPEPS_introduction}%
  \BibitemOpen
  \bibfield  {author} {\bibinfo {author} {\bibfnamefont {B.}~\bibnamefont
  {{Bruognolo}}}, \bibinfo {author} {\bibfnamefont {J.-W.}\ \bibnamefont
  {{Li}}}, \bibinfo {author} {\bibfnamefont {J.}~\bibnamefont {{von Delft}}},\
  and\ \bibinfo {author} {\bibfnamefont {A.}~\bibnamefont {{Weichselbaum}}},\
  }\href@noop {} {\bibfield  {journal} {\bibinfo  {journal} {arXiv e-prints}\
  ,\ \bibinfo {eid} {arXiv:2006.08289}} (\bibinfo {year} {2020})},\ \Eprint
  {https://arxiv.org/abs/2006.08289} {arXiv:2006.08289 [cond-mat.str-el]}
  \BibitemShut {NoStop}%
\bibitem [{\citenamefont {Jordan}\ \emph {et~al.}(2008)\citenamefont {Jordan},
  \citenamefont {Or\'us}, \citenamefont {Vidal}, \citenamefont {Verstraete},\
  and\ \citenamefont {Cirac}}]{iPEPS_original_2008}%
  \BibitemOpen
  \bibfield  {author} {\bibinfo {author} {\bibfnamefont {J.}~\bibnamefont
  {Jordan}}, \bibinfo {author} {\bibfnamefont {R.}~\bibnamefont {Or\'us}},
  \bibinfo {author} {\bibfnamefont {G.}~\bibnamefont {Vidal}}, \bibinfo
  {author} {\bibfnamefont {F.}~\bibnamefont {Verstraete}},\ and\ \bibinfo
  {author} {\bibfnamefont {J.~I.}\ \bibnamefont {Cirac}},\ }\href
  {https://doi.org/10.1103/PhysRevLett.101.250602} {\bibfield  {journal}
  {\bibinfo  {journal} {Phys. Rev. Lett.}\ }\textbf {\bibinfo {volume} {101}},\
  \bibinfo {pages} {250602} (\bibinfo {year} {2008})}\BibitemShut {NoStop}%
\bibitem [{\citenamefont {{Lubasch}}\ \emph
  {et~al.}(2014{\natexlab{b}})\citenamefont {{Lubasch}}, \citenamefont
  {{Cirac}},\ and\ \citenamefont {{Ba{\~n}uls}}}]{Lubasch_algorithms}%
  \BibitemOpen
  \bibfield  {author} {\bibinfo {author} {\bibfnamefont {M.}~\bibnamefont
  {{Lubasch}}}, \bibinfo {author} {\bibfnamefont {J.~I.}\ \bibnamefont
  {{Cirac}}},\ and\ \bibinfo {author} {\bibfnamefont {M.-C.}\ \bibnamefont
  {{Ba{\~n}uls}}},\ }\href {https://doi.org/10.1103/PhysRevB.90.064425}
  {\bibfield  {journal} {\bibinfo  {journal} {\prb}\ }\textbf {\bibinfo
  {volume} {90}},\ \bibinfo {eid} {064425} (\bibinfo {year}
  {2014}{\natexlab{b}})},\ \Eprint {https://arxiv.org/abs/1405.3259}
  {arXiv:1405.3259 [quant-ph]} \BibitemShut {NoStop}%
\bibitem [{\citenamefont {{Verstraete}}\ \emph {et~al.}(2008)\citenamefont
  {{Verstraete}}, \citenamefont {{Murg}},\ and\ \citenamefont
  {{Cirac}}}]{Verstraete_introduction}%
  \BibitemOpen
  \bibfield  {author} {\bibinfo {author} {\bibfnamefont {F.}~\bibnamefont
  {{Verstraete}}}, \bibinfo {author} {\bibfnamefont {V.}~\bibnamefont
  {{Murg}}},\ and\ \bibinfo {author} {\bibfnamefont {J.~I.}\ \bibnamefont
  {{Cirac}}},\ }\href {https://doi.org/10.1080/14789940801912366} {\bibfield
  {journal} {\bibinfo  {journal} {Advances in Physics}\ }\textbf {\bibinfo
  {volume} {57}},\ \bibinfo {pages} {143} (\bibinfo {year} {2008})},\ \Eprint
  {https://arxiv.org/abs/0907.2796} {arXiv:0907.2796 [quant-ph]} \BibitemShut
  {NoStop}%
\bibitem [{\citenamefont {{Zhao}}\ \emph {et~al.}(2010)\citenamefont {{Zhao}},
  \citenamefont {{Xie}}, \citenamefont {{Chen}}, \citenamefont {{Wei}},
  \citenamefont {{Cai}},\ and\ \citenamefont {{Xiang}}}]{SRG_original}%
  \BibitemOpen
  \bibfield  {author} {\bibinfo {author} {\bibfnamefont {H.~H.}\ \bibnamefont
  {{Zhao}}}, \bibinfo {author} {\bibfnamefont {Z.~Y.}\ \bibnamefont {{Xie}}},
  \bibinfo {author} {\bibfnamefont {Q.~N.}\ \bibnamefont {{Chen}}}, \bibinfo
  {author} {\bibfnamefont {Z.~C.}\ \bibnamefont {{Wei}}}, \bibinfo {author}
  {\bibfnamefont {J.~W.}\ \bibnamefont {{Cai}}},\ and\ \bibinfo {author}
  {\bibfnamefont {T.}~\bibnamefont {{Xiang}}},\ }\href
  {https://doi.org/10.1103/PhysRevB.81.174411} {\bibfield  {journal} {\bibinfo
  {journal} {Physical Review B}\ }\textbf {\bibinfo {volume} {81}},\ \bibinfo
  {eid} {174411} (\bibinfo {year} {2010})},\ \Eprint
  {https://arxiv.org/abs/1002.1405} {arXiv:1002.1405 [cond-mat.str-el]}
  \BibitemShut {NoStop}%
\bibitem [{MAT(2020)}]{MATLAB}%
  \BibitemOpen
  \href@noop {} {\emph {\bibinfo {title} {{MATLAB versions 9.8+
  (R2020a-21a)}}}},\ \bibinfo {organization} {The Mathworks, Inc.},\ \bibinfo
  {address} {Natick, Massachusetts} (\bibinfo {year} {2020})\BibitemShut
  {NoStop}%
\bibitem [{\citenamefont {Pfeifer}\ \emph {et~al.}(2015)\citenamefont
  {Pfeifer}, \citenamefont {Evenbly}, \citenamefont {Singh},\ and\
  \citenamefont {Vidal}}]{pfeifer2015ncon}%
  \BibitemOpen
  \bibfield  {author} {\bibinfo {author} {\bibfnamefont {R.~N.~C.}\
  \bibnamefont {Pfeifer}}, \bibinfo {author} {\bibfnamefont {G.}~\bibnamefont
  {Evenbly}}, \bibinfo {author} {\bibfnamefont {S.}~\bibnamefont {Singh}},\
  and\ \bibinfo {author} {\bibfnamefont {G.}~\bibnamefont {Vidal}},\
  }\href@noop {} {\bibinfo {title} {{NCON: A tensor network contractor for
  MATLAB}}} (\bibinfo {year} {2015}),\ \Eprint
  {https://arxiv.org/abs/1402.0939} {arXiv:1402.0939 [physics.comp-ph]}
  \BibitemShut {NoStop}%
\bibitem [{\citenamefont {{Murg}}\ \emph {et~al.}(2007)\citenamefont {{Murg}},
  \citenamefont {{Verstraete}},\ and\ \citenamefont {{Cirac}}}]{bMPS}%
  \BibitemOpen
  \bibfield  {author} {\bibinfo {author} {\bibfnamefont {V.}~\bibnamefont
  {{Murg}}}, \bibinfo {author} {\bibfnamefont {F.}~\bibnamefont
  {{Verstraete}}},\ and\ \bibinfo {author} {\bibfnamefont {J.~I.}\ \bibnamefont
  {{Cirac}}},\ }\href {https://doi.org/10.1103/PhysRevA.75.033605} {\bibfield
  {journal} {\bibinfo  {journal} {\pra}\ }\textbf {\bibinfo {volume} {75}},\
  \bibinfo {eid} {033605} (\bibinfo {year} {2007})},\ \Eprint
  {https://arxiv.org/abs/cond-mat/0611522} {arXiv:cond-mat/0611522
  [cond-mat.other]} \BibitemShut {NoStop}%
\bibitem [{\citenamefont {Corboz}\ \emph {et~al.}(2018)\citenamefont {Corboz},
  \citenamefont {Czarnik}, \citenamefont {Kapteijns},\ and\ \citenamefont
  {Tagliacozzo}}]{iPEPS_hubbard_2018}%
  \BibitemOpen
  \bibfield  {author} {\bibinfo {author} {\bibfnamefont {P.}~\bibnamefont
  {Corboz}}, \bibinfo {author} {\bibfnamefont {P.}~\bibnamefont {Czarnik}},
  \bibinfo {author} {\bibfnamefont {G.}~\bibnamefont {Kapteijns}},\ and\
  \bibinfo {author} {\bibfnamefont {L.}~\bibnamefont {Tagliacozzo}},\
  }\bibfield  {journal} {\bibinfo  {journal} {Physical Review X}\ }\textbf
  {\bibinfo {volume} {8}},\ \href {https://doi.org/10.1103/physrevx.8.031031}
  {10.1103/physrevx.8.031031} (\bibinfo {year} {2018})\BibitemShut {NoStop}%
\bibitem [{\citenamefont {Cristoforetti}\ \emph {et~al.}(2013)\citenamefont
  {Cristoforetti}, \citenamefont {Di~Renzo}, \citenamefont {Mukherjee},\ and\
  \citenamefont {Scorzato}}]{Cristoforetti:2013wha}%
  \BibitemOpen
  \bibfield  {author} {\bibinfo {author} {\bibfnamefont {M.}~\bibnamefont
  {Cristoforetti}}, \bibinfo {author} {\bibfnamefont {F.}~\bibnamefont
  {Di~Renzo}}, \bibinfo {author} {\bibfnamefont {A.}~\bibnamefont
  {Mukherjee}},\ and\ \bibinfo {author} {\bibfnamefont {L.}~\bibnamefont
  {Scorzato}},\ }\href {https://doi.org/10.1103/PhysRevD.88.051501} {\bibfield
  {journal} {\bibinfo  {journal} {Phys. Rev. D}\ }\textbf {\bibinfo {volume}
  {88}},\ \bibinfo {pages} {051501} (\bibinfo {year} {2013})},\ \Eprint
  {https://arxiv.org/abs/1303.7204} {arXiv:1303.7204 [hep-lat]} \BibitemShut
  {NoStop}%
\bibitem [{\citenamefont {Cristoforetti}\ \emph {et~al.}(2014)\citenamefont
  {Cristoforetti}, \citenamefont {Di~Renzo}, \citenamefont {Eruzzi},
  \citenamefont {Mukherjee}, \citenamefont {Schmidt}, \citenamefont
  {Scorzato},\ and\ \citenamefont {Torrero}}]{Cristoforetti:2014gsa}%
  \BibitemOpen
  \bibfield  {author} {\bibinfo {author} {\bibfnamefont {M.}~\bibnamefont
  {Cristoforetti}}, \bibinfo {author} {\bibfnamefont {F.}~\bibnamefont
  {Di~Renzo}}, \bibinfo {author} {\bibfnamefont {G.}~\bibnamefont {Eruzzi}},
  \bibinfo {author} {\bibfnamefont {A.}~\bibnamefont {Mukherjee}}, \bibinfo
  {author} {\bibfnamefont {C.}~\bibnamefont {Schmidt}}, \bibinfo {author}
  {\bibfnamefont {L.}~\bibnamefont {Scorzato}},\ and\ \bibinfo {author}
  {\bibfnamefont {C.}~\bibnamefont {Torrero}},\ }\href
  {https://doi.org/10.1103/PhysRevD.89.114505} {\bibfield  {journal} {\bibinfo
  {journal} {Phys. Rev. D}\ }\textbf {\bibinfo {volume} {89}},\ \bibinfo
  {pages} {114505} (\bibinfo {year} {2014})},\ \Eprint
  {https://arxiv.org/abs/1403.5637} {arXiv:1403.5637 [hep-lat]} \BibitemShut
  {NoStop}%
\bibitem [{\citenamefont {Ulybyshev}\ \emph {et~al.}(2020)\citenamefont
  {Ulybyshev}, \citenamefont {Winterowd},\ and\ \citenamefont
  {Zafeiropoulos}}]{Ulybyshev:2019fte}%
  \BibitemOpen
  \bibfield  {author} {\bibinfo {author} {\bibfnamefont {M.}~\bibnamefont
  {Ulybyshev}}, \bibinfo {author} {\bibfnamefont {C.}~\bibnamefont
  {Winterowd}},\ and\ \bibinfo {author} {\bibfnamefont {S.}~\bibnamefont
  {Zafeiropoulos}},\ }\href {https://doi.org/10.1103/PhysRevD.101.014508}
  {\bibfield  {journal} {\bibinfo  {journal} {Phys. Rev. D}\ }\textbf {\bibinfo
  {volume} {101}},\ \bibinfo {pages} {014508} (\bibinfo {year} {2020})},\
  \Eprint {https://arxiv.org/abs/1906.07678} {arXiv:1906.07678
  [cond-mat.str-el]} \BibitemShut {NoStop}%
\bibitem [{\citenamefont {{Alexandru}}\ \emph {et~al.}(2016)\citenamefont
  {{Alexandru}}, \citenamefont {{Basar}}, \citenamefont {{Bedaque}},
  \citenamefont {{Ridgway}},\ and\ \citenamefont
  {{Warrington}}}]{Alexandru:2015sua}%
  \BibitemOpen
  \bibfield  {author} {\bibinfo {author} {\bibfnamefont {A.}~\bibnamefont
  {{Alexandru}}}, \bibinfo {author} {\bibfnamefont {G.}~\bibnamefont
  {{Basar}}}, \bibinfo {author} {\bibfnamefont {P.~F.}\ \bibnamefont
  {{Bedaque}}}, \bibinfo {author} {\bibfnamefont {G.~W.}\ \bibnamefont
  {{Ridgway}}},\ and\ \bibinfo {author} {\bibfnamefont {N.~C.}\ \bibnamefont
  {{Warrington}}},\ }\href {https://doi.org/10.1007/JHEP05(2016)053} {\bibfield
   {journal} {\bibinfo  {journal} {Journal of High Energy Physics}\ }\textbf
  {\bibinfo {volume} {2016}},\ \bibinfo {eid} {53} (\bibinfo {year}
  {2016})}\BibitemShut {NoStop}%
\bibitem [{\citenamefont {Alexandru}\ \emph {et~al.}(2016)\citenamefont
  {Alexandru}, \citenamefont {Basar},\ and\ \citenamefont
  {Bedaque}}]{Alexandru:2015xva}%
  \BibitemOpen
  \bibfield  {author} {\bibinfo {author} {\bibfnamefont {A.}~\bibnamefont
  {Alexandru}}, \bibinfo {author} {\bibfnamefont {G.}~\bibnamefont {Basar}},\
  and\ \bibinfo {author} {\bibfnamefont {P.}~\bibnamefont {Bedaque}},\ }\href
  {https://doi.org/10.1103/PhysRevD.93.014504} {\bibfield  {journal} {\bibinfo
  {journal} {Phys. Rev. D}\ }\textbf {\bibinfo {volume} {93}},\ \bibinfo
  {pages} {014504} (\bibinfo {year} {2016})},\ \Eprint
  {https://arxiv.org/abs/1510.03258} {arXiv:1510.03258 [hep-lat]} \BibitemShut
  {NoStop}%
\bibitem [{\citenamefont {Wynen}\ \emph {et~al.}(2021)\citenamefont {Wynen},
  \citenamefont {Berkowitz}, \citenamefont {Krieg}, \citenamefont {Luu},\ and\
  \citenamefont {Ostmeyer}}]{leveragingML}%
  \BibitemOpen
  \bibfield  {author} {\bibinfo {author} {\bibfnamefont {J.-L.}\ \bibnamefont
  {Wynen}}, \bibinfo {author} {\bibfnamefont {E.}~\bibnamefont {Berkowitz}},
  \bibinfo {author} {\bibfnamefont {S.}~\bibnamefont {Krieg}}, \bibinfo
  {author} {\bibfnamefont {T.}~\bibnamefont {Luu}},\ and\ \bibinfo {author}
  {\bibfnamefont {J.}~\bibnamefont {Ostmeyer}},\ }\href
  {https://doi.org/10.1103/PhysRevB.103.125153} {\bibfield  {journal} {\bibinfo
   {journal} {Phys. Rev. B}\ }\textbf {\bibinfo {volume} {103}},\ \bibinfo
  {pages} {125153} (\bibinfo {year} {2021})}\BibitemShut {NoStop}%
\bibitem [{\citenamefont {Rams}\ \emph {et~al.}(2018)\citenamefont {Rams},
  \citenamefont {Czarnik},\ and\ \citenamefont
  {Cincio}}]{precise_extrapolation}%
  \BibitemOpen
  \bibfield  {author} {\bibinfo {author} {\bibfnamefont {M.~M.}\ \bibnamefont
  {Rams}}, \bibinfo {author} {\bibfnamefont {P.}~\bibnamefont {Czarnik}},\ and\
  \bibinfo {author} {\bibfnamefont {L.}~\bibnamefont {Cincio}},\ }\bibfield
  {journal} {\bibinfo  {journal} {Physical Review X}\ }\textbf {\bibinfo
  {volume} {8}},\ \href {https://doi.org/10.1103/physrevx.8.041033}
  {10.1103/physrevx.8.041033} (\bibinfo {year} {2018})\BibitemShut {NoStop}%
\bibitem [{\citenamefont {Levenberg}(1944)}]{levenberg}%
  \BibitemOpen
  \bibfield  {author} {\bibinfo {author} {\bibfnamefont {K.}~\bibnamefont
  {Levenberg}},\ }\href {http://www.jstor.org/stable/43633451} {\bibfield
  {journal} {\bibinfo  {journal} {Quarterly of Applied Mathematics}\ }\textbf
  {\bibinfo {volume} {2}},\ \bibinfo {pages} {164} (\bibinfo {year}
  {1944})}\BibitemShut {NoStop}%
\bibitem [{\citenamefont {Marquardt}(1963)}]{marquardt}%
  \BibitemOpen
  \bibfield  {author} {\bibinfo {author} {\bibfnamefont {D.~W.}\ \bibnamefont
  {Marquardt}},\ }\href {https://doi.org/10.1137/0111030} {\bibfield  {journal}
  {\bibinfo  {journal} {Journal of the Society for Industrial and Applied
  Mathematics}\ }\textbf {\bibinfo {volume} {11}},\ \bibinfo {pages} {431}
  (\bibinfo {year} {1963})}\BibitemShut {NoStop}%
\bibitem [{lm_(2019)}]{lm_explained_gsl}%
  \BibitemOpen
  \href {https://www.gnu.org/software/gsl/doc/html/nls.html} {\emph {\bibinfo
  {title} {{Nonlinear Least-Squares Fitting}}}},\ \bibinfo {organization} {GNU
  Scientific Library},\ \bibinfo {address}
  {www.gnu.org/software/gsl/doc/html/nls.html} (\bibinfo {year}
  {2019})\BibitemShut {NoStop}%
\bibitem [{\citenamefont {Xie}\ \emph {et~al.}(2012)\citenamefont {Xie},
  \citenamefont {Chen}, \citenamefont {Qin}, \citenamefont {Zhu}, \citenamefont
  {Yang},\ and\ \citenamefont {Xiang}}]{HOTRG_original}%
  \BibitemOpen
  \bibfield  {author} {\bibinfo {author} {\bibfnamefont {Z.~Y.}\ \bibnamefont
  {Xie}}, \bibinfo {author} {\bibfnamefont {J.}~\bibnamefont {Chen}}, \bibinfo
  {author} {\bibfnamefont {M.~P.}\ \bibnamefont {Qin}}, \bibinfo {author}
  {\bibfnamefont {J.~W.}\ \bibnamefont {Zhu}}, \bibinfo {author} {\bibfnamefont
  {L.~P.}\ \bibnamefont {Yang}},\ and\ \bibinfo {author} {\bibfnamefont
  {T.}~\bibnamefont {Xiang}},\ }\bibfield  {journal} {\bibinfo  {journal}
  {Physical Review B}\ }\textbf {\bibinfo {volume} {86}},\ \href
  {https://doi.org/10.1103/physrevb.86.045139} {10.1103/physrevb.86.045139}
  (\bibinfo {year} {2012})\BibitemShut {NoStop}%
\bibitem [{\citenamefont {Kadoh}\ and\ \citenamefont
  {Nakayama}(2019)}]{triad_renormalization}%
  \BibitemOpen
  \bibfield  {author} {\bibinfo {author} {\bibfnamefont {D.}~\bibnamefont
  {Kadoh}}\ and\ \bibinfo {author} {\bibfnamefont {K.}~\bibnamefont
  {Nakayama}},\ }\href@noop {} {\bibinfo {title} {Renormalization group on a
  triad network}} (\bibinfo {year} {2019}),\ \Eprint
  {https://arxiv.org/abs/1912.02414} {arXiv:1912.02414 [hep-lat]} \BibitemShut
  {NoStop}%
\bibitem [{\citenamefont {Ostmeyer}\ \emph {et~al.}(2021)\citenamefont
  {Ostmeyer}, \citenamefont {Berkowitz}, \citenamefont {Krieg}, \citenamefont
  {Lähde}, \citenamefont {Luu},\ and\ \citenamefont
  {Urbach}}]{more_observables}%
  \BibitemOpen
  \bibfield  {author} {\bibinfo {author} {\bibfnamefont {J.}~\bibnamefont
  {Ostmeyer}}, \bibinfo {author} {\bibfnamefont {E.}~\bibnamefont {Berkowitz}},
  \bibinfo {author} {\bibfnamefont {S.}~\bibnamefont {Krieg}}, \bibinfo
  {author} {\bibfnamefont {T.~A.}\ \bibnamefont {Lähde}}, \bibinfo {author}
  {\bibfnamefont {T.}~\bibnamefont {Luu}},\ and\ \bibinfo {author}
  {\bibfnamefont {C.}~\bibnamefont {Urbach}},\ }\href@noop {} {\bibinfo {title}
  {{The Antiferromagnetic Character of the Quantum Phase Transition in the
  Hubbard Model on the Honeycomb Lattice}}} (\bibinfo {year} {2021}),\ \Eprint
  {https://arxiv.org/abs/2105.06936} {arXiv:2105.06936 [cond-mat.str-el]}
  \BibitemShut {NoStop}%
\end{thebibliography}%

\end{document}